\pdfoutput=1 %arXiv admin added this line
\documentclass[12pt,twoside]{article}
\usepackage{etex,amsmath, amssymb, amsthm, mathrsfs, chemarrow, amsfonts, bbm, ctable,graphicx,color,epsfig,citesort, stmaryrd,textpos, upgreek, xy, extarrows,protosem,fixltx2e,fancybox}

\usepackage[text={7in,9in},centering,includehead,includefoot,headheight=14.5pt]{geometry}
\usepackage[version=3,arrows=pgf-filled]{mhchem} 
\usepackage[all,cmtip]{xypic}

\usepackage[abs]{overpic}

\usepackage{fancyhdr}
\renewcommand{\subsectionmark}[1]{}
\newlength\Li \newlength\Lii
\date{}

\makeatletter
\def\@seccntformat#1{}
\makeatother
\numberwithin{equation}{section}
\renewcommand{\numberline}[1]{}
 
\makeatletter
\def\blfootnote{\xdef\@thefnmark{}\@footnotetext}
\makeatother
 
\title{Adaptive hierarchic transformations for dynamically $p$-enriched slope-limiting over discontinuous Galerkin systems of generalized equations} 

\author{C.~Michoski\textsuperscript{\htmladdnormallink{\textproto{\AAsade}}{https://webspace.utexas.edu/michoski/Michoski.html}} , \ C.~Mirabito,  \ C.~Dawson, \\  \small{ \it Institute for Computational Engineering and Sciences (ICES), Computational Hydraulics Group (CHG)} \\ \small{\it University of Texas at Austin, Austin, TX, 78712} \\ \\ E.~J.~Kubatko,  \\  \small{ \it Department of Civil and Environmental Enineering and Geodetic Science}  \\ \small{\it The Ohio State University, Columbus, OH, 43210} \\ \\  D.~Wirasaet, \ J.~J.~Westerink \\ \small{\it Computational Hydraulics Laboratory, Department of Civil Engineering and Geological Sciences} \\ \small{\it University of Notre Dame, Notre Dame, IN, 46556}}

\usepackage[square,comma,numbers,sort&compress]{natbib}
\usepackage[pdftex,bookmarks]{hyperref}
\usepackage{hypernat}

\begin{document}
\maketitle
\begin{abstract} 
We study a family of generalized slope limiters in two dimensions for Runge-Kutta discontinuous Galerkin (RKDG) solutions of  advection--diffusion systems.  We analyze the numerical behavior of these limiters applied to a pair of model problems, comparing the error of the approximate solutions, and discuss each limiter's advantages and disadvantages.  We then introduce a series of coupled $p$-enrichment schemes that may be used as standalone dynamic $p$-enrichment strategies, or may be augmented via any in the family of variable-in-$p$ slope limiters presented. 
\end{abstract} 

\blfootnote{\textsuperscript{\htmladdnormallink{\textproto{\AAsade}}{https://webspace.utexas.edu/michoski/Michoski.html}}Corresponding author, {\it michoski@ices.utexas.edu}}

\tableofcontents

\section{\texorpdfstring{\protect\centering $\S 1$ Introduction}{\S 1  Introduction}}

Generally when solving advection-diffusion equations --- which are not strictly diffusion dominated --- by way of, for example, finite element or finite volume techniques, one observes the presence of spurious oscillations in the solution space often brought about by the existence of shocks in the space of approximate solutions as well as from the presence of sharp and/or discontinuous profiles in the physical domain itself.   Such ill-behaved approximate solutions have led to the development of numerous methods designed with the intent to consistently stabilize and ``limit'' the solution in order to deal with these oscillations, as they are seen to arise quite frequently in common scientific applications.  For example, slope limiters are known to be of central importance in storm surge modeling \cite{BKWD,MGB} in order to obtain, for example, well-behaved solutions in the presence of complicated free--boundary conditions along adapting shorelines.  Limiting regimes are also of substantial importance in quantum hydrodynamic systems \cite{MESV2,BalSac} and surface wave models \cite{LSY} where they are used to reduce the oscillations caused by mathematical dispersion terms (\emph{i.e.} nonlinear third order spatial derivative terms) that pervade, for example, tunneling solutions.  In fact, slope limiters are of fundamental importance in most applications in standard fluid dynamics, being employed commonly in compressible Navier--Stokes \cite{MESV}, Eulers\cite{WP}, and magnetofluid \cite{TGMG,ICCHSGSST} applications, not to mention the important role limiters play in the study of radiative transfer \cite{GoDes} and kinetic theory \cite{JX}; just to note a handful.

From a numerical perspective, it is clear that one should desire that even shock dominated solutions, like both their smooth and non-limited counterparts, converge in $p$ as $p\nearrow p_{\max}$.  However, such convergence is fundamentally coupled to the behavior of the error accumulation with respect to one's chosen slope limiting methodology, which, it turns out, must operate over a larger number of degrees of freedom, respectively, as $p$ increases.  For example, in a hierarchical basis (as shown explicitly below) the degrees of freedom grow nonlinearly as a function of $p$ and each degree of freedom ends up carrying information of potentially pathological (or undesirable) overshoots and undershoots which have developed over the native (or non-limited) solution space.  It turns out that this complication introduces a substantial technical difficulty in practice, which many papers on numerical shock capturing  \cite{HAMEP,GKMVG,Kuzmin,LBL,Abgrall2,LSTZ,BarthJesp,Bell,DEO} tend to avoid addressing directly.  Most noteworthy is the observation that slopelimiters tend to limit the coefficients in their chosen basis independently of each other, in the sense that each component is adjusted based on information about the surrounding solution on a relatively local submanifold of the total domain.  It then follows directly that the application of the limiter grows nonlinearly in each timestep as a function of $p$.  Since a limiter \emph{de jure} introduces error into the FEM solution space each time it operates on the FEM solution, more applications of it (iteratively) to the solution space should, as a general rule, lead to greater error accumulation assuming the first application always introduces approximately the same amount of error.  In fact, this is what we observe in each of our limiters \emph{de facto}.  However, we offer an alternative approach to this problem below which is both highly efficient and consistent with the more general setting of $hp$-adaptivity.

It perhaps comes as no surprise that the same type of complications do not arise with respect to the mesh size $h$.  That is, the convergence in $h$ as $h\searrow h_{\min}$ tends to arise as a natural consequence of the usual $h$ convergence, where convergence seems essentially guaranteed in most reasonable limiting regimes, while the order of convergence most certainly is not \cite{LSTZ}.  This issue raises another subtle technical difficulty which we will not address directly in this paper, though we will mention its importance in the proper context.  

Another important technicality pertaining to computational efficiency arises with respect to the well-known Courant-Friedrichs-Lewy (CFL) condition.  In this setting the temporal discretization is (partially) bounded from above by the spatial discretization.    That is, in order to reach a higher order accuracy at a fixed value of $h$ one must project onto a higher order polynomial basis in $p$, thus reducing the admissible timestep $\Delta t$ of the scheme --- which obeys an inverse relation by virtue of the CFL condition: $\Delta t \propto 1/p$ as discussed in \cite{ThomasJW}. 

Since this $p$-dependence on the solution accuracy runs counter to the CFL restriction in a practical computational sense, substantial effort has been invested in developing ``smart schemes'' which in some way are able to ``sense'' the appropriate place (\emph{e.g.} $\boldsymbol{x}\in\Omega$) within the solution domain to enrich the polynomial order $p$, while keeping other areas either unaffected or adaptively de-enriching areas of ``less importance.''  The ultimate goal of these schemes is to attempt to substantially improve the computational efficiency of the numerical scheme without ceding notable accuracy in the solution.  In fact, it is generally theoretically true that when one couples adaptive $h$-refinement to $p$-enrichment (\emph{i.e.} $hp$-adaptivity) an exponential improvement in the convergence scaling of the solution may be obtained \cite{Dem}.  However, dynamic adaptive $h$-refinement is beyond the current scope of this paper and will be addressed elsewhere.

On the other hand several different schemes have been developed for dynamic $p$-enrichment of solutions (independent of $h$-refinement), though many suffer the added complexity of being extremely system (PDE) dependent.  The advantage of system dependent regimes is that such schemes often display very close coupling to the physics of the solution (\emph{e.g.} energy methods as discussed in \cite{MES}).  The disadvantage is, of course, that the scheme is very system dependent and hence whenever a variable is added or changed the entire scheme must be recalculated; which is particularly troublesome for systems of equations which are not mathematically well-posed.  Other schemes rely on --- in the FEM setting for example --- the generalized features of numerical variational solutions and as a consequence often depend strongly on a relatively large array of user defined constants.  These schemes are obviously quite attractive from a meta-application perspective, where being able to deal with generalizable systems displaying complicated initial-boundary data generates great allure in itself.  In this paper we focus on the latter class of solutions, as we are interested in schemes which may apply to a large and generalized class of PDEs, without being bound, \emph{ab initio}, to any one particular system of equations.  

Nevertheless, in the present paper we restrict ourselves to the class of discontinuous Galerkin finite element methods, where the underlying basis is chosen such as to signify a ubiety of discontinuous solutions -- that is, we turn our focus in this paper to shock-dominated solutions.  In this setting we are interested in the situation where continuously adaptive $p$-enrichment is coupled to an adapting-in-$p$ slope limiting regime.  We view this setting as very attractive, since the discontinuity sensors for $p$-adaptation schemes are well established \cite{PMH,WanMa} to be good sensors for slope limiting methodologies as well, where the $p$-enrichment leads to stability and efficiency of the scheme while the slope-limiting further stabilizes the presence of spurious oscillations emerging near pathological discontinuities as so approximated to order $p$.  

The outline of this paper is as follows.  In \textsection{2} we present our generalized setting, which can be summarized as: given an advection-diffusion system of equations, consider the initial free boundary value problem recast into the weak formulation and spatially discretized.  We then take a temporal discretization via a RKSSP DG approach in which we obtain the form of our approximate solutions.  Our formulation is general, while our examples focus on problems of hyperbolic transport saving the more general applications for the sequel to this paper.  In \textsection{3} we introduce a number of slope limiters consistent with any order $p$ basis.  The first is the vertex limiter regime of \cite{Kuzmin}, the second the classical Barth--Jespersen limiter \cite{BarthJesp}, and the third and fourth are minor adaptations of the former two limiters made with an eye towards improving the $L^{2}$--error convergence by adjusting a so--called ``blind spot'' present in the previous schemes at higher order $p$.   The fifth approach is comprised of a family of hierarchical reconstruction approaches \cite{Abgrall2,LSTZ}, while the sixth regime is a linear restriction method that can be viewed as a generalization of a limiter originally sketched in \cite{Bell}.  The final limiting regime we present is a mixed extension of the previous limiters referred to here as a hierarchic recombination approach.  Section \textsection{4} then provides numerical experimentation using the schemes presented in \textsection{3} -- namely a classical advective scalar transport problem, and a stationary solution to a closely related problem with highly singular initial data.  We also show some convergence results on an analytic test case.  Finally, in \textsection{5} we present the adaptive $p$-enrichment schemes, which are fully coupled to the slope limiters from \textsection{4} \emph{ab initio}.  These come in two basic types, the first for (\emph{heuristically}) smooth solutions, and the second for solutions demonstrating (vaguely) ``appreciable gradients.''

\section{\texorpdfstring{\protect\centering $\S 2$ Advection--diffusion systems in the DG formalism}{\S 2   Advection--diffusion systems in the DG formalism}}

We are interested in solutions to an initial-boundary value problem for a generalized advection-diffusion system of arbitrarily mixed hyperbolic-parabolic type in $\Omega\times (0,T)$, where $\Omega\subset\mathbb{R}^{2}$ with boundary $\partial\Omega$, such that the system satisfies: \begin{equation}\label{system}\boldsymbol{U}_{t} + \boldsymbol{F}_{x} - \boldsymbol{G}_{x} = \boldsymbol{g}, \quad \mathrm{given \ initial \ conditions} \quad\boldsymbol{U}_{|t=0}=\boldsymbol{U}_{0},\end{equation} and generalized componentwise Robin boundary values \begin{equation} a_{i}U_{i} + \nabla_{x}U_{i,x}\left( b_{i}\cdot \boldsymbol{n} + c_{i}\cdot \boldsymbol{\tau}\right) - f_{i} = 0,\quad \mathrm{on} \ \partial\Omega.\end{equation} That is, the system is comprised of a generalized $m$-dimensional state vector $\boldsymbol{U}=\boldsymbol{U}(t,\boldsymbol{x}) = (U_{1},\ldots,U_{m})$, an advective flux matrix $\boldsymbol{F} = \boldsymbol{F}(\boldsymbol{U})$, a viscous flux matrix $\boldsymbol{G}=\boldsymbol{G}(\boldsymbol{U},\boldsymbol{U}_{x})$, and a source term $\boldsymbol{g}=\boldsymbol{g}(t,\boldsymbol{x})= (g_{1},\ldots,g_{m})$, where $\boldsymbol{x}\in\mathbb{R}^{2}$ and $t\in(0,T)$.  The vectors $\boldsymbol{a}$, $\boldsymbol{b}$, $\boldsymbol{c}$ and $\boldsymbol{f}$ are comprised of the $m$ functions,  $a_{i}=a_{i}(t,\boldsymbol{x})$, $b_{i}=b_{i}(t,\boldsymbol{x})$, $c_{i}=c_{i}(t,\boldsymbol{x})$ and $f_{i}=f_{i}(t,\boldsymbol{x})$ for $i=1,\ldots,m$, where $\boldsymbol{n}$ denotes the unit outward pointing normal and $\boldsymbol{\tau}$ the unit tangent vector.

In addition, because we are interested in approximate numerical solutions of the form of \cite{clint2,ABCM} restricted in part to the family of methods for elliptic equations, we rewrite (\ref{system}) as a coupled system in terms of an auxiliary variable $\boldsymbol{\Sigma}$, such that  \begin{equation}\begin{aligned}\label{system2}\boldsymbol{U}_{t} + \boldsymbol{F}_{x} - \boldsymbol{G}_{x} = \boldsymbol{g}, \quad\mathrm{and}\quad \boldsymbol{\Sigma} = \boldsymbol{U}_{x},\end{aligned}\end{equation} where we have substituted in the viscous flux matrix the auxiliary term, so that $\boldsymbol{G} = \boldsymbol{G}(\boldsymbol{U},\boldsymbol{\Sigma})$.

For notational completeness we adopt the following discretization scheme motivated by \cite{MESV,FFS}.  Take an open $\Omega\subset\mathbb{R}^{2}$ with boundary $\partial\Omega$, given $T>0$ such that $\mathcal{Q}_{T}=((0,T)\times\Omega)$. Let $\mathscr{T}_{h}$ denote the partition of the closure of the polygonal triangulation of $\Omega$, which we denote $\Omega_{h}$, into a finite number of polygonal elements denoted $\Omega_{e}$, such that  $\mathscr{T}_{h}= \{\Omega_{e_1},\Omega_{e_2}, \ldots,\Omega_{e_{ne}}\}$, for $ne\in\mathbb{N}$ the number of elements in $\Omega_{h}$.   In this work we define the mesh diameter $h$ to satisfy $h = \min_{ij}(d_{ij})$ for the distance function $d_{ij}= d(\boldsymbol{x}_{i},\boldsymbol{x}_{j})$ and elementwise edge vertices $\boldsymbol{x}_{i},\boldsymbol{x}_{j}\in\partial\Omega_{e}$ when the mesh is structured and regular.  For unstructured meshes we mean the average value of $h$ over the mesh.

Now, let $\Gamma_{ij}$ denote the edge shared by two neighboring elements $\Omega_{e_{i}}$ and $\Omega_{e_{j}}$, and for $i\in I\subset\mathbb{Z}^{+}=\{1,2,\ldots\}$ define the indexing set $r(i)=\{j \in I : \Omega_{e_{j}}$ is a neighbor of $\Omega_{e_{i}}\}$.  Let us denote all $\Omega_{e_{i}}$ containing the boundary $\partial\Omega_{h}$ by $S_{j}$ and letting $I_{B}\subset \mathbb{Z}^{-}=\{-1,-2,\ldots\}$ define $s(i)=\{j\in I_{B}:S_{j}$ is an edge of $\Omega_{e_{i}}\}$ such that $\Gamma_{ij}=S_{j}$ for $\Omega_{e_{i}}\in \Omega_{h}$ when $S_{j}\in\partial\Omega_{e_{i}}$, $j\in I_{B}$.  Then for $\Xi_{i}=r(i)\cup s(i)$, we have \[\partial\Omega_{e_{i}}=\bigcup_{j\in \Xi(i)}\Gamma_{ij},\quad\mathrm{and}\quad \partial\Omega_{e_{i}}\cap\partial\Omega_{h} = \bigcup_{j\in s(i)}\Gamma_{ij}.\] 

We are interested in obtaining an approximate solution to $\boldsymbol{U}$ at time $t$ on the finite dimensional space of discontinuous piecewise polynomial functions over $\Omega$ restricted to $\mathscr{T}_{h}$, given as \[S_{h}^{p}(\Omega_{h},\mathscr{T}_{h})=\{v:v_{|\Omega_{e_{i}}}\in \mathscr{P}^{p}(\Omega_{e_{i}}) \ \ \forall\Omega_{e_{i}}\in\mathscr{T}_{h}\}\] for $\mathscr{P}^{p}(\Omega_{e_{i}})$ the space of degree $\leq p$ polynomials over $\Omega_{e_{i}}$.      

Choosing a set of degree $p$ polynomial basis functions $N_{\ell}\in\mathscr{P}^{p}(\Omega_{e_{i}})$ for $\ell =1,\ldots, n_{p}$ corresponding to the degree of freedom, we can denote the state vector at time $t$ over $\Omega_{e_{i}}$, by
\begin{equation}\label{shapefunctions}
\boldsymbol{U}_{h}(t,\boldsymbol{x})=\sum_{\ell=1}^{n_{p}}\boldsymbol{U}_{\ell}^{i}(t)N^{i}_{\ell}(\boldsymbol{x}),\quad  \forall x\in\Omega_{e_{i}},
\end{equation}
 where the $N^{i}_{\ell}$'s are the finite element shape functions in the DG setting, and the $\boldsymbol{U}_{\ell}^{i}$'s correspond to the unknowns.   We characterize the finite dimensional test functions \[\begin{aligned}\boldsymbol{v}_{h}, \boldsymbol{\omega}_{h}\in W^{k,q}(\Omega_{h},\mathscr{T}_{h}),\quad\mathrm{by}\quad\boldsymbol{v}_{h}(x)=\sum_{\ell=1}^{n_{p}}\boldsymbol{v}_{\ell}^{i}N_{\ell}^{i}(x)\quad \mathrm{and}\quad\boldsymbol{\omega}_{h}(x)=\sum_{\ell=1}^{n_{p}}\boldsymbol{\omega}_{\ell}^{i}N_{\ell}^{i}(x) \end{aligned}\] where  $\boldsymbol{v}_{\ell}^{i}$ and  $\boldsymbol{\omega}_{\ell}^{i}$ are the coordinates of the test functions in each $\Omega_{e_{i}}$, and with the broken Sobolev space over the partition $\mathscr{T}_{h}$ defined by \[W^{k,q}(\Omega_{h},\mathscr{T}_{h})=\{w : w_{|\Omega_{e_{i}}}\in W^{k,q}(\Omega_{e_{i}}) \ \ \forall\Omega_{e_{i}}\in\mathscr{T}_{h}\}.\]   

Thus, for $\boldsymbol{U}$ a classical solution to (\ref{system2}), multiplying by $\boldsymbol{v}_{h}$ or $\boldsymbol{\omega}_{h}$ and integrating elementwise by parts yields the coupled system: \begin{equation}\begin{aligned}\label{approxsystem} & \frac{d}{dt}\int_{\Omega_{e_{i}}}\boldsymbol{U}\cdot\boldsymbol{v}_{h}dx +  \int_{\Omega_{e_{i}}} (\boldsymbol{F}\cdot\boldsymbol{v}_{h})_{x} dx - \int_{\Omega_{e_{i}}} \boldsymbol{F}:\boldsymbol{v}^{h}_{x}dx  \\ & \qquad\qquad - \int_{\Omega_{e_{i}}} ( \boldsymbol{G}\cdot\boldsymbol{v}_{h})_{x} dx  + \int_{\Omega_{e_{i}}}  \boldsymbol{G} :\boldsymbol{v}^{h}_{x}dx  =  \int_{\Omega_{e_{i}}}  \boldsymbol{v}_{h}\cdot\boldsymbol{g}dx, \\ & \int_{\Omega_{e_{i}}}\boldsymbol{\Sigma}\cdot \boldsymbol{\omega}_{h} dx - \int_{\Omega_{e_{i}}}(\boldsymbol{U}\cdot\boldsymbol{\omega}_{h})_{x} dx +\int_{\Omega_{e_{i}}}\boldsymbol{U}:\boldsymbol{\omega}^{h}_{x}dx=0,\end{aligned}\end{equation}  where $(:)$ denotes the scalar product.

Now, let $\boldsymbol{n}_{ij}$ be the unit outward normal to $\partial\Omega_{e_{i}}$ on $\Gamma_{ij}$, and let $v_{|\Gamma_{ij}}$ and  $v_{|\Gamma_{ji}}$ denote the values of $v$ on $\Gamma_{ij}$ considered from the interior and the exterior of $\Omega_{e_{i}}$, respectively.  Then by choosing componentwise approximations in (\ref{approxsystem}) by substituting in (\ref{shapefunctions}), we arrive with the approximate form of the first term of (\ref{approxsystem}) given by, \begin{equation}
\begin{aligned}\label{timeterm}
\frac{d}{dt}\int_{\Omega_{e_{i}}}\boldsymbol{U}_{h}\cdot \boldsymbol{v}_{h}dx  \approx  \frac{d}{dt}\int_{\Omega_{e_{i}}}\boldsymbol{U}\cdot\boldsymbol{v}_{h}dx,
\end{aligned}
\end{equation}  the second term using an inviscid numerical flux $\boldsymbol{\Phi}_{i}$, by
\begin{equation}
\begin{aligned}\label{invflux}
\tilde{\boldsymbol{\Phi}}_{i}(\boldsymbol{U}_{h}|_{\Gamma_{ij}},\boldsymbol{U}_{h}|_{\Gamma_{ji}}, \boldsymbol{v}_{h}) & = \sum_{j\in \Xi(i)}\int_{\Gamma_{ij}}\boldsymbol{\Phi}(\boldsymbol{U}_{h}|_{\Gamma_{ij}},\boldsymbol{U}_{h}|_{\Gamma_{ji}},\boldsymbol{n}_{ij})\cdot\boldsymbol{v}_{h}|_{\Gamma_{ij}} d\Xi \\ & \approx  \sum_{j\in \Xi(i)} \int_{\Gamma_{ij}}\sum_{l =1}^{2}(\boldsymbol{F})_{l}\cdot (n_{ij})_{l}\boldsymbol{v}_{h}|_{\Gamma_{ij}}d\Xi,
\end{aligned}
\end{equation} and the third term in (\ref{approxsystem}) by,
\begin{equation}\label{third}
\boldsymbol{\Theta}_{i}(\boldsymbol{U}_{h},\boldsymbol{v}_{h})= \int_{\Omega_{e_{i}}} \boldsymbol{F}_{h} : \boldsymbol{v}^{h}_{x} dx \approx \int_{\Omega_{e_{i}}} \boldsymbol{F} : \boldsymbol{v}^{h}_{x} dx.
\end{equation}

Next we approximate the boundary viscous term of (\ref{approxsystem}) using a generalized viscous flux $\hat{\mathscr{G}}$ such that,
\begin{equation}
\begin{aligned}
\label{viscous}
\mathscr{G}_{i}(\boldsymbol{\Sigma}_{h},\boldsymbol{U}_{h},\boldsymbol{v}_{h}) & = \sum_{j\in \Xi(i)}\int_{\Gamma_{ij}}\hat{\mathscr{G}}(\boldsymbol{\Sigma}_{h}|_{\Gamma_{ij}},\boldsymbol{\Sigma}_{h}|_{\Gamma_{ji}}, \boldsymbol{U}_{h}|_{\Gamma_{ij}}, \boldsymbol{U}_{h}|_{\Gamma_{ji}}, \boldsymbol{n}_{ij})\cdot\boldsymbol{v}_{h}|_{\Gamma_{ij}} d\Xi \\ & \approx  \sum_{j\in\Xi(i)}\int_{\Gamma_{ij}}\sum_{l=1}^{2}(\boldsymbol{G})_{l}\cdot (n_{ij})_{l}\boldsymbol{v}_{h}|_{\Gamma_{ij}}d\Xi,
\end{aligned}
\end{equation} 
 while the second viscous term is approximated by:
\begin{equation}\label{second}\mathscr{N}_{i}(\boldsymbol{\Sigma}_{h},\boldsymbol{U}_{h},\boldsymbol{v}_{h})=\int_{\Omega_{e_{i}}}\boldsymbol{G}_{h}:\boldsymbol{v}^{h}_{x}dx\approx \int_{\Omega_{e_{i}}}\boldsymbol{G}:\boldsymbol{v}^{h}_{x}dx.\end{equation}   Finally the source $\boldsymbol{g}$ term of (\ref{approxsystem}) is given to satisfy \begin{equation}\begin{aligned}\label{sourcrea}
\mathscr{H}_{i}(\boldsymbol{v}_{h},\boldsymbol{x}_{h},t)=\int_{\Omega_{e_{i}}}\boldsymbol{v}_{h}\cdot \boldsymbol{g}_{h} dx \approx \int_{\Omega_{e_{i}}}\boldsymbol{v}_{h}\cdot\boldsymbol{g} dx.
\end{aligned}
\end{equation} 

For the auxiliary equation in (\ref{approxsystem}) we expand it such that the approximate solution satisfies: \begin{equation} \begin{aligned}\label{penalty} \mathscr{Q}_{i}(\hat{\boldsymbol{U}},\boldsymbol{\Sigma}_{h},\boldsymbol{U}_{h},\boldsymbol{\omega}_{h},\boldsymbol{\omega}_{x}^{h}) & =\int_{\Omega_{e_{i}}} \boldsymbol{\Sigma}_{h}\cdot \boldsymbol{\omega}_{h}dx  + \int_{\Omega_{e_{i}}}\boldsymbol{U}_{h}:\boldsymbol{\omega}^{h}_{x}dx \\ & - \sum_{j\in \Xi(i)}\int_{\Gamma_{ij}}\hat{\boldsymbol{U}}(\boldsymbol{U}_{h}|_{\Gamma_{ij}},\boldsymbol{U}_{h}|_{\Gamma_{ji}},\boldsymbol{\omega}_{h}|_{\Gamma_{ij}},\boldsymbol{n}_{ij}) d\Xi = 0,\end{aligned}\end{equation} where, \[\begin{aligned} \sum_{i\in I}\sum_{j\in \Xi(i)}\int_{\Gamma_{ij}}\hat{\boldsymbol{U}}(\boldsymbol{U}_{h}|_{\Gamma_{ij}},\boldsymbol{U}_{h}|_{\Gamma_{ji}},\boldsymbol{\omega}_{h}|_{\Gamma_{ij}},\boldsymbol{n}_{ij}) d\Xi \approx \sum_{i\in I}\sum_{j\in \Xi(i)}\int_{\Gamma_{ij}}\sum_{l=1}^{2}(\boldsymbol{U})_{l}\cdot (n_{ij})_{l} \boldsymbol{\omega}_{h}|_{\Gamma_{ij}}d\Xi\end{aligned}\] given $\hat{\boldsymbol{U}}$ a generalized numerical flux, and where \[\int_{\Omega_{e_{i}}} \boldsymbol{\Sigma}_{h}\cdot \boldsymbol{\omega}_{h}dx \approx \int_{\Omega_{e_{i}}} \boldsymbol{\Sigma}\cdot \boldsymbol{\omega}_{h}dx, \quad\mathrm{and}\quad \int_{\Omega_{e_{i}}}\boldsymbol{U}_{h}:\boldsymbol{\omega}^{h}_{x}dx\approx  \int_{\Omega_{e_{i}}}\boldsymbol{U}: \boldsymbol{\omega}^{h}_{x}dx.\]

Combining the above approximations and setting, $\mathscr{X} = \sum_{\Omega_{e_{i}}\in\mathscr{T}_{h}}\mathscr{X}_{i}$, while denoting the inner product \[(\boldsymbol{a}_{h}^{n},\boldsymbol{b}_{h})_{\Omega_{\mathcal{G}}} = \sum_{\Omega_{e_{i}}\in\mathscr{T}_{h}}\int_{\Omega_{e_{i}}}\boldsymbol{a}_{h}^{n}\cdot\boldsymbol{b}_{h} dx,\] we arrive at our approximate solution to (\ref{system2}) as the pair of functions $(\boldsymbol{U}_{h},\boldsymbol{\Sigma}_{h})$ for all $t\in (0,T)$ satisfying: \begin{center}\underline{The Discontinuous Galerkin formulation}\end{center}
\begin{equation}
\begin{aligned}
\label{aprox}
& a) \ \boldsymbol{U}_{h}\in C^{1}((0,T); S_{h}^{p}), \ \ \boldsymbol{\Sigma}_{h}\in S_{h}^{p}, \\
& b) \ \frac{d}{dt}(\boldsymbol{U}_{h},\boldsymbol{v}_{h})_{\Omega_{\mathcal{G}}}+\tilde{\boldsymbol{\Phi}}(\boldsymbol{U}_{h},\boldsymbol{v}_{h}) -  \boldsymbol{\Theta}(\boldsymbol{U}_{h},\boldsymbol{v}_{h}) \\ & \qquad-\mathscr{G}(\boldsymbol{\Sigma}_{h},\boldsymbol{U}_{h},\boldsymbol{v}_{h})+ \mathscr{N}(\boldsymbol{\Sigma}_{h},\boldsymbol{U}_{h},\boldsymbol{v}_{h})=\mathscr{H}(\boldsymbol{v}_{h},\boldsymbol{x}_{h},t), \\  & c) \ \mathscr{Q}(\hat{\boldsymbol{U}},\boldsymbol{\Sigma}_{h},\boldsymbol{U}_{h},\boldsymbol{\omega}_{h},\boldsymbol{\omega}_{x}^{h}) = 0, \\
& d) \ \boldsymbol{U}_{h}(0)=\Pi_{h}\boldsymbol{U}_{0},
\end{aligned}
\end{equation} where $\Pi_{h}$ is a projection operator onto the space of discontinuous piecewise polynomials $S_{h}^{p}$, and where below we always utilize a standard $L^{2}$--projection, given for a function $\boldsymbol{f}_{0}\in L^{2}(\Omega_{e_{i}})$ such that our approximate projection $\boldsymbol{f}_{0,h}\in L^{2}(\Omega_{e_{i}})$ is obtained by solving, $\int_{\Omega_{e_{i}}}\boldsymbol{f}_{0,h}\boldsymbol{v}_{h} dx = \int_{\Omega_{e_{i}}}\boldsymbol{f}_{0}\boldsymbol{v}_{h} dx.$  We provide several explicit simplified examples of this generalized formalism below, though in the followup paper we address models motivated by more complicated dynamics (\emph{e.g.} see \cite{KubatkoWD,KBDWM,BKWD,KDW,KWD}) that employ the full system of (\ref{aprox}) including multicomponent reaction-advection-diffusion and free boundary conditions, \emph{etc.}

The discretization in time follows now directly from (\ref{aprox}), where we employ a family of SSP (strong stability preserving, or often ``total variation diminishing (TVD)'') Runge-Kutta schemes as discussed in \cite{Ruuth,SO}.  That is, for the generalized SSP  Runge-Kutta scheme we rewrite (\ref{aprox}$b$) in the form: $\mathbf{M}\boldsymbol{U}_{t} = \mathbf{R}$, where $\boldsymbol{U} = (\boldsymbol{U}_{1},\ldots,\boldsymbol{U}_{p})$ for each element from (\ref{shapefunctions}), where $\mathbf{R}=\mathbf{R}(\boldsymbol{U},\boldsymbol{\Sigma})$ is the  advection-diffusion contribution along with the source term, and where $\mathbf{M}$ is the usual mass matrix.  Then the generalized $s$ stage of order $\gamma$ SSP Runge-Kutta method (denoted SSP($s,\gamma$) or RKSSP($s,\gamma$)) may be written to satisfy: \begin{equation}\begin{aligned}\label{SSPRK} & \boldsymbol{U}^{(0)}  = \boldsymbol{U}^{n}, \\ & \boldsymbol{U}^{(i)} = \sum_{r = 0}^{i-1}\left(\alpha_{ir}\boldsymbol{U}^{r}+\Delta t\beta_{ir}\mathbf{M}^{-1}\mathbf{R}^{r}\right), \quad\mathrm{for} \ \ i=1,\ldots,s \\ & \boldsymbol{U}^{n+1} =\boldsymbol{U}^{(s)},\end{aligned}\end{equation} where $\mathbf{R}^{r} =\mathbf{R}\left(\boldsymbol{U}^{r},\boldsymbol{\Sigma}^{r},\boldsymbol{x},t^{n}+\delta_{r}\Delta t \right)$ and the solution at the $n$--th timestep is given as $\boldsymbol{U}^{n}=\boldsymbol{U}_{|t=t^{n}}$ and at the $n$--th plus first timestep by $\boldsymbol{U}^{n+1}=\boldsymbol{U}_{|t=t^{n+1}}$, with $t^{n+1}=t^{n}+\Delta t$.  The $\alpha_{ir}$ and $\beta_{ir}$ are the coefficients arising from the Butcher Tableau, and the third argument in $\mathbf{R}^{r}$ corresponds to the time-lag complication arising in the constraints of the TVD formalism.  That is $\delta_{r}=\sum_{l=0}^{r-1}\mu_{rl}$, where $\mu_{ir} = \beta_{ir}+\sum_{l=r+1}^{i-1}\mu_{lr}\alpha_{il}$, where we have taken that $\alpha_{ir}\geq 0$ satisfying $\sum_{r=0}^{i-1}\alpha_{ir}=1$.

It is often possible to optimize the generalized SSP schemes of (\ref{SSPRK}) by restricting to an optimization class of stage exceeding order SSP Runge--Kutta time discretizations of \cite{KWD} as long as $p\leq 3$.  This class of SSP Runge--Kutta schemes has the advantage of optimizing the polynomial order $p$ of the approximate solution $\boldsymbol{U}_{h}$ with respect to the $r$ stage of the SSP Runge--Kutta scheme (incidentally satisfying SSP$(r,p+1)$) in order to minimize the effect of the rigid constraint introduced by the CFL condition on the timestep $\Delta t$.  The limitation on $p$ (\emph{i.e.} requiring $p\leq 3$) is generally more restrictive than we encounter here, and thus, as will become apparent below, in the context of dynamic $p$-enriched slope limited solutions we are generally unable to exploit these optimization schemes directly.

\section{\texorpdfstring{\protect\centering $\S 3$ A dynamic--in--$p$ family of slope limiters}{\S 3 A dynamic--in--$p$ family of slope limiters}} 

\subsection{\texorpdfstring{$\S 3.1$ A transformation of basis}{\S$3.1$ A transformation of basis}}

Finite element approximate solutions are recovered with respect to any number of different finite element bases (\emph{e.g.} Legendre polynomials, Lagrange polynomials, Labotto polynomials, Jacobi polynomials, Gegenbauer polynomials, Chebyshev polynomials, Bernstein polynomials, Gram-Schmidt polynomials, NURBS, $T$-splines, Wachspress functions, \emph{etc.}).  As a consequence of this, it is often advantageous to develop a strategy to transform into a specific basis in order to limit the solution, and then to transform back into the native bases to perform the remainder of the calculations.  This occurs because some slope limiting regimes use fundamental properties of a certain choice of basis in order to develop a limiting strategy.  We provide an explicit example of this procedure below, in the case of transforming between the Dubiner basis and the Taylor basis; or as we denote it below: by way of the invertible Dubiner--Taylor transform $\mathcal{L}$.  We also not here that for the sake of providing explicit calculations, we restrict below to triangular meshes, though the formalism can be easily extended to a more general framework.

Take a solution vector $\boldsymbol{U}$ with approximate form $\boldsymbol{U}_{h}\approx \boldsymbol{U}$ as given by (\ref{shapefunctions}), and project it onto the degree $p$ Dubiner basis such that: \begin{equation}\label{dub}\boldsymbol{U}_{h}(\boldsymbol{x},t)_{|\Omega_{e}}=\sum_{0<i+j\leq p}\boldsymbol{U}_{ij}(t)\phi_{ij}(\boldsymbol{x}),\quad \forall\boldsymbol{x}\in\Omega_{e},\end{equation} where the $\phi_{ij}(\boldsymbol{x})$ are the Dubiner basis functions for each degree of freedom in the solution vector.

%(see the appendix \textsection{a} for an explicit example of the first few monomials).

It is our aim to take this approximate solution $\boldsymbol{U}_{h}$ and limit it with respect to the $k$--th order Taylor basis via, for example, the vertex slope limiter of \cite{Kuzmin} and the hierarchical reconstruction of \cite{Abgrall2,LSTZ}, \emph{etc.}  Now, the Taylor basis in two dimensions is given to arbitrary differential order $k\geq (i+j)$ by the Taylor polynomial centered at $c$ via: \begin{equation}\label{taylor}\boldsymbol{U}_{h}(x,y) = \boldsymbol{U}_h|_{c} + \sum_{0<i+j\leq k}\frac{(x-x_{c})^{i}(y-y_{c})^{j}}{i!j!}\left(\frac{\partial^{i + j}\boldsymbol{U}_{h}}{\partial x^{i}\partial y^{j}}\right)\bigg|_{c},\end{equation} where $x_{c}$ and $y_{c}$ are explicitly chosen as the values at the centroid $c = (x_{c},y_{c})$ of each finite element $\Omega_{e}$ in the physical space $\Omega$ --- that is, each  $\Omega_{e_{i}}$ taking coordinates $\boldsymbol{x}\in\Omega$ --- where it is clear that $i+j\geq 1$ in the sum denotes the differential order of the basis expansion (\emph{i.e.} the indices satisfy $i,j\in\mathbb{N}$).

Now, for cell averages satisfying $\bar{\boldsymbol{U}} = |\Omega_{e}|^{-1}\int_{\Omega_{e}}\boldsymbol{U}_{h} d\boldsymbol{x}$, the average of (\ref{taylor}) may be simply written by \begin{equation}\label{avetay}\bar{\boldsymbol{U}}_{h}(x,y) = \boldsymbol{U}_{h}|_{c} + \sum_{0<i+j\leq k}\overline{\left(\frac{(x-x_{c})^{i}(y-y_{c})^{j}}{i!j!}\right)}\left(\frac{\partial^{i + j}\boldsymbol{U}_{h}}{\partial x^{i}\partial y^{j}}\right)\bigg|_{c}\end{equation} such that subtracting (\ref{avetay}) from (\ref{taylor}) formally yields:  \begin{equation}\label{taylor2}\boldsymbol{U}_{h} = \bar{\boldsymbol{U}}_{h}+ \sum_{0<i+j\leq k}\left(\frac{(x-x_{c})^{i}(y-y_{c})^{j}}{i!j!}- \overline{\frac{(x-x_{c})^{i}( y- y_{c})^{j}}{i!j!}}\right)\left(\frac{\partial^{i + j}\boldsymbol{U}_{h}}{\partial x^{i}\partial y^{j}}\right)\bigg|_{c}.\end{equation}   

Additional analysis (also see \cite{LBL}) has shown empirically that the conditioning of the system in the Taylor basis (with respect to, for example, the invertibility of the Taylor mass matrix) is improved by rescaling with respect to the cell averages over the local bounds, given by $\psi\Delta x = (x_{\max} - x_{\min})$ and $\psi\Delta y = (y_{\max} - y_{\min})$ where $\psi = p$ for $p>2$, and $\psi=2$ for $p\leq 2$.  It is useful to note here that in the master element representation these scalings are merely a pair of constants, while in the physical element representation they will in general vary elementwise.  

Then we are interested in implementing a locally renormalized Taylor basis prescribed with respect to the physical space $\Omega$ given componentwise via the explicit formulation: \begin{equation}\label{compbas}\varphi_{ij}(x,y)=\left(\frac{(x-x_{c})^{i}}{i!\Delta x^{i}} \right)\left(\frac{(y-y_{c})^{j}}{j!\Delta y^{j}} \right)- \overline{\left(\frac{(x-x_{c})^{i}}{i!\Delta x^{i}} \right)\left(\frac{( y- y_{c})^{j}}{j!\Delta y^{j}} \right)},\end{equation} where again cell averages are chosen to satisfy, \[  \overline{\left(\frac{(x-x_{c})^{i}}{i!\Delta x^{i}} \right)\left(\frac{( y- y_{c})^{j}}{j!\Delta y^{j}} \right)} =\frac{1}{|\Omega_{e}|}\int_{\Omega_{e}}\left(\frac{(x-x_{c})^{i}}{i!\Delta x^{i}} \right)\left(\frac{(y-y_{c})^{j}}{j!\Delta y^{j}} \right) dx dy.\]  

Notice also that the constant terms of (\ref{taylor2}) vanish with respect to the barycenter $c$, which is just to say that the value of the centroid is by definition the cell average.  Moreover, note that the renormalization vanishes for linear terms, since the average value is achieved at the centroid $c$ (see \cite{Kuzmin} for more examples at order $p\leq 2$).

Now we see that (\ref{taylor2}) satisfies in vector form that:  \begin{equation}\label{taylor3}\boldsymbol{U}_{h} = \bar{\boldsymbol{U}}_{h} \varphi_{00} + \sum_{0<i+j\leq k} \varphi_{ij}\bigg\{\left(\frac{\partial^{i + j}\boldsymbol{U}_{h}}{\partial x^{i}\partial y^{j}}\right)\bigg|_{c}\Delta x^{i}\Delta y^{j}\bigg\},\end{equation} where we have denoted our effective Taylor basis $\varphi_{ij}\in\mathbb{R}[\Omega]$, such that $\varphi_{ij}=\varphi_{ij}(\boldsymbol{x})$ in the polynomial ring $\mathbb{R}[\Omega]$ such that $\boldsymbol{x}\in\Omega$.  By the polynomial ring $\mathbb{R}[\Omega]$ we simply mean the set of all polynomials with coefficients in $\mathbb{R}$ centered at a particular $\boldsymbol{x}\in\Omega$.  The bracketed terms in (\ref{taylor3}) here represent our effective scaled coefficients, and from here forward the scaling parameters will generally be suppressed for notational simplicity.

 We will further make use of the fact that (\ref{taylor3}) may be viewed as the $k$-jet over $\mathbb{R}^{2}$.   That is, for $\Omega\subset\mathbb{R}^{2}$ and components of the approximate solution vector $\boldsymbol{U}_{h}$ the Taylor basis functions $\varphi_{ij}$ comprise the abstract indeterminates of the $k$-jet $(J_{c}^{k}\boldsymbol{U}_{h})(\varphi_{ij})$ centered at $c$, in that by definition \begin{equation}\label{jets}\boldsymbol{U}_{h}|_{\Omega_{e_{j}}}:=(J_{c}^{k}\boldsymbol{U}_{h})(\varphi_{ij}),\end{equation} such that our approximate solutions are elements of the abstract jet space $\boldsymbol{U}_{h}|_{\Omega_{e_{j}}}\in J^{k}_{c}(\mathbb{R}^{2},\Omega)$.   The jet space $ J^{k}_{c}(\mathbb{R}^{2},\Omega)$ is simply defined as the set of equivalence classes of $k$-jets which agree to order $k$ and map between the Cartesian plane and an element of $\Omega$, as clearly our approximate solutions in the Taylor (polynomial) basis do.  By the set of equivalence classes of $k$-jets which agree to order $k$, we mean for any two solutions $\boldsymbol{V}_{h}|_{\Omega_{e_{j}}}$ and $\boldsymbol{U}_{h}|_{\Omega_{e_{j}}}$ in the Taylor basis restricted to $\Omega_{e_{j}}$ --- that is $k$-jets --- the equivalence relation $\boldsymbol{U}_{h}|_{\Omega_{e_{j}}}- \boldsymbol{V}_{h}|_{\Omega_{e_{j}}} \sim ~ 0$ holds to order $k$.   

In this sense, an effective slope limiter may be viewed as a stabilization rescaling of the jet by the $k$ coefficients $\alpha^{(i+j)}$ (as derived in \textsection{4}), such that the slope limited approximate solution vector $\boldsymbol{U}^{\mathrm{v}}_{h}$ is formally the same as the stabilized $k$--jet centered at $c$; that is $\boldsymbol{U}^{\mathrm{v}}_{h}|_{\Omega_{e_{j}}}:= (J_{c}^{k}\boldsymbol{\alpha}\boldsymbol{U}_{h})(\varphi_{ij})$ where both the approximate solution and the corresponding limited approximate solution are each, respectively, elements of the same abstract jet space $\boldsymbol{U}_{h}|_{\Omega_{e_{j}}},\boldsymbol{U}^{\mathrm{v}}_{h}|_{\Omega_{e_{j}}}\in J^{k}_{c}(\mathbb{R}^{2},\Omega)$ when letting the equivalence relation $\sim$ be approximate $\sim_{h}$ (\emph{i.e.} approximate with respect to the solution order accuracy but with vanishing asymptotics).

Now, in order to work between the Taylor basis representation $\varphi_{ij}$ and the Dubiner basis representation $\phi_{ij}$, we must construct a transformation between the physical element space $\Omega$ and the master element space $\mathcal{M}$, as well as a transformation between the two (abstract) polynomial bases.  Below we make these mappings explicit, and refer to them collectively in this work as the Dubiner--Taylor transform, which is given by the invertible mapping $\mathcal{L}\colon\mathbb{R}[\mathcal{M}]\to J_{c}^{k}(\mathbb{R}^{2},\Omega).$

First consider the usual Dubiner basis functions in the master element space componentwise $\phi_{ij}\in\mathbb{R}[\mathcal{M}]$ for $\phi_{ij}=\phi_{ij}(\boldsymbol{x})$, and $\mathbb{R}[\mathcal{M}]$ the polynomial ring in coordinates $\boldsymbol{x}\in\mathcal{M}$ given by: \begin{equation}\label{dubiner}\phi_{ij}=P_{i}^{0,0}(\psi_{1})\left(\frac{1-\psi_{2}}{2}\right)^{i}P_{j}^{2i+1,0}(\psi_{2}),\end{equation} using $p$-th order Jacobi polynomials with weights $\alpha,\beta$, such that $P_{p}^{\alpha,\beta}(\cdot)$ is evaluated with respect to the coordinates $\boldsymbol{x}=(\xi,\eta)$ of the master triangle element, where the master element quadrilateral transformation in the Dubiner mapping provides that: $\psi_{1}=\left(\frac{2(1+\xi)}{(1-\eta)}-1\right)$ and $\psi_{2}=\eta$, such that $\psi_{1}=\psi_{1}(\boldsymbol{x})$ and $\psi_{2}=\psi_{2}(\boldsymbol{x})$.  

\begin{figure}[t!]
\[\xymatrixcolsep{5pc}
\xymatrix{
\sum_{ij}\boldsymbol{U}_{ij}\phi_{ij}\ar@/^1pc/[dr]^{\mathbf{N}}\ar@/^1pc/[d]^{\mathcal{L}} & \\ \bar{\boldsymbol{U}}_{h} \varphi_{00} + \sum_{ij}\frac{\partial \boldsymbol{U}_{h}^{i+j}}{\partial x^{i}\partial y^{j}}\big|_{c}\varphi_{ij} \ar[r]^{\mathbf{S}}  \ar@/^-1pc/[d] \ar@/^1pc/[u]^{\mathcal{L}^{-1}} &  \tilde{\boldsymbol{U}}_{h} \varsigma_{00} + \sum_{ij}\frac{\partial \boldsymbol{U}_{h}^{i+j}}{\partial \xi^{i}\partial \eta^{j}}\big|_{c}\varsigma_{ij} \ar@/^1pc/[d] \ar@/^.2pc/[ul]^{\mathbf{N}^{-1}}  \ar@<1ex>[l]^{\mathbf{S}^{-1}} \\ \mathscr{L}_{\Omega} \ar@/^-1pc/[u] &  \mathscr{L}_{\mathcal{M}} \ar@/^1pc/[u]
}
\]
\caption{We look at the maps $\mathbf{N}\colon \mathbb{R}[\mathcal{M}] \to J^{k}_{c}(\mathbb{R}^{2},\mathcal{M})$, $\mathbf{S}\colon J^{k}_{c}(\mathbb{R}^{2},\mathscr{P})\to J^{k}_{c}(\mathbb{R}^{2},\mathcal{M})$, and  $\mathcal{L}\colon\mathbb{R}[\mathcal{M}]\to J_{c}^{k}(\mathbb{R}^{2},\Omega)$, where $\mathscr{L}_{\Omega}$ and $\mathscr{L}_{\mathcal{M}}$ are the abstract operators that limit in either the physical element space $\Omega$ or the master element space $\mathcal{M}$.}
\label{fig:com}
\end{figure}
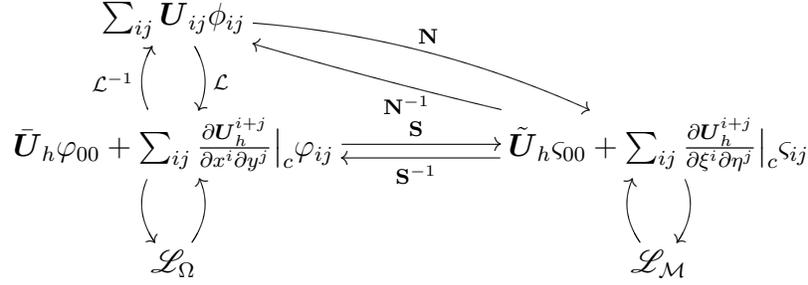

Now, consider the two state vectors, $\boldsymbol{\phi}=(\phi_{00},\phi_{10},\ldots,\phi_{cd})^{T}$ and $\boldsymbol{\varphi}=(\varphi_{00},\varphi_{10},\ldots,\varphi_{cd})^{T}$, where in the lexicographic ordering (described in detail in \textsection{3.2}) we have $c+d\leq p$.  Now, we may transform between the master and physical element representations of our components $\varphi=\varphi(x,y)$ and $\phi=\phi(\xi,\eta)$ using the following affine mapping: \begin{equation}\begin{aligned} \label{xytransform}& x=-\frac{1}{2}\Big\{\xi(x_{1}-x_{2})+&\eta(x_{1}-x_{3})-x_{2}-x_{3}\Big\}, \quad  y = -\frac{1}{2}\Big\{\xi(y_{1}-y_{2})+&\eta(y_{1}-y_{3})-y_{2}-y_{3}\Big\},\end{aligned}\end{equation} with inverse given by \begin{equation} \label{xytransform2} \begin{aligned}& \xi=\chi\bigg\{(y_{3}-y_{1})(x-\frac{1}{2}(x_{2}+x_{3}))+(x_{1}-x_{3})(y-\frac{1}{2}(y_{2}+y_{3}))\bigg\}, \\ & \eta =\chi\bigg\{(y_{1}-y_{2})(x-\frac{1}{2}(x_{2}+x_{3}))+(x_{2}-x_{1})(y-\frac{1}{2}(y_{2}+y_{3}))\bigg\}. \end{aligned}\end{equation}  Here $\{(x_{1},y_{1}),(x_{2},y_{2}),(x_{3},y_{3})\}$ are the vertices of the triangles in the physical space, and the area $\chi^{-1}$ of the physical element $\Omega_{e}$ is given from the cross product of two of the triangle edge vectors, via the usual formula \[\chi=2\left(x_{2}y_{3}-x_{3}y_{2}+x_{3}y_{1}-x_{1}y_{3}+x_{1}y_{2}-x_{2}y_{1}\right)^{-1}.\]

Then by substitution of (\ref{xytransform}) and (\ref{xytransform2}), we may easily construct the invertible mapping  $\mathbf{S}\colon J_{c}^{k}(\mathbb{R}^{2},\Omega)\to J_{c}^{k}(\mathbb{R}^{2},\mathcal{M})$, such that $\boldsymbol{\varsigma}=\mathbf{S}(\boldsymbol{\varphi})$ represents the Taylor basis in the master element space $\mathcal{M}$.  That is, to construct $\mathbf{S}$ explicitly we take the constant first order transformation rules for the derivatives in the base coordinates, given by \begin{equation}\begin{aligned} \label{xider} & \partial_{x}\xi = \chi (y_{3}-y_{1}), \quad \partial_{y}\xi = \chi (x_{1}-x_{3}),\quad \partial_{x}\eta = \chi (y_{1}-y_{2}), \quad \partial_{y}\eta = \chi (x_{2}-x_{1}), \end{aligned}\end{equation} in the master element representation $\hat{\Omega}_{e_{i}}\in\mathcal{M}$, and \begin{equation}\begin{aligned}\label{xyder} &\partial_{\xi }x= (x_{2}-x_{1})/2, \quad \partial_{\xi}y= (y_{2}-y_{1})/2, \quad \partial_{\eta}x=(x_{1}-x_{3})/2,\quad  \partial_{\eta}y=(y_{1}-y_{3})/2\end{aligned}\end{equation} in the physical element representation $\Omega_{e_{i}}\in\Omega$.

Thus provided the coordinate pair $(\xi,\eta)$ in the master element representation $\hat{\Omega}_{e_{i}}\in\mathcal{M}$ we may use (\ref{xytransform}) evaluated at the element quadrature points $\ell$ to fully determine $\mathbf{S}$, where the evaluation at the quadrature points allows for explicit computation of the integral averages in the Taylor basis components (\ref{compbas}), or, more explicitly, where we compute: \[ \overline{\left(\frac{(x-x_{c})^{i}}{i!\Delta x^{i}} \right)\left(\frac{( y- y_{c})^{j}}{j!\Delta y^{j}} \right)}  \approx \frac{1}{|\Omega_{e}|}\sum_{\ell} w_{\ell} \left(\frac{(x_{\ell}-x_{c})^{i}}{i!\Delta x_{\ell}^{i}} \right)\left(\frac{(y_{\ell}-y_{c})^{j}}{j!\Delta y_{\ell}^{j}} \right) |\mathrm{det}\mathbf{J}|\] for $w_{\ell}$ the quadrature weights and the determinant of the Jacobian matrix $\mathbf{J}$ satisfying $|\mathrm{det}\mathbf{J}| = \big|\frac{\partial x}{\partial \xi}\frac{\partial y}{\partial \eta} - \frac{\partial x}{\partial \eta} \frac{\partial y}{\partial \xi}\big|.$

All that remains then is to find the coefficient matrix which constructs the change of polynomial basis mapping $\mathbf{N}\colon\mathbb{R}[\mathcal{M}]\to J_{c}^{k}(\mathbb{R}^{2},\mathcal{M})$, such that we may write the components of the transformed Taylor basis $\varsigma_{ij}$, given by terms $T_{ij}\varsigma_{ij}$, with respect to the components of the master element frame Dubiner basis $\phi_{ij}$, given by terms $D_{ij}\phi_{ij}$; or such that we recover the matrices \begin{equation}\label{transform}\mathbf{T}=\mathbf{N}(\boldsymbol{\phi}),\quad\mathrm{and \ \emph{vice} \ \emph{versa}}\quad\mathbf{D} = \mathbf{N}^{-1}(\boldsymbol{\varsigma}). \end{equation} 

But in light of (\ref{dub}) and (\ref{taylor2}) it follows that for the $\kappa$-th component of the $m$-th size solution vector $\boldsymbol{U}_{h}$ in $\boldsymbol{\phi}$ we may solve for the Taylor coefficients $T_{ij}$ using the system: \begin{equation}\label{matco}\begin{pmatrix} \int_{\hat{\Omega}_{e_{i}}}  \varsigma_{00} U d\eta d\xi \\  \int_{\hat{\Omega}_{e_{i}}}  \varsigma_{10} U d\eta d\xi \\ \vdots \\  \int_{\hat{\Omega}_{e_{i}}}\varsigma_{cd} U d\eta d\xi \end{pmatrix} = \begin{pmatrix} \int_{\hat{\Omega}_{e_{i}}}\varsigma_{00}^{2}d\eta d\xi &  \int_{\hat{\Omega}_{e_{i}}}\varsigma_{00}\varsigma_{10} d\eta d\xi & \hdots & \int_{\hat{\Omega}_{e_{i}}}\varsigma_{00}\varsigma_{cd} d\eta d\xi \\   \int_{\hat{\Omega}_{e_{i}}}\varsigma_{00}\varsigma_{10} d\eta d\xi &  \int_{\hat{\Omega}_{e_{i}}}\varsigma_{10}^{2} d\eta d\xi & \hdots & \int_{\hat{\Omega}_{e_{i}}}\varsigma_{10}\varsigma_{cd} d\eta d\xi   \\ \vdots & \vdots & \ddots  & \vdots   \\ \int_{\hat{\Omega}_{e_{i}}}\varsigma_{00}\varsigma_{cd} d\eta d\xi &  \int_{\hat{\Omega}_{e_{i}}}\varsigma_{10}\varsigma_{cd} d\eta d\xi & \hdots & \int_{\hat{\Omega}_{e_{i}}}\varsigma_{cd}^{2} d\eta d\xi\end{pmatrix} \begin{pmatrix} T_{00} \\ T_{11} \\ \vdots \\ T_{cd} \end{pmatrix},  \end{equation} for the $\kappa$--th component of $\boldsymbol{U}_{h}$.    Note that for the convenience of the reader, we suppress the component index $\kappa$ here and below, though it should be understood that the slope limiting operations are generally performed componentwise over the elements of the solution state vector.

Now, extending (\ref{matco}) over all the components, the Taylor mass matrix tensor $\mathbf{M}_{\boldsymbol{\varsigma}}$ on the right and the inner product matrix $\mathbf{P}_{\boldsymbol{\varsigma}}$ on the left serve to define the desired transformation: \[\mathbf{N}(\boldsymbol{\phi})=\mathbf{M}_{\boldsymbol{\varsigma}}^{-1}\circ\mathbf{P}_{\boldsymbol{\varsigma}}.\]  Its inverse is simply given by forming the Dubiner mass matrix tensor $\mathbf{M}_{\boldsymbol{\phi}}$ and the  inner product matrix in $\boldsymbol{\phi}$ denoted $\mathbf{P}_{\boldsymbol{\phi}}$, such that:  \[\mathbf{N}(\boldsymbol{\varsigma})^{-1}=\mathbf{M}_{\boldsymbol{\phi}}^{-1}\circ\mathbf{P}_{\boldsymbol{\phi}}.\]  Then we have fully constructed the invertible Dubiner--Taylor transform $\mathcal{L}\colon\mathbb{R}[\mathcal{M}]\to J_{c}^{k}(\mathbb{R}^{2},\Omega)$ as satisfying  \begin{equation}\label{TayDub}\mathcal{L}(\boldsymbol{\phi})=\mathbf{S}^{-1}\circ\mathbf{N}=\mathbf{S}^{-1}\circ\mathbf{M}_{\boldsymbol{\varsigma}}^{-1}\circ\mathbf{P}_{\boldsymbol{\varsigma}} = \mathbf{T}\circ\boldsymbol{\varphi}.\end{equation} with inverse satisfying : \[ \mathcal{L}^{-1}(\boldsymbol{\varphi})= \mathbf{N}(\boldsymbol{\varsigma})^{-1}\circ\mathbf{S}(\boldsymbol{\varphi})=\mathbf{M}_{\boldsymbol{\phi}}^{-1}\circ\mathbf{P}_{\boldsymbol{\phi}}\circ\mathbf{S}(\boldsymbol{\varphi}) = \mathbf{D}\circ\boldsymbol{\phi}.\]

\subsection{\texorpdfstring{\protect\centering $\S 3.2$ The formal vertex based hierarchical limiters}{\S 3.2 The formal vertex based hierarchical limiters}}

\begin{figure}[t!]
\centering
\includegraphics[width=8cm]{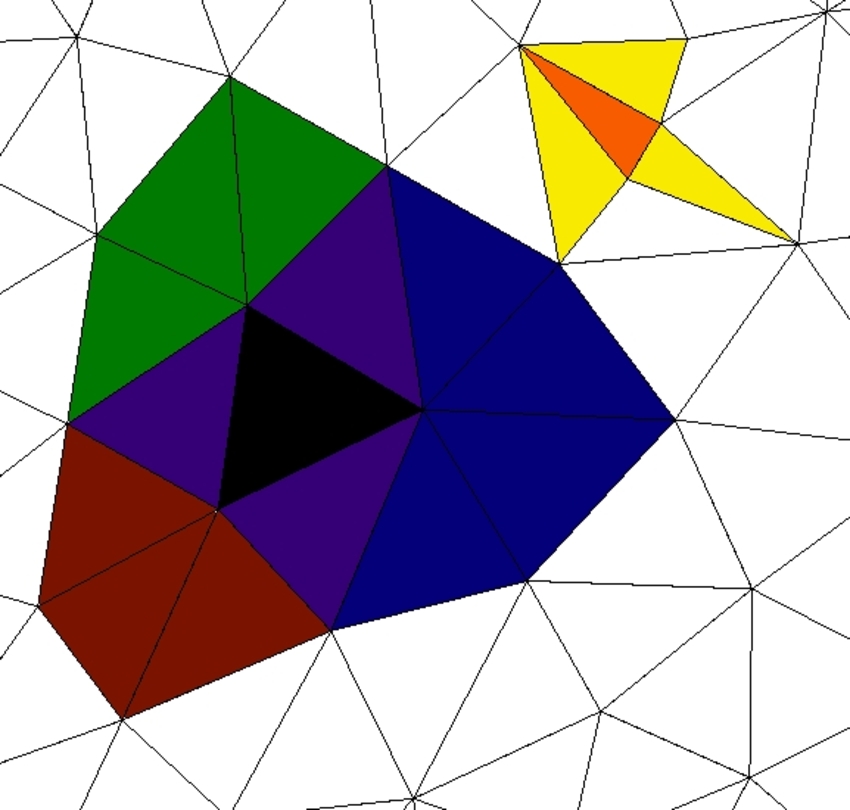}
\caption{ Here we show the \emph{focal neighborhood} $\Omega_{f}$ of a base element $\Omega_{e_{i}}$ filled in black.  Green, red and blue are the three \emph{focal neighborhood} groups based at vertices $\boldsymbol{x}_{i}$ of the black base cell, while purple are cells contained in more than one of the two \emph{focal neighbor stencils} (incidentally comprising the \emph{edge neighborhoood} of  $\Omega_{e_{i}}$).  In a contrasting geometric locale, the orange base cell's \emph{edge neighbors} $\Omega_{E_{j}}$ are each filled in yellow, comprising the \emph{edge neighborhood} $\Omega_{E}$.  See Figure \ref{fig:mesh2} for details.}
\label{fig:mesh}  
\end{figure}

We now formally construct the generalized vertex-based slope limiter based off the Barth--Jespersen limiter \cite{Kuzmin,BarthJesp}.  In this context we define a neighborhood as comprised of those elements that share a common vertex $\boldsymbol{x}_{i}$, indexed with respect to every vertex of each finite element cell $\Omega_{e_{j}}$.   More clearly, we define the \emph{focal neighborhood} $\Omega_{f} = \{\Omega_{e_{j}}\}_{i}$ (in the sense of the \emph{foci} of geometric optics, as shown in Figure \ref{fig:mesh}) as the collection of elements such that $\boldsymbol{x}_i\in\Omega_{e_{j}}$  --- where $\{\Omega_{e_{j}}\}_{i}$ includes the base element $\Omega_{e_{i}}$ --- such that $i=1,2,3$ over triangular elements.  

We now note that one must choose a base space in which to implement this slope limiter (\emph{e.g.} the physical $\Omega$ or master $\mathcal{M}$ element spaces, \emph{etc.}).  A fairly common choice (\emph{viz.} \cite{LBL,Kuzmin}) is to limit with respect to the full physical space $\Omega$.  However, in the context of the local DG formulation this choice is not always so clearly taken.  That is, given our transformations from \textsection{3.1}, it is clear that we may not require the full Dubiner--Taylor transform $\mathcal{L}$ but rather have the option to restrict to the master element space $\mathcal{M}$ by simply using the invertible map $\mathbf{N}$.  More clearly, since local DG formulations often exploit computational efficiency by working over a master element representation $\mathcal{M}$, we are presented with a choice of composition maps to limit in the master or physical element spaces as shown in Figure \ref{fig:com}, and given either by $\mathbf{N}^{-1}\circ\mathscr{L}_{\mathcal{M}}\circ\mathbf{N}$ over $\mathcal{M}$, or by $\mathcal{L}^{-1}\circ\mathscr{L}_{\Omega}\circ\mathcal{L}$ over $\Omega$.  However, since (\ref{TayDub}) shows that $\mathcal{L}$ requires the extra algorithmic step of transforming back into the physical coordinate frame $\Omega$, in the name of computational efficiency, we clearly prefer the former composition given the context of a relatively standard local DG method.  However, when working in a global DG formulation where one elects, for example, a global linear solve, it may be more beneficial to limit with respect to $\Omega$, which as shown in Figure \ref{fig:com} may also be easily accomplished.

\begin{figure}[t!]
\centering
\includegraphics[width=8.7cm]{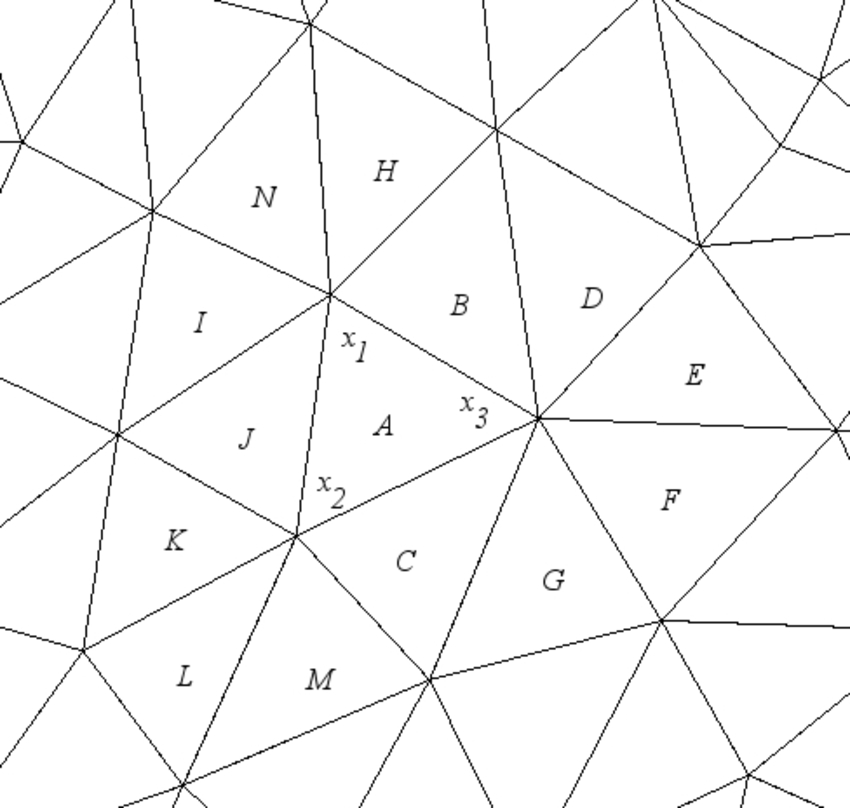}
\caption{ Here we show the \emph{focal stencil} $\Omega_{f_{i}}$ and the \emph{edge stencil} $\Omega_{E_{i}}$ of a base element $\Omega_{e_{l}}=A$.  The stencils are defined with respect to the base elements vertices $\boldsymbol{x}_{i}$ for $i=1,2,3$, such that the  \emph{focal stencil} at $\boldsymbol{x}_{1}$ is $\Omega_{f_{1}}=\{J,I,N,H,B,A\}$, and likewise  $\Omega_{f_{2}}=\{J,K,L,M,C,A\}$ and $\Omega_{f_{3}}=\{C,G,F,E,D,B,A\}$.  Similarly the \emph{edge stencils} are given by:  $\Omega_{E_{1}}=\{J,B,A\}$,  $\Omega_{E_{2}}=\{J,C,A\}$ and  $\Omega_{E_{3}}=\{B,C,A\}$.  Notice that the union of sets recovers the \emph{focal neighborhood} ($\Omega_{f} = \cup_{i}\Omega_{f_{i}}$) and the \emph{edge neighborhood} ($\Omega_{E} = \cup_{i}\Omega_{E_{i}}$), while the restriction of the symmetric difference of sets defines the \emph{focal neighborhood group} ($\ominus_{i}\Omega_{f_{i}}|_{\Omega_{f_{j}}}$) and  \emph{edge neighborhood group} ($\ominus_{i}\Omega_{E_{i}}|_{\Omega_{E_{j}}}$) for any vertex $j$.}
\label{fig:mesh2}  
\end{figure}

Now, we may define the explicit role of the vertex slope limiter as: a method of finding the \emph{limiter matrix} $\boldsymbol{\alpha}=(\boldsymbol{\alpha}_{1},\ldots,\boldsymbol{\alpha}_{m})^{T}$ such that for the solution vector satisfying $\boldsymbol{U}_{h}= (U_{1},\ldots,U_{m})^{T}$, with $m$ the number of unknowns in the system of equations, a vector defined by $\boldsymbol{\alpha}=(\alpha^{(0)},\ldots,\alpha^{(k)})^{T}$ for each order derivative $i+j\leq k$, the limiter coefficients $\alpha^{(i+j)}\in [0,1]$ allow for a recasting of the renormalized solution in (\ref{taylor3}) componentwise in the vertex slope limited form with respect to a \emph{focal stencil}, that is $\Omega_{f_{i}}\subset\Omega_{f}$ for a fixed vertex $\boldsymbol{x}_i$ (see Figure \ref{fig:mesh2} for more detail).

In fact, regardless of the initial location containing the state vector (\emph{i.e.} with respect to $\mathcal{M}$ or with respect to $\Omega$) by simply using our transformations $\mathbf{S}$ and $\mathbf{N}$ from \textsection{2} we can recast (\ref{taylor3}) in the master element space $\mathcal{M}$ such that componentwise we have the vertex slope limited approximate solution $U^{\mathrm{v}}$ which satisfies: \begin{equation}\label{limit1}U^{\mathrm{v}} = \bar{U}\varsigma_{00} + \sum_{0<i+j\leq k}\alpha^{(i+j)}\varsigma_{ij}\left(\frac{\partial^{i + j}U}{\partial \xi^{i}\partial \eta^{j}}\right)\bigg|_{c},\end{equation} where $ \bar{U}$ and $U$ correspond to the approximate solution vector $\boldsymbol{U}_{h}$ transformed to the master element frame in the Taylor basis representation.

Now, notice that above there exists only one $\alpha^{(i+j)}$ for each top $k$--th order mixed derivative in $\xi$ and $\eta$.  In order to recover the $\alpha^{(i+j)}$'s in the polynomial basis expansion, we must decompose our solution Taylor expansion into mixed order linear reconstructions.  To do this, we first order our Taylor polynomial into a hierarchical basis such that each monomial index $b=b(i,j)$ is provided using the lexicographic ordering with ordered lattice pairs $(i,j)$ given by the sequence $(0,0)<(0,1)<(1,0)<(0,2)<(1,1)<\ldots = (i,j)$ corresponding to indices $b$, respectively; that is by the sequence $(1)<(2)<(3)<(4)<(5)<\ldots <(b)\ldots < (s) $ in the Taylor expansion.  In fact, the monomial index in the hierarchy may be determined by the diophantine equation: \begin{equation}\label{monomial} b = \frac{j}{2}\left( j+1\right) + ij + \frac{i}{2}\left(i+3\right) + 1. \end{equation} Then we generate the hierarchical triangular sequence $s=s(p)$, where $p=p(i,j)$ satisfies $p=(i+j)$, such that $s$ determines the upper bound on the degrees of freedom in the polynomial expansion, \begin{equation}\label{invmap}s = \frac{1}{2}(p+1)(p+2),\quad\mathrm{given \ inverse} \ g=g(s)\ \mathrm{such \ that} \ g= \bigg\lfloor\frac{1}{2} + \sqrt{2 s}\bigg\rfloor,\end{equation} where $\lfloor\cdot\rfloor$ is the usual floor function.    Note that in particular we may use $g=g(s)$ or $g=g(b)$ for $g(b)\in\mathfrak{l}$ corresponding to \emph{level} $\mathfrak{l}\neq\mathfrak{l}_{top}$ (defined below) since by virtue of the mapping (\ref{invmap}) both return the same value.

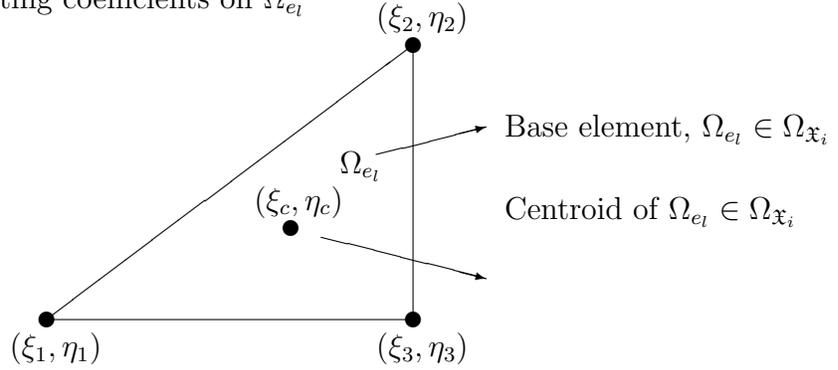
\begin{figure}[!t]
\centering
{\setlength{\unitlength}{4544sp}%
{\begin{picture}(5043,3600)(300,-800)
{\thinlines
}
{\put(  2000,-400){\circle*{90}}}
{\put(  1800,-600){$(\xi_{1},\eta_{1})$}}
{\put(  4000,-400){\circle*{90}}}
{\put(  3800,-600){$(\xi_{3},\eta_{3})$}}
{\put(  4000,1100){\circle*{90}}}
{\put(  3800,1200){$(\xi_{2},\eta_{2})$}}
\put(2000,-400){\line(1,0){2000}}
\put(4000,-400){\line(0,1){1500}}
{\put(2000,-400){\line(4,3){2000}}}
\put(3600,400){$\Omega_{e_{l}}$}
\put(3800,500){\vector(4,1){600}}
\put(4500,600){Base element, $\Omega_{e_{l}}\in\Omega_{\mathfrak{X}_{i}}$}
\put(3333,100){\circle*{90}}
\put(3133,200){$(\xi_{c},\eta_{c})$}
\put(3500,050){\vector(4,-1){900}}
\put(600,2800){ \large{\underline{\bf A schematic of the vertex-based methods}}}
\put(4500,150){Centroid of $\Omega_{e_{l}}\in\Omega_{\mathfrak{X}_{i}}$}
\put(500,2500){$ \bullet \ \ i$ -- the first index of the Taylor expansion}
\put(500,2300){$\bullet \ \ j$ -- the second index of the Taylor expansion}
\put(500,2100){$\bullet \ \ b(i,j)$ --  the monomial index}
\put(500,1900){$\bullet \ \ \mathscr{C}_{b}$ -- the Taylor monomial of index $b$}
\put(500,1700){$\bullet \ \  U_{b,i}^{(i+j)}$ -- the linear reconstruction at $(\xi_{i},\eta_{i})$}
\put(500,1500){$\bullet \ \  U_{i,b}^{\max},U_{i,b}^{\min}$ -- the extrema over the \emph{stencil} $\Omega_{\mathfrak{X}_{i}}$}
\put(500,1300){$\bullet \ \  \alpha^{(q)}$ -- the limiting coefficients on $\Omega_{e_{l}}$}
\end{picture}}}
\caption{Here we provide a key for the vertex and Barth--Jespersen limiters of \textsection{3.2}--\textsection{3.2.1}.  Generally theses limiting procedures depend on the chosen \emph{stencil} $\Omega_{\mathfrak{X}_{i}}$ and a local linear reconstruction of the solution in order to develop the limiting coefficients from (\ref{limit1}).}
\label{fig:ver}
\end{figure}

 Then letting $U^{e_{j}}_{i,c,b}$ be the value of the $b$-th term in the polynomial basis of $\boldsymbol{U}_{h}$ at the centroid $c$ of element $\hat{\Omega}_{e_{j}}$ containing $\boldsymbol{x}_{i}=(\xi_{i},\eta_{i})$ in the master element representation, we define the maximum $U_{i,b}^{\max}$ and minimum $U_{i,b}^{\min}$ values for each unknown monomial at $\boldsymbol{x}_{i}$ over the \emph{focal stencil} $\Omega_{f_{i}}$ situated with respect to the master element frame $\hat{\Omega}_{f_{i}}$ as \begin{equation}\label{vertex1}U_{i,b}^{\max}=\max_{\hat{\Omega}_{{e}_{j}}\in\hat{\Omega}_{f_{i}}} \big\{U^{e_{j}}_{i,b,c}\big\} \quad\mathrm{and}\quad U_{i,b}^{\min}=\min_{\hat{\Omega}_{{e}_{j}}\in\hat{\Omega}_{f_{i}}} \big\{U^{e_{j}}_{i,b,c}\big\}. \end{equation}

Now, we are able to define the $(i+j)$-th linear reconstructions $U_{b,i}^{(i+j)}$ over the vertices $\boldsymbol{x}_{i}$  of any element by taking derivations with respect to the monomial coefficients of (\ref{limit1}).  That is, the linear perturbation of the constant term is constructed such that,  \begin{equation} \label{linear} U_{b,i}^{(1)} = \bar{U} + \frac{\partial U_{i}}{\partial \xi}\bigg|_{c}(\xi_{i} - \xi_{c}) +  \frac{\partial U_{i}}{\partial \eta}\bigg|_{c}(\eta_{i} - \eta_{c}),  \qquad \mathrm{for} \ s =3. \end{equation}  Moreover, it is now direct to construct the higher order terms whereby setting \[\mathscr{C}_{b} =  \left(\frac{\partial^{i+j}U}{\partial \xi^{i}\partial \eta^{j}}\right)\bigg|_{c} \quad \mathrm{for} \ \ b(i,j)>1,\] such that for any mixed derivative order in the hierarchical basis --- as a property of the lexicographic ordering --- we can write: \begin{equation}\label{higher} U_{b,i}^{(i+j)}=\mathscr{C}_{b}  +\mathscr{C}_{b+g}(\eta_{i} -\eta_{c})+ \mathscr{C}_{b+g+1}(\xi_{i} -\xi_{c}), \end{equation} for any polynomial order $k$. Proceeding, we can now define the correction factors $\alpha_{b}^{(i+j)}$ for each element $\Omega_{e_{l}}$, where the vertex-based condition is simply defined as \begin{equation}\label{alphabig}\alpha_{b}^{(i+j)} = \min_{\boldsymbol{x}_{i}\in\hat{\Omega}_{e_{l}}} \begin{cases} \min\Bigg\{1,\left(\frac{ U_{i,b}^{\max} - U^{e_{l}}_{i,c,b}}{ U_{b,i}^{(i+j)} - U_{i,c,b}^{e_{l}}}\right)\Bigg\}, & \mathrm{for}   \ \ U_{b,i}^{(i+j)} > U^{e_{l}}_{i,c,b} \\ \qquad 1, & \mathrm{for} \ \ U_{b,i}^{(i+j)} = U^{e_{l}}_{i,c,b} \\  \min\Bigg\{1,\left(\frac{ U_{i,b}^{\min} - U^{e_{l}}_{i,c,b}}{ U_{b,i}^{(i+j)} - U_{i,c,b}^{e_{l}}}\right)\Bigg\}, & \mathrm{for}   \ \ U_{b,i}^{(i+j)} < U^{e_{l}}_{i,c,b} \end{cases} \end{equation} which, again, is determined separately for each monomial represented in the master frame.

These  $ \alpha_{b}^{(i+j)}$ determine a set of limiting constraints for every hierarchical monomial in the Taylor expansion, but as in \cite{Kuzmin}, we minimize over derivatives of similar top order, such that we recover the components: \begin{equation}\label{alphafirs}\alpha_{\mathfrak{l}(p)}^{(i+j)}= \min_{g(b) =\mathfrak{l}(p_{0})}\alpha_{b}^{(i+j)}.\end{equation}  Notice that these limiting coefficients span the \emph{level} $\mathfrak{l}(p_{0})$, not the hierarchical index $b$ corresponding to \emph{level} $\mathfrak{l}(p)$ (where $p$ and $p_{0}$ are fully explained below).   That is, in the hierarchical basis the linear reconstructions from the perturbation at the \emph{level} below (\emph{i.e.} \emph{level} $(\mathfrak{l}-1)$) are what effectively determine the limiting coefficient at \emph{level} $\mathfrak{l}$ (\emph{e.g.} the gradient terms).  More precisely, the \emph{level} $\mathfrak{l}=\mathfrak{l}(p_{0})$ is determined with respect to the sequence of integers starting at $p_{0}(p_{0}-1)/2+1$ and increasing by one until reaching $p_{0}(p_{0}+1)/2$.   Then the \emph{level} is defined by $\mathfrak{l}=\sup\{g(p_{0}(p_{0}-1)/2+1),\ldots,g(p_{0}(p_{0}+1)/2)\}$, where $p_{0}=1$ for the strictly linear case, and in general is a positive integer such that $p_{0}\leq p$ and is fully determined by $g(b(i,j))$.  In general however, the \emph{level} $\mathfrak{l}(p)$ spans $\mathfrak{l}=\sup\{g(p(p+1)/2)+1,\ldots,g((p+1)(p+2)/2)\}$ such that the \emph{level} below $\mathfrak{l}(p_{0})$ simply corresponds to setting $p=p_{0}-1$.

Finally we limit the magnitude of the correction by the maximum value of every correction factor of greater than or equal order.  In other words, we do not allow a higher order correction to demonstrate greater regularity than a lower order correction, and in fact empirical experimentation has found this to be a necessary constraint.  That is, setting $q = (i+j)$ and $r=(i'+j')$ for $i'$ and $j'$ indices, then we determine an upper bound on the correction parameter by resetting: \begin{equation}\label{alphasec}\alpha^{(q)} := \max_{q\leq r,\mathfrak{l}\leq \mathfrak{l}_{\mathrm{top}}} \alpha_{\mathfrak{l}}^{(r)},\quad \forall q\geq 1,\ \forall r \geq q. \end{equation}   The top \emph{level} $\mathfrak{l}_{\mathrm{top}}$ simply corresponds to the \emph{level} whose upper bound is determined by $g(s)=g(b)$.  Also notice that the derivative order $(i+j)$ is fundamentally coupled to the \emph{level} $\mathfrak{l}$, and so is in some ways redundant notation which we have used in order to emphasize this coupling.

It is also worth noting, that as a consequence of the above construction we are now easily able to implement an arbitrarily higher-order extension of the Barth--Jespersen limiter \cite{Kuzmin,BarthJesp}, where we may perform the exact steps as above, but simply exchange (\ref{vertex1}) with \begin{equation}\label{BJ1}U_{i,b}^{\max}=\max_{\hat{\Omega}_{{e}_{j}}\in\hat{\Omega}_{E_{i}}} \big\{U^{e_{j}}_{i,b,c}\big\} \quad\mathrm{and}\quad U_{i,b}^{\min}=\min_{\hat{\Omega}_{{e}_{j}}\in\hat{\Omega}_{E_{i}}} \big\{U^{e_{j}}_{i,b,c}\big\},\end{equation} where $\hat{\Omega}_{E_i}$ is the \emph{edge stencil} of $\hat{\Omega}_{e_{j}}$ at $\boldsymbol{x}_{i}$ in the master element representation --- or the corresponding set of those physical elements sharing an edge with $\Omega_{e_{j}}$ at vertex $\boldsymbol{x}_{i}$ such that the base element $\Omega_{e_{j}}\in\Omega_{E_{i}}$ (see Figure \ref{fig:mesh}).  

A schematic is provided in Figure \ref{fig:ver} which is meant to simplify the notation and unify the basic principles underlying both the vertex and Barth--Jespersen limiters (as well as the adapted vertex-based limiters of \textsection{3.2.1}).

\subsubsection{\texorpdfstring{$\S 3.2a$ On adapted vertex based limiters}{\S$3.2a$ On adapted vertex based limiters}}

Both the vertex limiter and the Barth--Jespersen limiter from \textsection{3.2} demonstrate a similar --- though often times non-ideal --- behavior.  That is, notice that in both the definition of (\ref{vertex1}) and \ref{BJ1}) that we have found a maximum or minimum with respect to a local neighborhood of the mesh.  Hence, in either case, when we compute the limiting coefficients in (\ref{alphabig}) a local bound (\emph{e.g.} (\ref{BJ1})) is always achieved, even in the degenerate case of when $U_{i,b}^{\min}=U_{i,b}^{\max}$.

As a consequence of this, the numerator in the quotients of (\ref{alphabig}) vanish on elements admitting a local extremum, leading to persistent and excessive diffusivity (\emph{i.e.} limiting $\alpha=0$ at each such timestep) arising at all orders in each local extrema of the mesh, even when those extrema are neither spurious nor potentially unstable; and moreover, this behavior compounds in $p$ since as $p$ increases the number of degrees of freedom (\emph{i.e.} monomials) in the solution which have local extrema also increases nonlinearly.

This behavior over values of local extrema can become quite dominant depending on the mesh geometry.   In particular, since the vertex-based limiter has a larger local neighborhood (\emph{i.e.} the \emph{focal neighborhood}) than the Barth--Jespersen limiter, in principle it should provide more information from which to glean a more accurate approximate local reconstruction.  However, due to this ``diffusivity,'' the larger local neighborhood actually lends itself towards increasing the nonlocality of the diffusive effects of the neighborhood-wise extrema as $p\nearrow p_{\max}$, and hence in practice can actually precipitate greater diffusion in the vertex limiter than the native Barth--Jespersen limiter as $p$ increases (up to the mesh geometry). 

In order to reduce this so-called ``blind diffusion'' in both limiters we introduce a simple functional which attempts to treat a portion of this special case separately.  That is, we simply replace (\ref{alphabig}) with:  \begin{equation}\label{alphabigadapted}\alpha_{b}^{(i+j)} =  \min_{\boldsymbol{x}_{i}\in\Omega_{e_{l}}} \begin{cases} \  \min\Bigg\{1,\left(\frac{ U_{i,b}^{\max} - U^{e_{l}}_{i,c,b}}{ U_{b,i}^{(i+j)} - U_{i,c,b}^{e_{l}}}\right)\Bigg\}, & \mathrm{for}   \ \ U_{b,i}^{(i+j)} > U^{e_{l}}_{i,c,b} \\ \min\Bigg\{f_{\max},\bigg|\frac{ U_{i,b}^{\max} -  U_{i,b}^{\min}}{ U_{b,i}^{(i+j)} - U_{i,c,b}^{e_{l}}}\bigg|\Bigg\}, &  \mathrm{for}   \ \ U^{e_{l}}_{i,c,b}=  U_{i,b}^{\max} \\ \qquad\qquad\qquad 1, & \mathrm{for} \ \    U_{b,i}^{(i+j)} = U^{e_{l}}_{i,c,b}  \\  \min\Bigg\{f_{\min},\bigg|\frac{ U_{i,b}^{\min} -  U_{i,b}^{\max}}{ U_{b,i}^{(i+j)} - U_{i,c,b}^{e_{l}}}\bigg|\Bigg\}, &  \mathrm{for}   \ \ U^{e_{l}}_{i,c,b}=  U_{i,b}^{\min} \\  \ \min\Bigg\{1,\left(\frac{ U_{i,b}^{\min} - U^{e_{l}}_{i,c,b}}{ U_{b,i}^{(i+j)} - U_{i,c,b}^{e_{l}}}\right)\Bigg\}, & \mathrm{for}   \ \ U_{b,i}^{(i+j)} < U^{e_{l}}_{i,c,b} \end{cases} \end{equation} where $f_{\max},f_{\min}\in(0,1)$ are constants used to limit the rate at which the extrema diffuse (that is, reduce the rate at which error is introduced into the solution), and when $f_{\max}=f_{\min}$ we denote them by $f_{d}$.  

We find when setting $f_{d}=1$ we generally get a very moderate improvement in the limiting error behavior of both the vertex and Barth--Jespersen limiters.  Nevertheless, clearly (\ref{alphabigadapted}) has only accounted partially for the degenerate local extrema cases, in particular it still fails to properly account for the case of $U_{b,i}^{\min}=U_{b,i}^{\max}$, and the absolute value is used to account for the fact that the signs have not been separately controlled.  We have developed strategies for adopting fixes for these issues into the limiter, but in general find even the augmented regimes to still demonstrate substantially more diffuse behavior than the restricted regime presented in \textsection{3.4}, and so will suppress any further comment on the subject at present, simply noting that it is possible to improve upon the basic behavior of the limiter in $p$ by developing selection strategies to deal with the many special cases which arise over solutions locally, and where alternatively one is often also able to improve the error behavior by tuning $f_{\max}$ and $f_{\min}$.

It should be additionally noted here that in \cite{Kuzmin} a mass lumping strategy is implemented with respect to the triangular meshes in order to prevent the formation of undershoots and overshoots caused in the presence of the non-orthogonal Taylor mass matrix.  It was also demonstrated in \cite{Kuzmin} that this strategy can have a measurably beneficial effect on the error behavior for $p\leq 2$, and is thus clearly worth further examination at higher $p$.  We will return to this issue briefly in \textsection{4} as a note of comparison between the implementational strategies.

\subsection{\texorpdfstring{$\S 3.3$ The hierarchical reconstruction via MUSCL or ENO}{\S$3.3$ The hierarchical reconstruction via MUSCL or ENO}}

We now consider the hierarchical reconstruction scheme presented in \cite{Abgrall2} and \cite{LSTZ}.  Formally in this setting we simply take derivatives of (\ref{taylor}) in the master element frame, and work locally over the averages and differences of these differential reconstructions.  The method is presented as a two step process, where we start in step 1 at the highest order derivatives and work down to the lowest, with the caveat that the linear and constant terms are dealt with separately in step 2.  

{\bf Step 1.}  Starting at the top order coefficient $k$, a linearization of the $(k-1)$--st derivative of (\ref{taylor}) is given by (\ref{higher}) in the Taylor basis for $i+j=k-1$.  Here, however, we recover the entire higher order component including the nonlinear terms, so that we must employ our monomial index function $b(i,j)$ given in (\ref{monomial}).  

That is, beginning at the top \emph{level} $\mathfrak{l}(k)$ for $i+j=k$ we define the linear part as satisfying:  \begin{equation}\label{topavg}\bar{U}_{b_{linear}, \Omega_{e_{l}}}^{(i+j)}:= \mathscr{C}_{b(i,j)},\quad \forall  b\in \mathfrak{l}(k) \ \land \ \forall \Omega_{e_{l}}\in\Omega_{\mathfrak{X}}, \end{equation}   where $\mathfrak{X}$ here and below may be $f$ or $E$ (\emph{i.e.} the \emph{focal} and the \emph{edge neighborhoods}, respectively, as shown in both Figure \ref{fig:mesh} and Figure \ref{fig:mesh2}), and where here and below $\land$ is the logical conjunction operator and $\lor$ is the corresponding logical disjunction operator.  

At the lower (nonlinear) \emph{levels} (\emph{i.e.} the \emph{levels} $\mathfrak{l}$ such that $\mathfrak{l}(1) <\mathfrak{l}<\mathfrak{l}(k)$) by expansion -- after recovering the $i$ and $j$ indices of the base $b$--th component ---then setting $\tilde{b} =  b(i+i',j+j')$ and integrating locally over each cell in the neighborhood, we have that:   \begin{equation}\label{avg}\bar{U}_{b,\Omega_{e_{l}}}^{(i+j)}=\mathscr{C}_{b(i,j)} +\Omega_{e_{l}}^{-1}\int_{\Omega_{e_{l}}}\bigg\{\sum_{i'+j'>0}\frac{1}{i'!j'!} \mathscr{C}_{\tilde{b}}(\eta -\eta_{c})^{i'}(\xi -\xi_{c})^{j'}\bigg\}d\eta d\xi,\quad \forall \tilde{b}\leq s \ \land \ \forall \Omega_{e_{l}}\in\Omega_{\mathfrak{X}}. \end{equation}  Likewise for each \emph{level} $\mathfrak{l}$ we integrate the higher order perturbative terms such that:  \begin{equation}\label{slopeavg}\bar{U}_{b_{slope},\Omega_{e_{l}}}^{(i+j)}=\Omega_{e_{l}}^{-1}\int_{\Omega_{e_{l}}}\bigg\{\sum_{i'+j'>0}\frac{1}{i'!j'!} \mathscr{C}_{\tilde{b}}(\eta -\eta_{c})^{i'}(\xi -\xi_{c})^{j'}\bigg\}d\eta d\xi,\quad \forall \tilde{b}\leq s \ \land \ \forall \Omega_{e_{l}}\in\Omega_{\mathfrak{X}}.\end{equation}  It is then these two averages which serve to limit the \emph{level} $\mathfrak{l}$ components of the Taylor basis by way of the linear type average $\bar{U}_{b_{linear},\Omega_{e_{l}}}^{(i+j)}$ of the difference of (\ref{avg}) with (\ref{slopeavg}) in each $b$:   \begin{equation}\label{linavg}\bar{U}_{b_{linear},\Omega_{e_{l}}}^{(i+j)} :=\left(\bar{U}_{b,\Omega_{e_{l}}}^{(i+j)}-\bar{U}_{b_{slope},\Omega_{e_{l}}}^{(i+j)}\right), \quad \forall \Omega_{e_{l}}\in\Omega_{\mathfrak{X}}.\end{equation}

Now the linear terms of (\ref{linavg}) will be used to determine the \emph{candidates} for the updated values of the base cell $\Omega_{base}$; which is to say that the $(k-1)$--st component of the $k$-th order jet $(J_{c}^{k}\boldsymbol{\alpha}\boldsymbol{U}_{h})(\varsigma_{ij})$ is limited by filtering a set of \emph{candidates} through a family of minmod functions, such that: \begin{equation}\label{minmod}U^{e_{l}}_{b,\Omega_{base}}:=\mathrm{minmod}^{*}_{\forall \Omega_{e_{l}}\in\Omega_{\mathfrak{X}}}\left(\bar{U}_{b_{linear},\Omega_{e_{l}}}^{(i+j)}\right).\end{equation}

Notice that we may also choose to find candidates over restricted subsets of the full neighborhood $\Omega_{\mathfrak{X}}$ in order to try and more effectively \emph{localize} our limiting.  For example, we may choose to find the minmod function over the local \emph{stencil} $\Omega_{\mathfrak{X}_{i}}$ centered about a vertex of the cell and then perform a different selection rule over that set of candidates; or, alternatively, we may compute the integral averages over the local \emph{stencil} $\Omega_{\mathfrak{X}_{i}}$ in (\ref{topavg})--(\ref{slopeavg}) and then perform a minmod with respect to the full neighborhood $\Omega_{\mathfrak{X}}$.  We have implemented and tested a number of these different regimes, and consider each of them in this paper to live under the general heading of ``hierarchical reconstruction schemes,'' though for the sake of brevity we focus only on (\ref{minmod}) below.

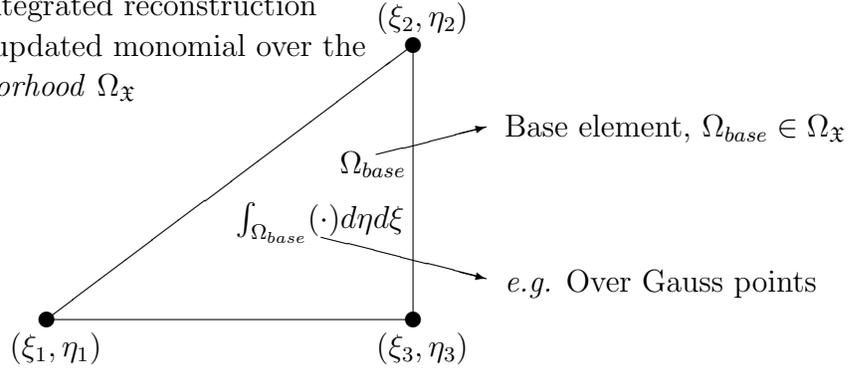
\begin{figure}[!t]
\centering 
{\setlength{\unitlength}{4544sp}%
{\begin{picture}(5043,3600)(300,-800)
{\thinlines
}
{\put(  2000,-400){\circle*{90}}}
{\put(  1800,-600){$(\xi_{1},\eta_{1})$}}
{\put(  4000,-400){\circle*{90}}}
{\put(  3800,-600){$(\xi_{3},\eta_{3})$}}
{\put(  4000,1100){\circle*{90}}}
{\put(  3800,1200){$(\xi_{2},\eta_{2})$}}
\put(2000,-400){\line(1,0){2000}}
\put(4000,-400){\line(0,1){1500}}
{\put(2000,-400){\line(4,3){2000}}}
\put(3600,400){$\Omega_{base}$}
\put(3800,500){\vector(4,1){600}}
\put(4500,600){Base element, $\Omega_{base}\in\Omega_{\mathfrak{X}}$}
\put(3033,100){$\int_{\Omega_{base}}(\cdot)d\eta d\xi$}
%\put(2833,100){\circle*{90}}
%\put(3333,500){\circle*{90}}
%\put(3133,600){$(\xi_{c},\eta_{c})$}
\put(3500,050){\vector(4,-1){900}}
\put(4500,-250){\emph{e.g.} Over Gauss points}
\put(100,2800){ \large{\underline{\bf A schematic of the hierarchical reconstruction method}}}
\put(500,2500){$ \bullet \ \ i$ -- the first index of the Taylor expansion}
\put(500,2300){$\bullet \ \ j$ -- the second index of the Taylor expansion}
\put(500,2100){$\bullet \ \ b,\tilde{b}$ --  the monomial/hierarchical indices}
\put(500,1900){$\bullet \ \ \mathscr{C}_{b}$ -- the Taylor monomial of index $b$}
\put(500,1700){$\bullet \ \  \bar{U}_{b_{linear},\Omega_{e_{l}}}^{(i+j)}$ -- the \emph{candidates} of the reconstruction}
\put(500,1480){$\bullet \ \  \bar{U}_{b_{slope},\Omega_{e_{l}}}^{(i+j)}$ -- the higher order terms}
\put(500,1260){$\bullet \ \  \bar{U}_{b,\Omega_{e_{l}}}^{(i+j)}$ -- the integrated reconstruction}

\put(500,1040){$\bullet \ \  U_{b,\Omega_{base}}^{e_{l}}$ -- the updated monomial over the}
\put(1200,820){ \emph{neighborhood} $\Omega_{\mathfrak{X}}$}
\end{picture}}}
\caption{Here we provide a key for the hierarchical reconstruction method developed in \textsection{3.3}.  The limiting procedure depends on the entire \emph{neighborhood} $\Omega_{\mathfrak{X}}$, the fully integrated solution, and a choice of minmod functions in order to reconstruct the limited form of the monomial coefficients on the base cell.}
\label{fig:fullrecon}
\end{figure}

Note that we perform Step 1 for each \emph{level} $\mathfrak{l}(\jmath)$ where $\jmath < k$, and recursing down to the \emph{level} corresponding to the $\mathfrak{l}$ associated to the quadratic components at $p=2$; where first we limit the difference (\ref{linavg}) across the \emph{neighborhood} of a base element in order to reconstruct the values on the base cell proper.  For these purposes, we employ the following set of minmod$^*_{\mathfrak{X}}=\Phi_{\mathfrak{X}}^{*}$ functions.  The MUSCL reconstruction method relies on the function: \[\Phi^{\mathfrak{m}}_{\mathfrak{X}}\left(\bar{U}_{b_{linear},\Omega_{e_{l}}}^{(i+j)}\right)=  \begin{cases}  \min_{i}\left( \bar{U}_{b_{linear},\Omega_{e_{l}}}^{(i+j)}\right), \quad & \mathrm{if} \quad \bar{U}_{b_{linear},\Omega_{e_{l}}}^{(i+j)} > 0 \quad \forall \Omega_{e_{l}}\in\Omega_{\mathfrak{X}}, \\   \max_{i}\left( \bar{U}_{b_{linear},\Omega_{e_{l}}}^{(i+j)}\right), \quad & \mathrm{if} \quad \bar{U}_{b_{linear},\Omega_{e_{l}}}^{(i+j)} < 0  \quad \forall  \Omega_{e_{l}}\in\Omega_{\mathfrak{X}}, \\ \qquad\qquad 0, \quad & \qquad\qquad\qquad \mathrm{otherwise}, \end{cases}\] while the ENO reconstruction is given by \[\Phi^{\mathfrak{e}}_{\mathfrak{X}}\left(\bar{U}_{b_{linear},\Omega_{e_{l}}}^{(i+j)}\right) = \bar{U}_{b_{linear},\Omega_{e_{l}}}^{(i+j)}\quad \mathrm{if}\quad  \bar{U}_{b_{linear},\Omega_{e_{l}}}^{(i+j)} = \min_{ \forall \Omega_{e_{l}}\in\Omega_{\mathfrak{X}}}\Big|\bar{U}_{b_{linear},\Omega_{e_{l}}}^{(i+j)}\Big|.\]  Additionally, following \cite{LSTZ}, the minmod$^*_{\mathfrak{X}}$ function may be set as a center bias scheme given by \begin{equation}\label{wm1}\Phi^{\mathfrak{c}}_{\mathfrak{X}}  = \Phi^{\mathfrak{m}}_{\mathfrak{X}}\left((1+\epsilon)\cdot\Phi^{\mathfrak{m}}_{\mathfrak{X}}\left(\bar{U}_{b_{linear},\Omega_{e_{l}}}^{(i+j)}\right),\frac{1}{r}\sum_{k=1}^{r}\bar{U}_{b_{linear},\Omega_{e_{k}}}^{(i+j)}\right),\end{equation} or the weighted ENO scheme, \begin{equation}\label{wm2} \Phi^{\mathfrak{e}_{2}}_{\mathfrak{X}} = \Phi^{\mathfrak{e}}_{\mathfrak{X}}\left((1+\epsilon)\cdot\Phi^{\mathfrak{e}}_{\Omega_{\mathfrak{X}}}\left(\bar{U}_{b_{linear},\Omega_{e_{l}}}^{(i+j)}\right),\frac{1}{r}\sum_{k=1}^{r}\bar{U}_{b_{linear},\Omega_{e_{k}}}^{(i+j)}\right), \end{equation} where in either case $r$ is the total number of neighboring cells of the base cell $\Omega_{e_{base}}$, $\epsilon$ is a user defined constant, and $(\cdot)$ in both (\ref{wm1}) and (\ref{wm2}) is merely standard multiplication.  It is known that setting $\epsilon$ large helps to achieve the expected order of accuracy over triangular meshes.

{\bf Step 2.}  Now we address the case of how to limit the solution with respect to the linear $i+j =1$ and constant $i= j = 0$ cases.  For the linear case, we simply choose to limit with respect to a subset of limiting regimes, including those in \textsection{3.2} \textsection{3.3} \textsection{3.4} and \textsection{3.5}.  We choose this, in particular, in order to electively replace the MUSCL and ENO schemes from Step 1, which are relatively speaking more diffuse in our experiments at \emph{level} $\mathfrak{l}(1)$ than some other possible alternatives.  

Finally, the constant terms at \emph{level} $\mathfrak{l}(0)$ are simply set equal to the average value on their base cell, $\bar{U}\varsigma_{00}|_{\Omega_{base}}$ in order to enforce invariance of the cell averages.  In other words, the constant terms remain unchanged.

We should also note that recent improvements have been made in the context of hierarchical reconstructuion techniques.  In particular, recent work has been done to extend the formalism above to include WENO-type linear reconstructions.   That is, the WENO-type formalism of \cite{zhu08,qiu05} has been extended to the context of hierarchical reconstruction based limiter regimes in \cite{Xu2009} specifically in order to address the fact that the MUSCL and ENO type approaches have been shown to often fail to give the desired order of accuracy on triangular meshes.  These techniques rely on the conditioning of a local ($\Omega_{\mathfrak{X}}$-restricted ) recontruction matrix, and are beyond the scope of the present paper.   We direct the interested reader to \cite{Xu2009}.

\subsection{\texorpdfstring{$\S 3.4$ On a dynamically adaptive linear restriction}{\S$3.4$ On a dynamically adaptive linear restriction}}

In this subsection we generalize and update a version of the BDS limiter that had its foundations initially seeded in \cite{Bell} for linear polynomials over uniform structured meshes.  We present the formal construction of a substantially more general form of this limiter, to act over an arbitrary order $p$ basis by way of a linear restriction technique over unstructured triangular meshes.   This limiter is developed with an eye towards $p$-enrichment schemes, and in particular $hp$-adaptive schemes, where in areas of high (jump) variability one generally wants to reduce the order of $p$ while refining the mesh parameter $h$.  In this section we first restrict back to the Dubiner basis $\phi_{ij}\in\mathbb{R}[\mathcal{M}]$, in part to compare to the same implementation carried out in the Taylor basis as a consequence of the formulation presented in \textsection{3.5}, which for linears turns out to be equivalent.

Let us first restrict to the sub-quadratic terms of the basis for any order $p$, such that we are only concerned initially with the terms corresponding to $i+j\leq 1$.  Then, similar to (\ref{vertex1}), setting  $U^{e_{j}}_{i}$ as the constant piece of the Dubiner basis in $\boldsymbol{U}_{h}$ of the base element $\hat{\Omega}_{e_{j}}$ containing $\boldsymbol{x}_{i}=(\xi_{i},\eta_{i})$ in the master element representation, we define the maximum $U_{i}^{\max}$ and minimum $U_{i}^{\min}$ values for each unknown at every $\boldsymbol{x}_{i}\in\hat{\Omega}_{e_{j}}$ over the chosen \emph{stencil} $\hat{\Omega}_{\mathfrak{X}_{i}}$ as \begin{equation}\label{res1}U_{i}^{\max}=\max_{\forall \hat{\Omega}_{{e}_{j}}\in\hat{\Omega}_{\mathfrak{X}_{i}}} \big\{U^{e_{j}}_{i}\big\} \quad\mathrm{and}\quad U_{i}^{\min}=\min_{\forall\hat{\Omega}_{e_{j}}\in\hat{\Omega}_{\mathfrak{X}_{i}}} \big\{U^{e_{j}}_{i}\big\}. \end{equation}

Next we take the full approximate solution restricted to its sub-quadratic part and evaluated at the three vertices of the cell, denoted by the three values $U(\boldsymbol{x}_{\ell})|_{i+j\leq 1}$ for $\ell=1,2,3$ corresponding to the vertices, while $i+j$ corresponds to the polynomial order.  Then at each vertex we employ the following minmod function $\Phi_{\boldsymbol{x}_{\ell}} = \Phi_{\boldsymbol{x}_{\ell}}(U(\boldsymbol{x}_{\ell})|_{i+j\leq 1})$: \begin{equation}\label{res2}\Phi_{\boldsymbol{x}_{\ell}} = \max\Big\{\min \big\{(U(\boldsymbol{x}_{\ell})|_{i+j\leq 1},U_{\ell}^{\max}\big\}, U_{\ell}^{\min}\Big\},\end{equation} where we subsequently reset the vertex value to $U(\boldsymbol{x}_{\ell})|_{i+j\leq 1} := \Phi_{\boldsymbol{x}_{\ell}}(U(\boldsymbol{x}_{\ell})|_{i+j\leq 1})$.

\begin{figure}[!t]
\centering
{\setlength{\unitlength}{4544sp}%
{\begin{picture}(5043,3800)(300,-1000)
{\thinlines
}
{\put(  2000,-600){\circle*{90}}}
{\put(  1800,-800){$(\xi_{1},\eta_{1})$}}
{\put(  4000,-600){\circle*{90}}}
{\put(  3800,-800){$(\xi_{3},\eta_{3})$}}
{\put(  4000,900){\circle*{90}}}
{\put(  3800,1000){$(\xi_{2},\eta_{2})$}}
\put(2000,-600){\line(1,0){2000}}
\put(4000,-600){\line(0,1){1500}}
{\put(2000,-600){\line(4,3){2000}}}
\put(3600,200){$\Omega_{e_{j}}$}
\put(3800,300){\vector(4,1){600}}
\put(4500,400){Base element, $\Omega_{e_{j}}\in\Omega_{\mathfrak{X}_{i}}$}
\put(3333,-100){/}
\put(-300,2800){ \large{\underline{\bf A schematic of the dynamic adaptive linear restriction method}}}
%\put(3133,600){$(\xi_{c},\eta_{c})$}
\put(3500,-150){\vector(4,-1){900}}
\put(4500,-450){The redistributed slope of}
\put(4500,-750){the solution on $\Omega_{e_{j}}\in\Omega_{\mathfrak{X}_{i}}$}
\put(500,2500){$ \bullet \ \ (\xi_{\ell},\eta_{\ell})$ -- vertices of $\Omega_{e_{j}}$}
\put(500,2300){$\bullet \ \  U_{i}^{\max},U_{i}^{\min}$ -- the extrema over the \emph{stencil} $\Omega_{\mathfrak{X}_{i}}$}
\put(500,2100){$\bullet \ \  \Phi_{\boldsymbol{x}_{\ell}}$ -- the minmod function at $(\xi_{\ell},\eta_{\ell})$}
\put(500,1900){$\bullet \ \  W_{\ell}$ -- the vertex-weighted difference of averages}
\put(500,1700){$\bullet \ \  \mathscr{R}_{\ell}$ -- the redistribution factor}
\put(500,1500){$\bullet \ \  U(\boldsymbol{x}_{\ell})|_{i+j\leq 1}$ -- the updated solution at $(\xi_{\ell},\eta_{\ell})$}
\put(500,1300){$\bullet \ \  U_{ij}|_{i+j\leq 1}$ -- the updated monomial coefficents}
\end{picture}}}
\caption{Here we provide a key for the adaptive linear restriction method from \textsection{3.4}.  Again this limiting procedure depends on the \emph{stencil} $\Omega_{\mathfrak{X}_{i}}$, the linear part of the full solution of order $p$, and on a redistribution strategy that can be thought of heuristically as depending on a ``consistent redistribution of the slopes of the linear coefficients.''}
\label{fig:linres}
\end{figure}
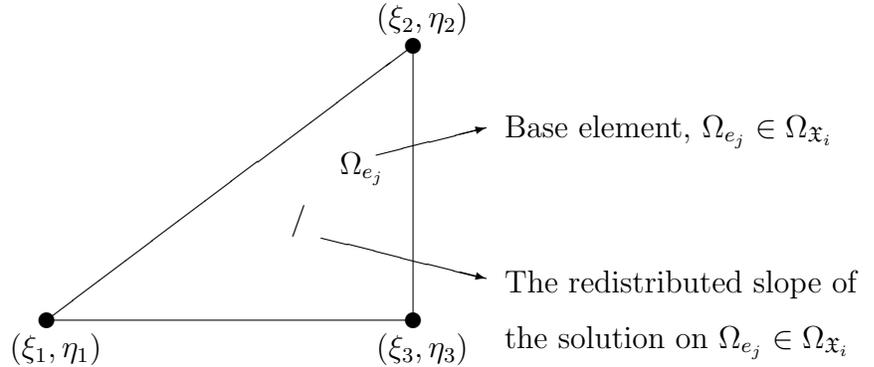

Proceeding, we estimate the average vertex value over the \emph{stencil} to its value on the minmod'ed \emph{neighborhood} by computing, $\mathrm{Avg}_{\ell}(U(\boldsymbol{x}_{\ell})|_{i+j\leq 1})=\frac{1}{3}\sum_{\ell}U(\boldsymbol{x}_{\ell})|_{i+j\leq 1}$, and then we calculate a vertex-weighted difference between this average and $U^{e_{j}}_{\ell}$, which is given by: \begin{equation}\label{res3}\mathrm{W}_{\ell} = 3\left(\mathrm{Avg}_{\ell}(U(\boldsymbol{x}_{\ell})|_{i+j\leq 1})-U^{e_{j}}_{\ell}\right).\end{equation}  The restricted difference functions $\mathfrak{D}_{\ell}$ are then given with respect to each vertex $\boldsymbol{x}_{\ell}$, \begin{equation}\label{res4}\mathfrak{D}_{\ell}=\left(U(\boldsymbol{x}_{\ell})|_{i+j\leq 1}-U^{e_{j}}_{\ell}\right)\mathrm{sgn}\mathrm{W}_{\ell}\end{equation} where $\mathrm{sgn}(\cdot)$ is the usual signum function except that $\mathrm{sgn}(0):=1$.  Then, if $\mathfrak{D}_{\ell}$ is positive, which means that either both the average and the approximate solution at the vertex are each larger than $U^{e_{j}}_{\ell}$, or similarly that they are both smaller than $U^{e_{j}}_{\ell}$, then we set: \begin{equation}\label{res5}\mathcal{D} = \max\left(1,\sum_{m=0}^{I} 1\right), \quad \mathrm{where} \ I = \sum_{\ell}\mathrm{sgn}\mathfrak{D}_{\ell}, \quad \mathrm{for \ each} \ \boldsymbol{x}_{\ell} \ \mathrm{restricted \ such \ that} \ \mathfrak{D}_{\ell}>0.\end{equation}

This allows us now to generate a vertex-wise redistribution factor $\mathscr{R_{\ell}}$ over each element, defined simply by setting \begin{equation}\label{res6}\mathscr{R}_{\ell} = \begin{cases} (W_{\ell}\mathrm{sgn}W_{\ell})/\mathcal{D}, & \mathrm{if} \ \mathfrak{D}_{\ell} > 0, \\ 0, & \mathrm{otherwise}, \end{cases}\end{equation} where the maximum allowed value $\mathscr{R}_{\ell}^{\max}$ is determined by: \begin{equation}\label{res7}\mathscr{R}_{\ell}^{\max} =\begin{cases} \left(U(\boldsymbol{x}_{\ell})|_{i+j\leq 1} - U_{\ell}^{\min}\right) & \mathrm{if} \quad \mathrm{sgn}W_{\ell}>0, \\  \left( U_{\ell}^{\max} - U(\boldsymbol{x}_{\ell})|_{i+j\leq 1}\right) & \mathrm{otherwise}.\end{cases}\end{equation}

The approximate values at the vertices are then updated, where we make sure the maximum redistribution amount is not exceeded, $\mathscr{R}_{\ell} = \min(\mathscr{R}_{\ell},\mathscr{R}_{\ell}^{\max})$.   The redistributed vertex value is updated explicitly to satisfy: \begin{equation}\label{res8} U(\boldsymbol{x}_{\ell})|_{i+j\leq 1} :=  U(\boldsymbol{x}_{\ell})|_{i+j\leq 1}-\mathscr{R}_{\ell}\mathrm{sgn}W_{\ell}.\end{equation}  

As an optional step, we add the ability to adapt our limiter to sense areas where substantial overshoots and/or undershoots have occurred, thus marking the presence of potential shock fronts.  We check back to determine that if the redistribution at a specific vertex passes a given tolerance $\varepsilon\in\mathbb{R}^{+}$, then we either zero out the higher order terms if in a fixed order $p$ solution, or we lower our polynomial order from $p$ to $p_{\lim}$ (where $p_{\lim}$ may be $p-1$ or $p_{\min}$, \emph{etc.}) if in a $p$-adaptive context (which will be fully addressed in \textsection{5}).  That is, we define a restriction function $\mathfrak{R}=\mathfrak{R}(\mathscr{P}^{k}, U|_{i+j > 1} )$ that operates either on the restricted solution $U|_{i+j > 1}$ or the local polynomial order $ \mathscr{P}^{k}(\Omega_{e_{l}})$ over the entire cell: \[\mathfrak{R} = \begin{cases} U|_{i+j > 1} = 0 & \mathrm{if}\quad \left(U(\boldsymbol{x}_{\ell})|_{i+j\leq 1} - \Phi_{\boldsymbol{x}_{\ell}}|\leq \varepsilon\right) \land \mathscr{P}^{i+j>1}(\Omega_{e_{l}}), \\ \mathscr{P}^{k}(\Omega_{e_{l}})\to \mathscr{P}^{k-1}(\Omega_{e_{l}}) & \mathrm{if} \quad  \left(U(\boldsymbol{x}_{\ell})|_{i+j\leq 1} - \Phi_{\boldsymbol{x}_{\ell}}|\leq \varepsilon\right) \land \left(p\mathrm{-adaptive}\right) \land \mathscr{P}^{i+j>1}(\Omega_{e_{l}})\end{cases}\] if any of the vertex values exceed the tolerance.  We also note that clearly $\varepsilon$ should have an implicit dependence on $h$.

Finally we make sure that the difference is properly re-weighted for the next computation at the elements next vertex (if one exists) by determining the amount available to redistribute by computing: $W_{\ell} := (W_{\ell}-\mathscr{R}_{\ell}\mathrm{sgn}W_{\ell})$.  This proceeds until no vertices are left to evaluate in the cell.

Thus we arrive with the sub-quadratic approximate solution, but what we need are the coefficients on $\phi_{10}$ and $\phi_{01}$ in the basis.  To get these we must simply invert the following local constant matrix: \begin{equation}\label{res9}\begin{pmatrix} \phi_{00}(\boldsymbol{x}_{1}) & \phi_{10}(\boldsymbol{x}_{2}) & \phi_{01}(\boldsymbol{x}_{3}) \\ \phi_{00}(\boldsymbol{x}_{1}) & \phi_{10}(\boldsymbol{x}_{2}) & \phi_{01}(\boldsymbol{x}_{3}) \\ \phi_{00}(\boldsymbol{x}_{1}) & \phi_{10}(\boldsymbol{x}_{2}) & \phi_{01}(\boldsymbol{x}_{3}) \end{pmatrix}\begin{pmatrix} U_{00} \\ U_{10} \\ U_{01}\end{pmatrix} = \begin{pmatrix}  U(\boldsymbol{x}_{1})|_{i+j\leq 1} \\ U(\boldsymbol{x}_{2})|_{i+j\leq 1} \\ U(\boldsymbol{x}_{3})|_{i+j\leq 1} \end{pmatrix},\end{equation} which provides the unknowns.

\subsection{\texorpdfstring{$\S 3.5$ The hierarchic linear recombination}{\S$3.5$ The hierarchic linear recombination}}

Now, we develop a new slope limiting strategy based on the limiter presented in \textsection{3.4}, but transformed into the Taylor basis $\varsigma_{ij}\in J_{c}^{k}(\mathbb{R}^{2},\mathcal{M})$, and generalized over linear recombinations of linear reconstructions.  

\begin{figure}[!t]
\centering
{\setlength{\unitlength}{4544sp}%
{\begin{picture}(5043,3800)(300,-1000)
{\thinlines
}
{\put(  2000,-600){\circle*{90}}}
{\put(  1800,-800){$(\xi_{1},\eta_{1})$}}
{\put(  4000,-600){\circle*{90}}}
{\put(  3800,-800){$(\xi_{3},\eta_{3})$}}
{\put(  4000,900){\circle*{90}}}
{\put(  3800,1000){$(\xi_{2},\eta_{2})$}}
\put(2000,-600){\line(1,0){2000}}
\put(4000,-600){\line(0,1){1500}}
{\put(2000,-600){\line(4,3){2000}}}
\put(3600,200){$\Omega_{e_{j}}$}
\put(3800,300){\vector(4,1){600}}
\put(4500,400){Base element, $\Omega_{e_{j}}\in\Omega_{\mathfrak{X}_{i}}$}
\put(3333,-100){/}
%\put(3133,600){$(\xi_{c},\eta_{c})$}
\put(3500,-150){\vector(4,-1){900}}
\put(4500,-450){The redistributed slope of}
\put(4500,-650){the solution on $\Omega_{e_{j}}\in\Omega_{\mathfrak{X}_{i}}$}
\put(500,2500){$ \bullet \ \ (\xi_{\ell},\eta_{\ell})$ -- vertices of $\Omega_{e_{j}}$}
\put(-200,2800){ \large{\underline{\bf A schematic of the hierarchic linear recombination method}}}
\put(500,2300){$\bullet \ \ b(i,j)$ -- the monomial indices}
\put(500,2100){$\bullet \ \  U_{i}^{\max},U_{i}^{\min}$ -- the extrema over the \emph{stencil} $\Omega_{\mathfrak{X}_{i}}$}
\put(500,1900){$\bullet \ \  \Phi_{\boldsymbol{x}_{\ell}}$ -- the minmod function at $(\xi_{\ell},\eta_{\ell})$}
\put(500,1700){$\bullet \ \  W_{\ell}$ -- the vertex-weighted difference of averages}
\put(500,1500){$\bullet \ \  \mathscr{R}_{\ell}$ -- the redistribution factor}
\put(500,1300){$\bullet \ \   U(\boldsymbol{x}_{\ell})|_{i-i'+j-j'\leq 1}$ -- the updated}
\put(500,1100){\ \ \ \ \ linear recombination at $(\xi_{\ell},\eta_{\ell})$}
\put(500,900){$\bullet \ \  U_{(i-i')(j-j')}|_{i-i'+j-j'\leq 1}$ -- the updated}
\put(500,700){\ \ \ \ \  monomial coefficents at \emph{level} $\mathfrak{l}(i+j)$}
\end{picture}}}
\caption{Here we provide a key for the hierarchic linear recombination method of \textsection{3.5}.  This procedure depends on the chosen \emph{stencil} $\Omega_{\mathfrak{X}_{i}}$, a collection of linear recombinations of restricted subsets of monomial coefficients from the total solution, and an application of the method developed in \textsection{3.4} to these linear recombinations in order to recover the limited solution.}
\label{fig:linrecom}
\end{figure}
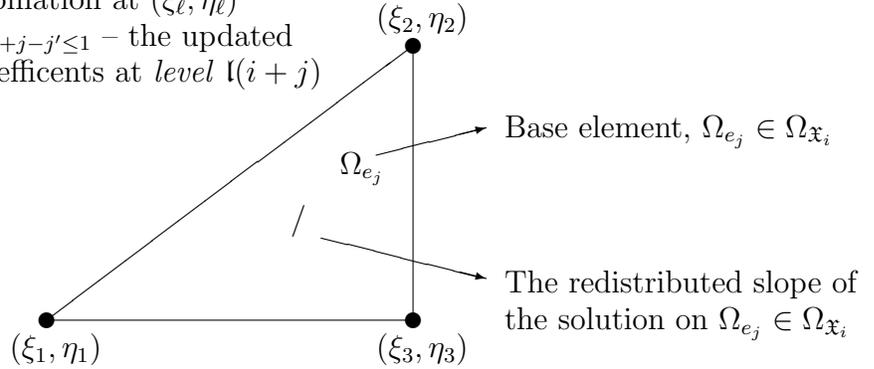

More clearly, we take our transformed solutions (\ref{limit1}) such that in the Taylor basis we can extract the hierarchical basis at any \emph{level} $\mathfrak{l}$, independently of cell vertices $\boldsymbol{x}_{i}$, by simply extracting for any hierarchical index $b$ the set $\{\mathscr{C}_{b},\mathscr{C}_{b+g},\mathscr{C}_{b+g+1}\}$ from \textsection{3.2}.  Notice that this set is entirely determined by its indices $i$ and $j$ by way of $b(i,j)$.   That is, we can simply denote $\{\mathscr{C}_{b},\mathscr{C}_{b+g},\mathscr{C}_{b+g+1}\}$ as the first three coefficients of the $(i+j)$--th derivative of $U^{\mathrm{v}}_{h}$.  As in \textsection{3.2} this provides our linear reconstruction, such that equation (\ref{higher}) becomes our effective sub-quadratic restriction of the $(i+j)$--th derivative of $U^{\mathrm{v}}_{h}$ which we substitute into the formalism of \textsection{3.4}.  That is we set \[ U(\boldsymbol{x}_{\ell})|_{i-i'+j-j'\leq 1}=\mathscr{C}_{b}  +\mathscr{C}_{b+g}(\eta_{i} -\eta_{c})+ \mathscr{C}_{b+g+1}(\xi_{i} -\xi_{c}),\] where $i'$ and $j'$ correspond to the sub-quadratic polynomial basis in the derivation of $U^{\mathrm{v}}_{h}$ with coefficients at \emph{level} $\mathfrak{l}(i+j)$; or, correspond to the coefficients of the linear recombination at \emph{level} $\mathfrak{l}(i+j)$.

Then (\ref{res1}) is calculated, where we evaluate over every $\mathscr{C}_{b}$ in decreasing order.  That is, for $b+g+1\leq s$, we compute starting at the top $(k-1)$-st order derivative  steps (\ref{res1})--(\ref{res9}) from  \textsection{3.4} with respect to each base coefficient $b$ at that \emph{level} $\mathfrak{l}(k-1)$.   Then, due to the redundacy of representation for the mixed terms as discussed \textsection{3.3}, we employ any of our minmod functions $\Phi_{\mathfrak{X}}^{*}$ from \textsection{3.3} (note that in the experiments below we always use the MUSCL minmod function).  This is performed until we reach the \emph{level} corresponding to $b=1$, at which point we perform the calculation one more time identically to that presented in \textsection{3.4} except in the Taylor basis.

Notice here that when the top order is linear, or when $p=k=1$ the strategy from \textsection{3.4} is equivalent to \textsection{3.5} up to a change of basis (for example in (\ref{res9}) the $\phi_{ij}$'s become $\varsigma_{ij}$'s), which provides for identical error behavior at $p=k=1$.

\section{\texorpdfstring{\protect\centering $\S 4$ Slope limiting: numerical results}{\S 4 Slope limiting: numerical results}}

In this section we solve two example problems for an advected scalar quantity $\iota=\iota(t,\boldsymbol{x})$.   All of our solutions have been run in parallel using an upwinding scheme for the choice of flux.

\subsection{\texorpdfstring{$\S 4.1$ Convergence of solutions}{$\S 4.1$ Convergence of solutions}}

\begin{figure}[ht]
\centering
\includegraphics[width=11cm]{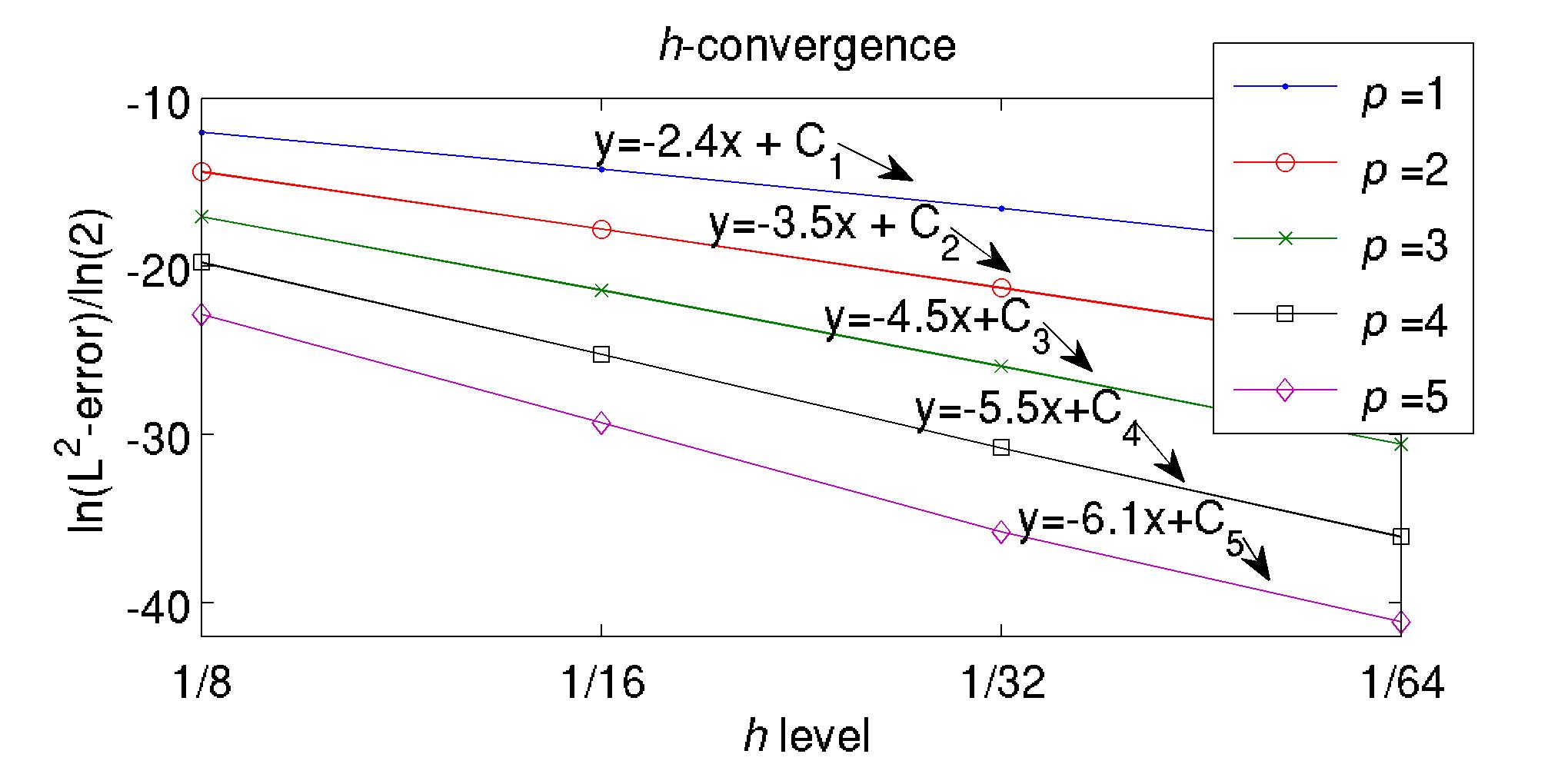}
\caption{The regression rates of convergence for the $p\in\{1,\ldots,5\}$ cases are given by the slope of a linear regression line taken from the data in Table \ref{convchart}.} 
\label{hconv} 
\end{figure}

\begin{figure}[ht]
\centering
\includegraphics[width=11cm]{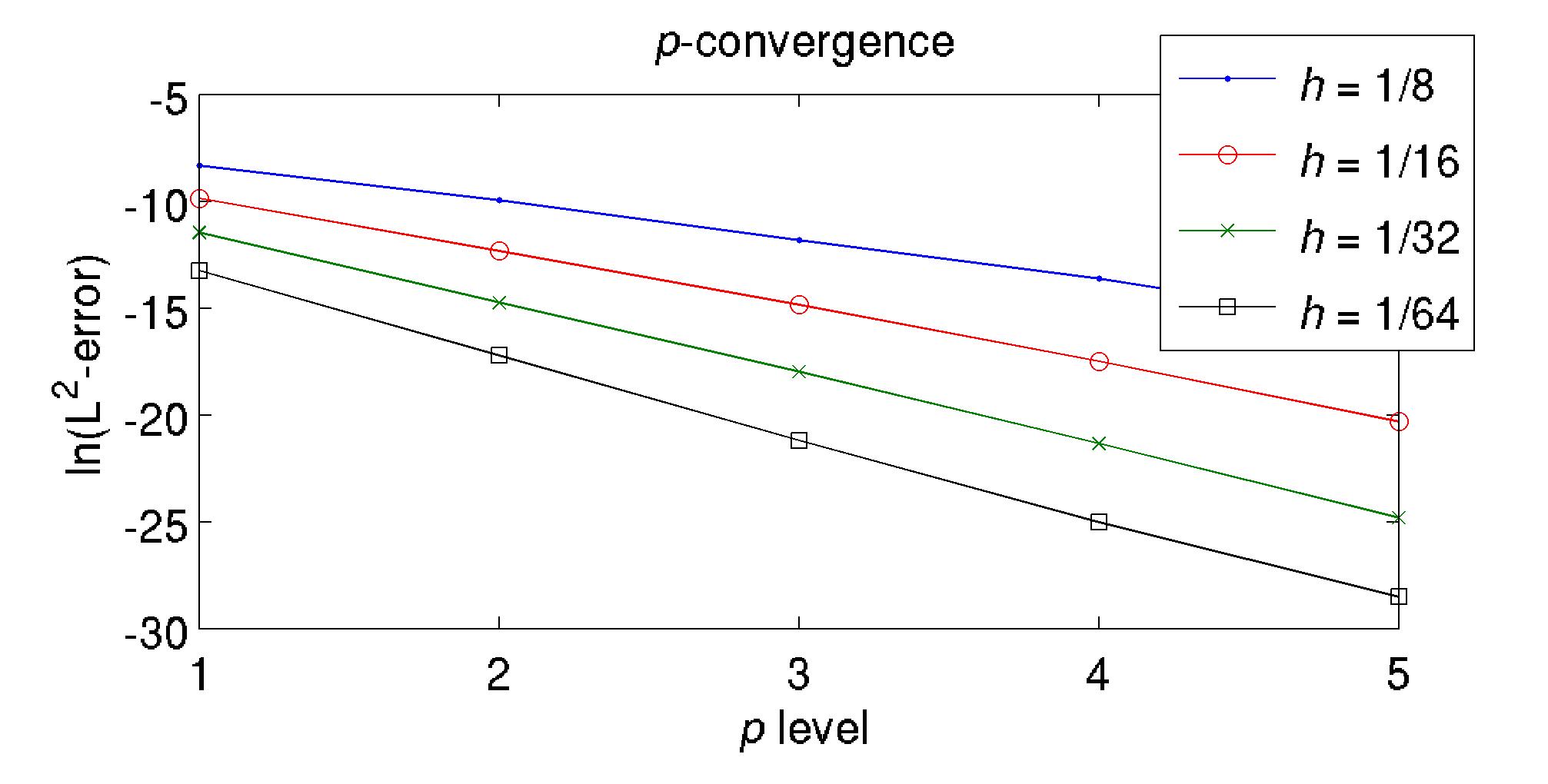}
\caption{The convergence in $p\in\{1,\ldots,5\}$ for the different mesh sizes, as taken from the data in Table \ref{convchart}.} 
\label{pconv} 
\end{figure}

 The examples developed in \textsection{4.2} and \textsection{4.3} both display discontinuities that have meaningful affects on the theoretical rates of convergence.  Thus first we simply restrict to a smooth solution.  That is, we use the same formalism of a scalar transport equation (\ref{rota}) developed in detail in \textsection{4.2}, though in this case we change the initial conditions to a smooth Gaussian centered at the origin, given by $\iota_{0}= a_{0}e^{-(x^{2}+y^{2})/25}$, where $a_{0}=1$ and the boundary condition is the standard transmissive condition on both $\iota_{b}$ and $\boldsymbol{u}_{b}$.  This is a steady state Gaussian field that ``rotates'' about the origin by way of a pseudo--timestepping.  The convergence results are shown in Table \ref{convchart}, Figure \ref{hconv} and Figure \ref{pconv}.

\begin{table}[t!]
\centering
\begin{tabular}{|c | c | c | c |}
%\toprule[0.1em]
\hline
$p$ &  $L^{2}/L^{\infty}$--error  & $L^{2}_{loc}$ projection error & $x=1/h$ \rule{0pt}{3ex} \rule[0ex]{0pt}{0pt} \\ 
\hline\hline

1 &  $ 1.76\times 10^{-6}/5.31\times 10^{-8}$  &  $4.57 \times 10^{-7}$ & 64 \rule{0pt}{3ex} \rule[0ex]{0pt}{0pt} \\
\hline
2 &  $ 3.34\times 10^{-8}/4.75\times 10^{-10}$ &  $1.63 \times 10^{-11}$ & 64 \rule{0pt}{3ex} \rule[0ex]{0pt}{0pt} \\
\hline
3 &  $ 6.24\times 10^{-10}/3.19\times 10^{-11}$ &  $8.61 \times 10^{-15}$ & 64 \rule{0pt}{3ex} \rule[0ex]{0pt}{0pt} \\
\hline
4 &  $ 1.36\times 10^{-11}/1.63\times 10^{-12}$ &  $8.22 \times 10^{-19}$ & 64 \rule{0pt}{3ex} \rule[0ex]{0pt}{0pt} \\
\hline
5 &  $ 4.14\times 10^{-13}/6.54\times 10^{-14}$ &  $5.31 \times 10^{-22}$ & 64 \rule{0pt}{3ex} \rule[0ex]{0pt}{0pt} \\
\hline
%5 &  $ 1.01\times 10^{-13}/2.27\times 10^{-14}$ & exact in basis & 64 \rule{0pt}{3ex} \rule[0ex]{0pt}{0pt} \\
%\hline
%6 &  $ 6.97\times 10^{-15}/1.57\times 10^{-15}$ & exact in basis & 64 \rule{0pt}{3ex} \rule[0ex]{0pt}{0pt} \\
\hline
1 &  $ 1.05\times 10^{-5}/6.84\times 10^{-7}$  & $7.12 \times 10^{-6}$ & 32 \rule{0pt}{3ex} \rule[0ex]{0pt}{0pt} \\
\hline
2 &  $ 3.94\times 10^{-7}/1.10\times 10^{-8}$ & $1.07 \times 10^{-9}$ & 32 \rule{0pt}{3ex} \rule[0ex]{0pt}{0pt} \\
\hline
3 &  $ 1.55\times 10^{-8}/3.79\times 10^{-10}$ & $2.16 \times 10^{-12}$& 32 \rule{0pt}{3ex} \rule[0ex]{0pt}{0pt} \\
\hline
4 &  $ 5.35\times 10^{-10}/2.49\times 10^{-11}$ & $8.52 \times 10^{-16}$& 32 \rule{0pt}{3ex} \rule[0ex]{0pt}{0pt} \\
\hline
5 &  $ 1.67\times 10^{-11}/1.42\times 10^{-12}$ & $2.12 \times 10^{-18}$& 32 \rule{0pt}{3ex} \rule[0ex]{0pt}{0pt} \\
\hline
%5 &  $ 8.92\times 10^{-13}/2.08\times 10^{-13}$ & exact in basis & 32 \rule{0pt}{3ex} \rule[0ex]{0pt}{0pt} \\
%\hline
%6 &  $ 7.61\times 10^{-14}/1.40\times 10^{-14}$ & exact in basis & 32 \rule{0pt}{3ex} \rule[0ex]{0pt}{0pt} \\
\hline
1 &  $ 5.24\times 10^{-5}/6.17\times 10^{-6}$  & $1.07 \times 10^{-4}$& 16 \rule{0pt}{3ex} \rule[0ex]{0pt}{0pt} \\
\hline
2 &  $ 4.44\times 10^{-6}/2.78\times 10^{-7}$ & $6.85 \times 10^{-8}$& 16 \rule{0pt}{3ex} \rule[0ex]{0pt}{0pt} \\
\hline
3 &  $ 3.56\times 10^{-7}/2.31\times 10^{-8} $ & $5.12 \times 10^{-10}$& 16 \rule{0pt}{3ex} \rule[0ex]{0pt}{0pt} \\
\hline
4 &  $ 2.55\times 10^{-8}/5.73\times 10^{-10}$ & $8.78 \times 10^{-13}$ & 16 \rule{0pt}{3ex} \rule[0ex]{0pt}{0pt} \\
\hline
5 &  $ 1.50\times 10^{-9}/4.36\times 10^{-11}$ & $7.89 \times 10^{-15}$ & 16 \rule{0pt}{3ex} \rule[0ex]{0pt}{0pt} \\
\hline\hline
%5 &  $ 2.52\times 10^{-11}/4.91\times 10^{-12}$ & exact in basis & 16 \rule{0pt}{3ex} \rule[0ex]{0pt}{0pt} \\
%\hline
%6 &  $ 2.14\times 10^{-12}/5.93\times 10^{-13}$ & exact in basis & 16 \rule{0pt}{3ex} \rule[0ex]{0pt}{0pt} \\
%\hline
1 &  $ 2.39\times 10^{-4}/5.17\times 10^{-5}$  & $1.30 \times 10^{-3}$ & 8 \rule{0pt}{3ex} \rule[0ex]{0pt}{0pt} \\
\hline
2 &  $ 4.73\times 10^{-5}/6.89\times 10^{-6}$ & $3.54 \times 10^{-6}$& 8 \rule{0pt}{3ex} \rule[0ex]{0pt}{0pt} \\
\hline
3 &  $ 7.36\times 10^{-6}/7.58\times 10^{-7}$ & $9.55 \times 10^{-8}$ & 8 \rule{0pt}{3ex} \rule[0ex]{0pt}{0pt} \\
\hline
4 &  $  1.21\times 10^{-6}/6.41\times 10^{-8}$ & $5.98 \times 10^{-10}$ & 8 \rule{0pt}{3ex} \rule[0ex]{0pt}{0pt} \\
\hline
5 &  $  1.32\times 10^{-7}/6.07\times 10^{-9}$ & $2.19 \times 10^{-11}$ & 8 \rule{0pt}{3ex} \rule[0ex]{0pt}{0pt} \\
\hline
%5 &  $  1.93\times 10^{-9}/1.32\times 10^{-10}$ & exact in basis & 8 \rule{0pt}{3ex} \rule[0ex]{0pt}{0pt} \\
%\hline
%6 &  $  2.33\times 10^{-10}/2.12\times 10^{-11}$ & exact in basis & 8 \rule{0pt}{3ex} \rule[0ex]{0pt}{0pt} \\
%\hline
\end{tabular}
\caption{We show the convergence results for the $h$ and $p$ levels whose errors are bounded by machine precision after 64 timesteps.  The $L^{2}_{loc}$ projection error into the basis is also included, though, as is clear, these errors are often \emph{below} machine double precision ($\sim 1.11\times 10^{-16}$), and hence not particularly meaningful.}
\label{convchart}
\end{table}

\subsection{\texorpdfstring{$\S 4.2$ The rotating half annular crest, cone, and hill solution}{$\S 4.2$ The rotating half annular crest, cone, and hill solution}}

Here we solve a standard rotating landscape solution to a scalar transport equation.  That is, consider the hyperbolic advection problem: \begin{equation} \label{rota}\partial_{t}\iota+\boldsymbol{u}\cdot\nabla_{x}\iota = 0,\end{equation} with initial-boundary data given by \[ \iota_{|t=0}=\iota_{0},\quad\mathrm{and}\quad  \iota_{b} = 0,\] corresponding to vanishing boundary data, given a time-independent velocity vector field $\boldsymbol{u}=\boldsymbol{u}(\boldsymbol{x})$ with the transported scalar quantity $\iota=\iota(t,\boldsymbol{x})$ in dimension two, such that $\boldsymbol{x} = (x,y)$ and $\boldsymbol{u} = (u,v)$.  

\begin{table}[t]
\centering
\begin{tabular}{|c | c | c | c | c |}
%\toprule[0.1em]
\hline
$p$ &  \bf{Limiter type} &  $\frac{L^{2}\mathrm{error}}{L^{\infty}\mathrm{error}}$  & \bf{Limiter type} &  $\frac{L^{2}\mathrm{error}}{L^{\infty}\mathrm{error}}$ \rule{0pt}{3ex} \rule[0ex]{0pt}{0pt} \\ 
\hline\hline
1 & BJ limiter\textsuperscript{\cite{BarthJesp},\textsection{3.2}}  & $\left(\frac{1.5\times 10^{-3}}{0.73}\right)$ & Adapted BJ\textsuperscript{\textsection{3.2.1}} &  $\left(\frac{1.5\times 10^{-3}}{0.73}\right)$ \rule{0pt}{3ex} \rule[0ex]{0pt}{0pt}  \\ 
\hline
1 & DEO limiter\textsuperscript{\cite{DEO}} & $\left(\frac{1.1\times 10^{-3}}{0.71} \right)$
 & BDS limiter\textsuperscript{\cite{Bell},\textsection{3.4}} & $\left(\frac{5.9\times 10^{-4}}{0.67}\right)$ \rule{0pt}{3ex} \rule[0ex]{0pt}{0pt}  \\
\hline
1 & Vertex\textsuperscript{\cite{Kuzmin,LBL},\textsection{3.2}} & $\left(\frac{1.1\times 10^{-3}}{0.73} \right)$  & Adapted vertex\textsuperscript{\textsection{3.2.1}} & $\left(\frac{1.0\times 10^{-3}}{0.72} \right)$   \rule{0pt}{3ex} \rule[0ex]{0pt}{0pt} \\ 
\hline
1 &  Recombination\textsuperscript{\textsection{3.5}} &  $\left(\frac{5.9\times 10^{-4}}{0.67}\right)$  & Reconstruction\textsubscript{MUSCL}\textsuperscript{ \cite{Abgrall2,LSTZ},\textsection{3.3}} &  $\left(\frac{5.9\times 10^{-4}}{0.67}\right)$ \rule{0pt}{3ex} \rule[0ex]{0pt}{0pt} \\
\hline\hline
2 & BJ limiter\textsuperscript{\cite{BarthJesp},\textsection{3.2}}  & $\left(\frac{2.3\times 10^{-3}}{0.74}\right)$  & Restriction \textsuperscript{\cite{Bell},\textsection{3.4}} & $\left(\frac{5.0\times 10^{-4}}{0.64}\right)$   \rule{0pt}{3ex} \rule[0ex]{0pt}{0pt} \\ 
\hline
2 & Vertex \textsuperscript{\cite{Kuzmin,LBL},\textsection{3.2}} &  $\left(\frac{2.3\times 10^{-3}}{0.73} \right)$  & Adapted vertex\textsuperscript{\textsection{3.2.1}} &  $\left(\frac{2.3\times 10^{-3}}{0.74} \right)$    \rule{0pt}{3ex} \rule[0ex]{0pt}{0pt} \\ 
\hline
2 &  Recombination\textsuperscript{\textsection{3.5}} &  $\left(\frac{2.2\times 10^{-3}}{0.74}\right)$  & Reconstruction\textsubscript{ENO}\textsuperscript{ \cite{Abgrall2,LSTZ},\textsection{3.3}} &  $\left(\frac{2.3\times 10^{-3}}{0.73}\right)$ \rule{0pt}{3ex} \rule[0ex]{0pt}{0pt} \\

\hline\hline

3 & BJ limiter\textsuperscript{\cite{BarthJesp},\textsection{3.2}}  & $\left(\frac{2.6\times 10^{-3}}{0.72}\right)$  & Restriction \textsuperscript{\cite{Bell},\textsection{3.4}} & $\left(\frac{4.7\times 10^{-4}}{0.73}\right)$   \rule{0pt}{3ex} \rule[0ex]{0pt}{0pt} \\ 
\hline
3 & Vertex \textsuperscript{\cite{Kuzmin,LBL},\textsection{3.2}} & $ \left(\frac{2.6\times 10^{-3}}{0.72}\right)$  & Adapted vertex\textsuperscript{\textsection{3.2.1}} & $\left(\frac{2.5\times 10^{-3}}{0.73}\right)$   \rule{0pt}{3ex} \rule[0ex]{0pt}{0pt} \\ 
\hline
3 &  Recombination\textsuperscript{\textsection{3.5}} &  $\left(\frac{2.6\times 10^{-3}}{0.72}\right)$  & Reconstruction\textsubscript{ENO}\textsuperscript{ \cite{Abgrall2,LSTZ},\textsection{3.3}} &  $\left(\frac{2.2\times 10^{-3}}{0.75}\right)$ \rule{0pt}{3ex} \rule[0ex]{0pt}{0pt} \\
\hline\hline
4 & BJ limiter\textsuperscript{\cite{BarthJesp},\textsection{3.2}}  & $\left(\frac{2.8\times 10^{-3}}{0.72}\right)$  & Restriction \textsuperscript{\cite{Bell},\textsection{3.4}} & $\left(\frac{4.8\times 10^{-4}}{0.69}\right)$   \rule{0pt}{3ex} \rule[0ex]{0pt}{0pt} \\ 
\hline
4 & Vertex \textsuperscript{\cite{Kuzmin,LBL},\textsection{3.2}} & $\left(\frac{2.9\times 10^{-3}}{0.72}\right)$  & Adapted vertex\textsuperscript{\textsection{3.2.1}} & $\left(\frac{2.8\times 10^{-3}}{0.72}\right)$  \rule{0pt}{3ex} \rule[0ex]{0pt}{0pt} \\ 
\hline
4 &  Recombination\textsuperscript{\textsection{3.5}} &  $\left(\frac{2.9\times 10^{-4}}{0.72}\right)$  & Reconstruction\textsubscript{ENO}\textsuperscript{ \cite{Abgrall2,LSTZ},\textsection{3.3}} &  $\left(\frac{2.6\times 10^{-3}}{0.73}\right)$ \rule{0pt}{3ex} \rule[0ex]{0pt}{0pt} \\
\hline 
\end{tabular}
\caption{We give the $L^{2}$ and $L^{\infty}$-errors of the approximate solutions after one full rotation with respect to (\ref{rota}), setting $h=1/256$, $\Delta t = 1\times 10^{-3}$ and using Runge--Kutta SSP$(5,3)$.  The error ratio for the solution with no limiter at $p=1$ is $L^{2}/L^{\infty}= 2.55\times 10^{-4}/0.61$, at $p=2$ is  $L^{2}/L^{\infty}= 2.28\times 10^{-4}/0.44$, at $p=3$ is  $L^{2}/L^{\infty}= 1.71\times 10^{-4}/0.35$, and for $p>3$ is unstable.  Though, as expected, the error in the stable unlimited solutions concentrate along the discontinuities demonstrating sharp ($\geq 10 \%$ cell-wise in $\iota$) overshoots and undershoots.}
\label{table:errortab1}
\end{table}

\begin{figure}[t!]\centering\includegraphics[width=7.2cm]{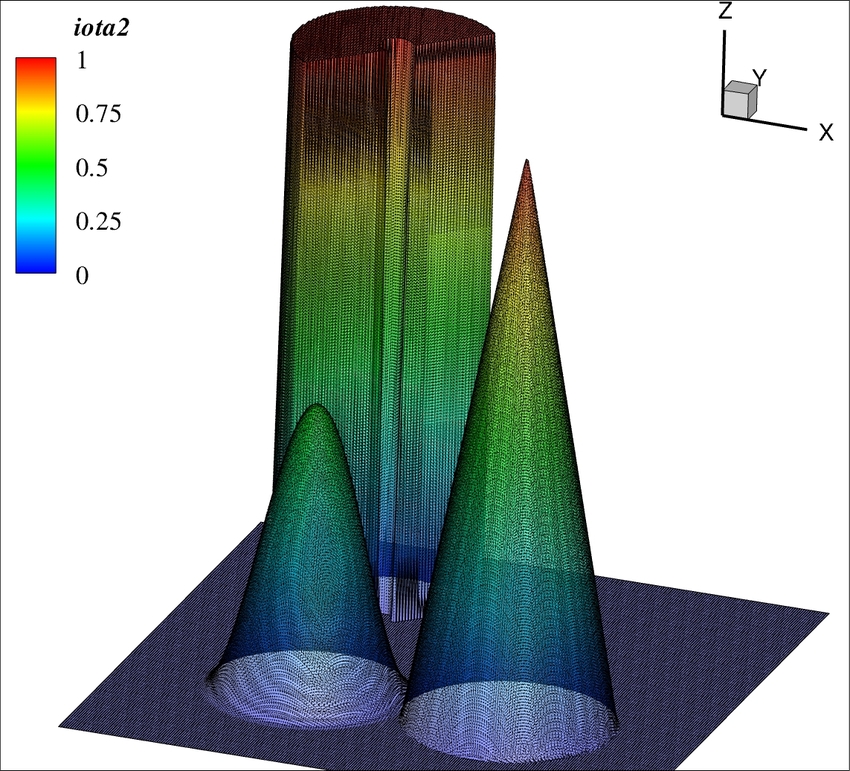}\includegraphics[width=7.2cm]{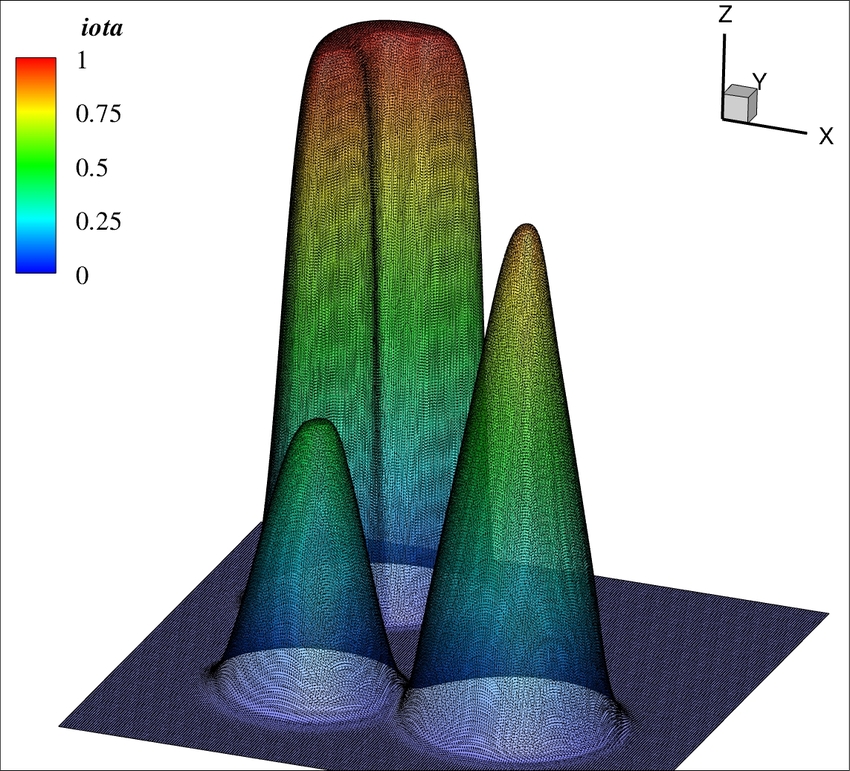} \\ \includegraphics[width=7.2cm]{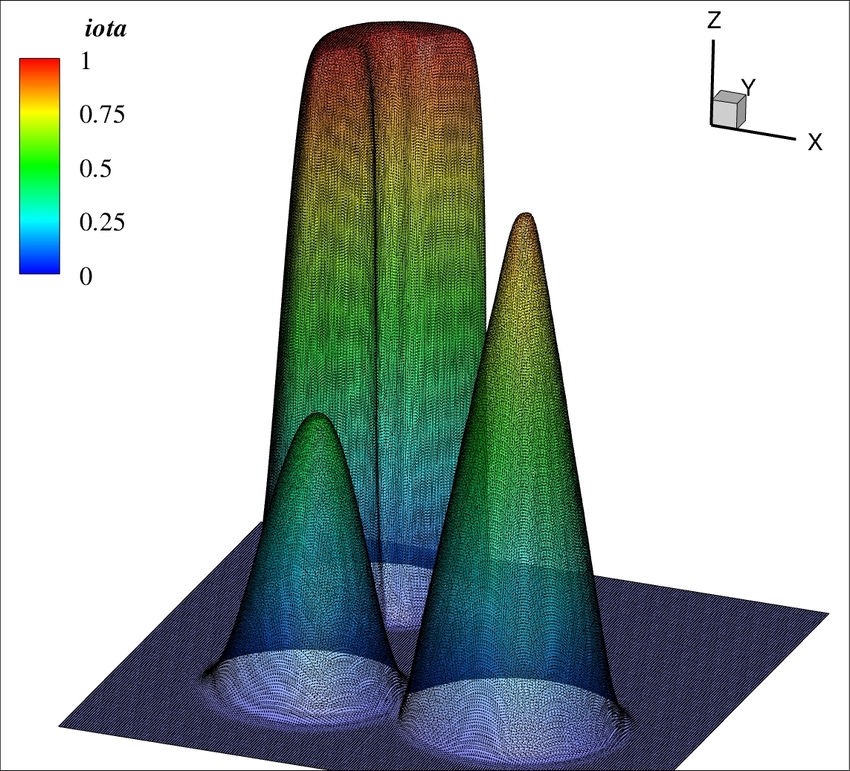}\includegraphics[width=7.2cm]{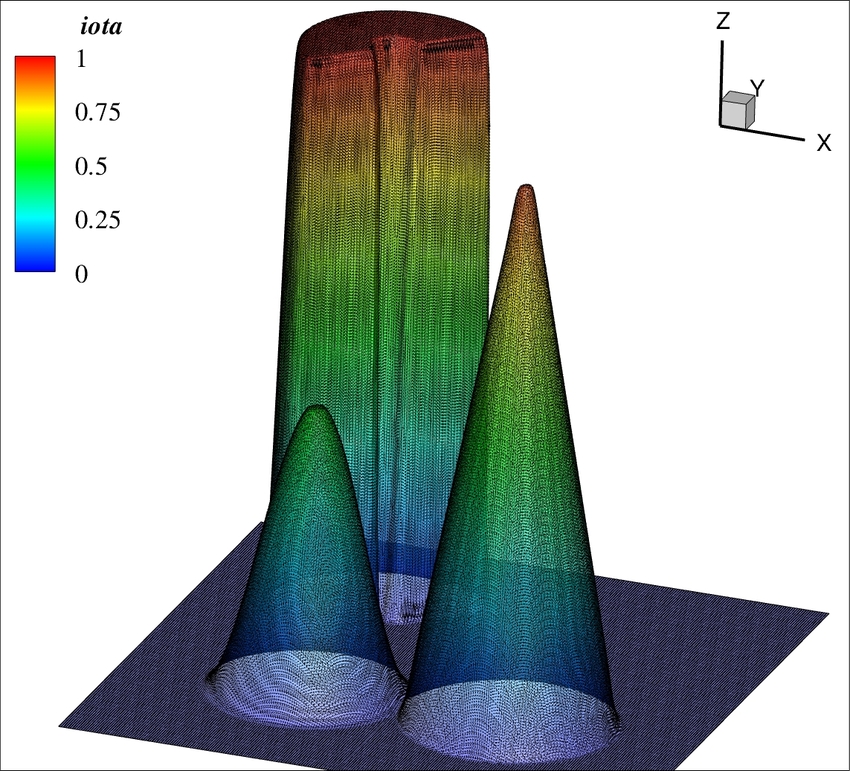}\\ \includegraphics[width=7.2cm]{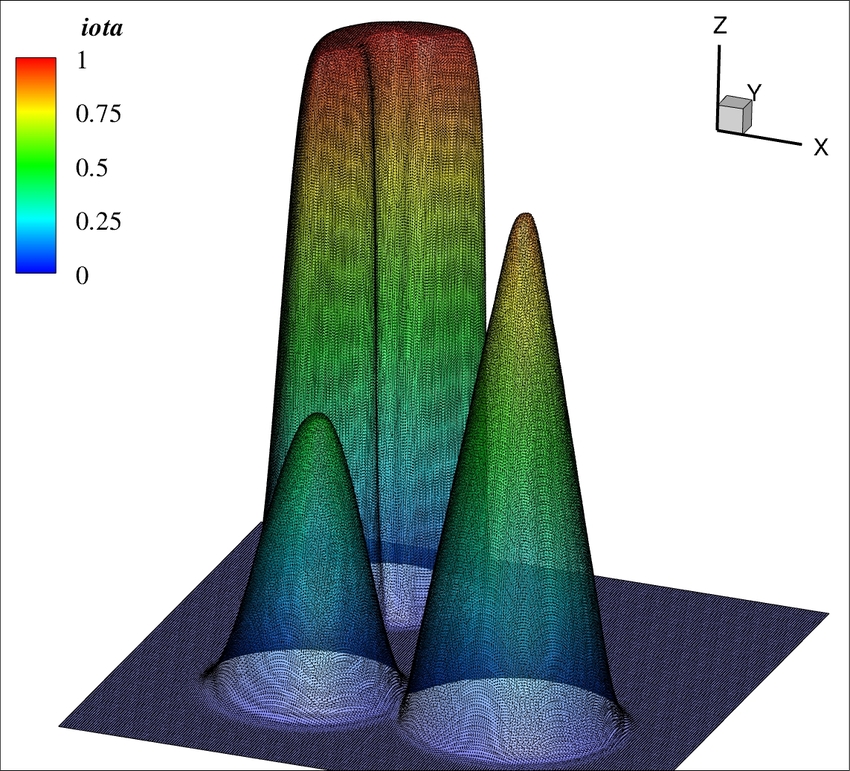}\includegraphics[width=7.2cm]{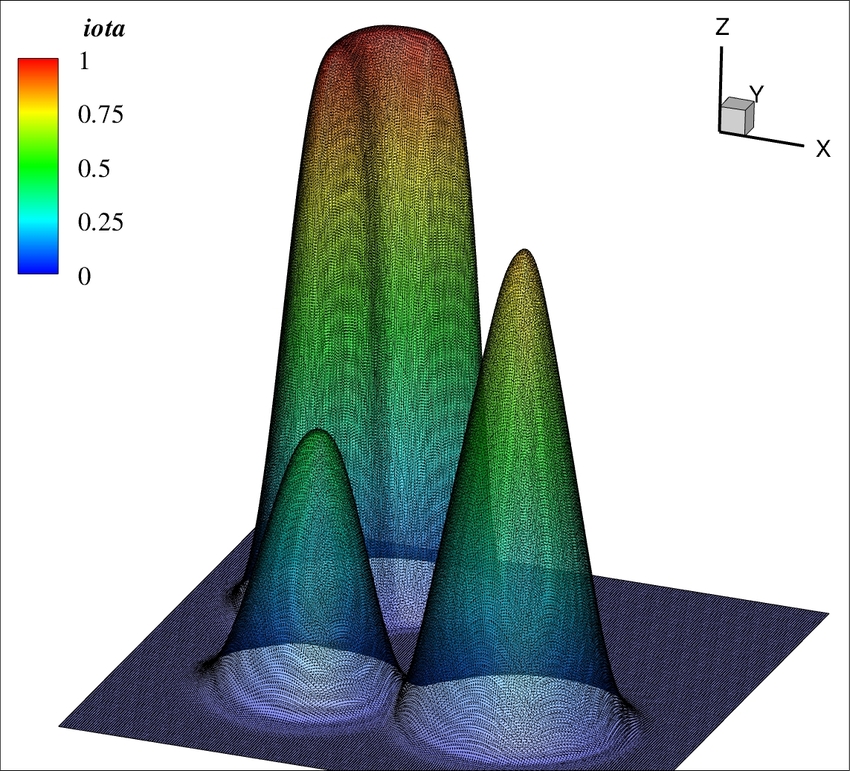}\caption{Here we show the $p=1$ results from Table \ref{table:errortab1} after one full revolution.  The upper left is the exact $L^{2}$ projection at $p=1$, the top right is the DEO limiter\textsuperscript{\cite{DEO}}, the middle left is the vertex limiter\textsuperscript{\cite{Kuzmin,LBL},\textsection{3.2}}, the middle right the BDS limiter\textsuperscript{\cite{Bell},\textsection{3.4}}, the bottom left the adapted vertex limiter\textsuperscript{\textsection{3.2.1}}, and the bottom right the BJ limiter\textsuperscript{\cite{BarthJesp},\textsection{3.2}}.}\label{fig:rotconst1}\end{figure}

We choose a simple square domain $\Omega = \big[-\frac{1}{2},\frac{1}{2}\big]^2$, with velocity field $\boldsymbol{u} = (y,-x)$.  Then letting $\tau_{\mathcal{O}}=\pi/4$ and defining the auxiliary variables \[\mathcal{O}_{x} = x\cos\tau_{\mathcal{O}} - y\sin\tau_{\mathcal{O}}\quad \mathrm{and} \quad \mathcal{O}_{y}= y\cos\tau_{\mathcal{O}}+x\sin\tau_{\mathcal{O}},\] we take initial data satisfying: \begin{equation}\label{rotain}\iota_{0} = \left\{\begin{matrix}  1 , & \mathrm{if} \ A, \\  1 - Ba^{-1} , & \mathrm{if} \  B \leq a,  \\ \frac{1}{4}\left( 1 + \cos{\pi r} \right), & \mathrm{otherwise},\end{matrix}\right.\end{equation} where \[\begin{aligned} A = & \left(a_{0}\leq B\leq a \right)\land \left( \mathcal{O}_{x} \leq a_{1} \right), \qquad B = \sqrt{\left( \mathcal{O}_{x}-\frac{1}{4}\right)^{2} + \mathcal{O}_{y}^{2}}, \end{aligned}\] and \[ r = a^{-1}\min\left(a,\sqrt{\mathcal{O}_{x}^{2} + (\mathcal{O}_{y}+1/4)^{2}}\right),\] taking $a=0.18$, $a_{0}=0.025$ and $a_{1}=-0.23$. 

The exact solution may be determined by noticing that since for any $F(x,y)$, where $x = x(t)$ and $y=y(t)$, that \[\frac{dF}{dt} = \partial_{t}F + \begin{pmatrix} x' \\ y' \end{pmatrix}\nabla F = 0,\] which implies that for \[\boldsymbol{u} = \begin{pmatrix} u \\ v \end{pmatrix} = \begin{pmatrix} y \\ -x \end{pmatrix},\quad \mathrm{we \ have \ the \ system}\quad x ' = y \quad \mathrm{and}\quad y' = -x,\] that may be solved by recombining such that the solution to the second order ODE, $y'' + y = 0$ can be viewed as a generator of the rotation matrix $R$ about the origin.  That is, we obtain the clockwise transformation \begin{equation}\label{clock}R = \begin{pmatrix}\cos t & -\sin t \\ \sin t & \cos t \end{pmatrix},\end{equation} such that $R\boldsymbol{x}$ yields the exact solution.  

\begin{figure}[t!]\centering\includegraphics[width=7.2cm]{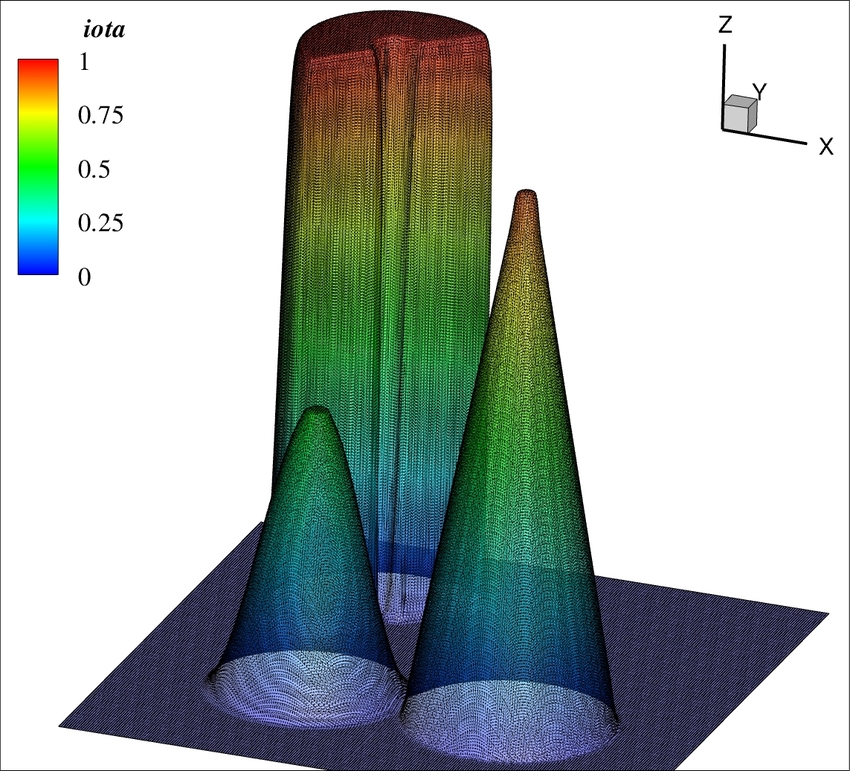}\includegraphics[width=7.2cm]{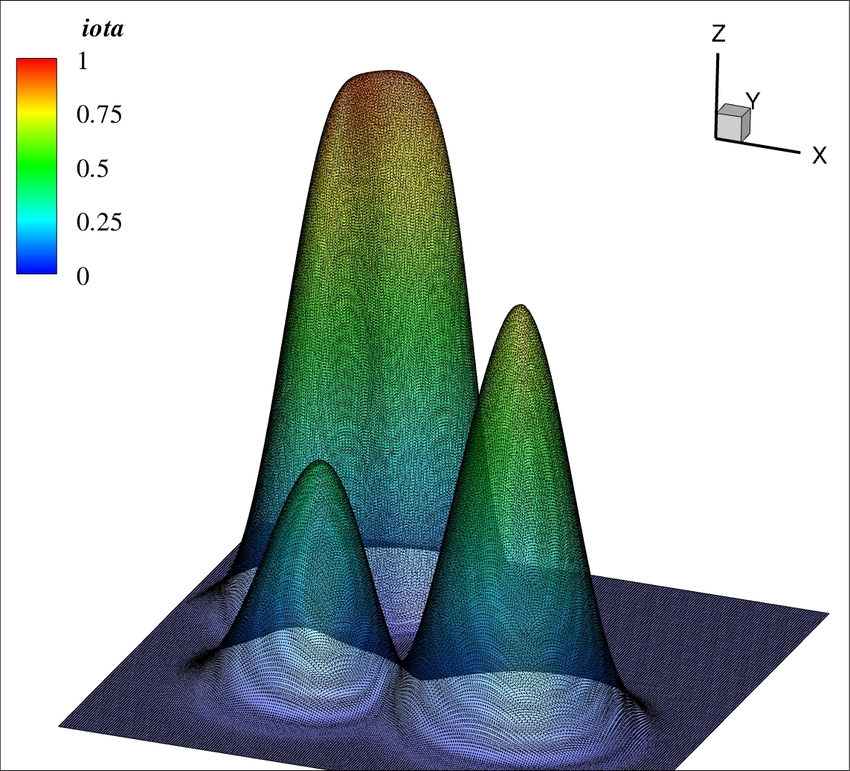} \\ \includegraphics[width=7.2cm]{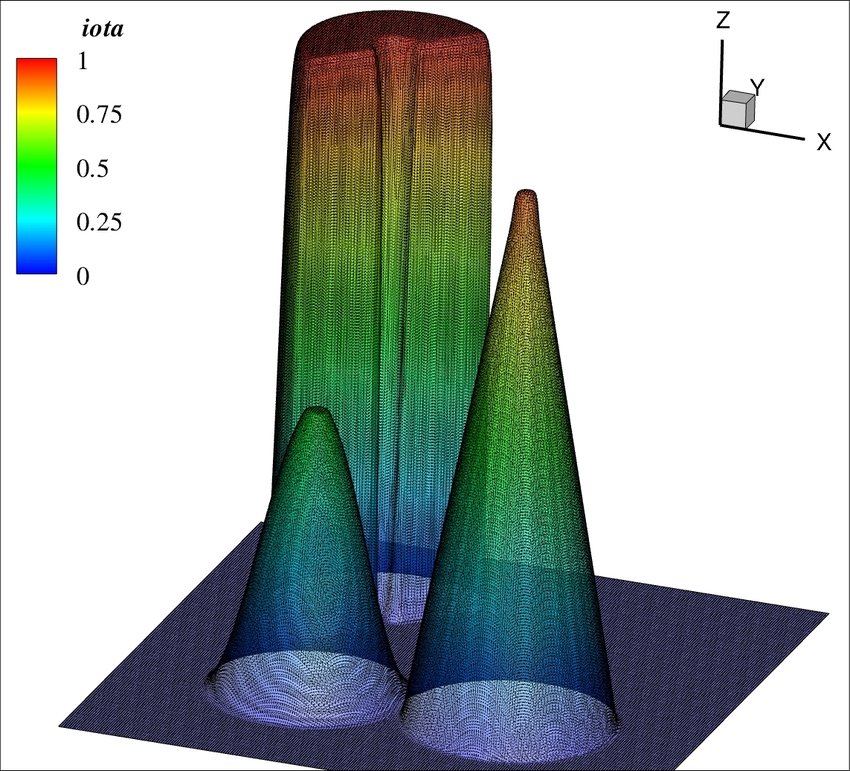}\includegraphics[width=7.2cm]{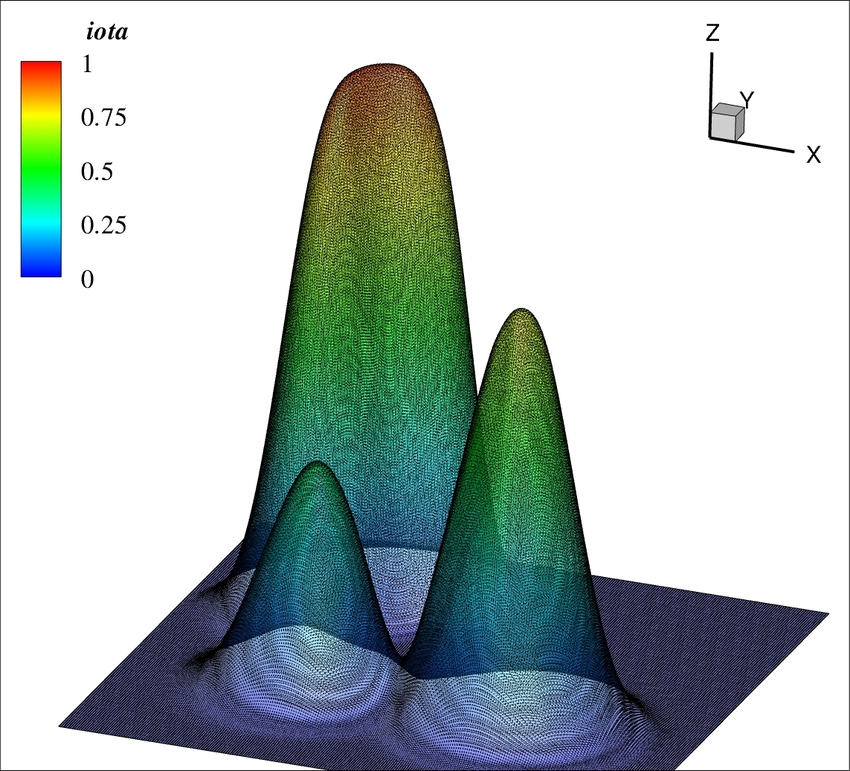}\\ \includegraphics[width=7.2cm]{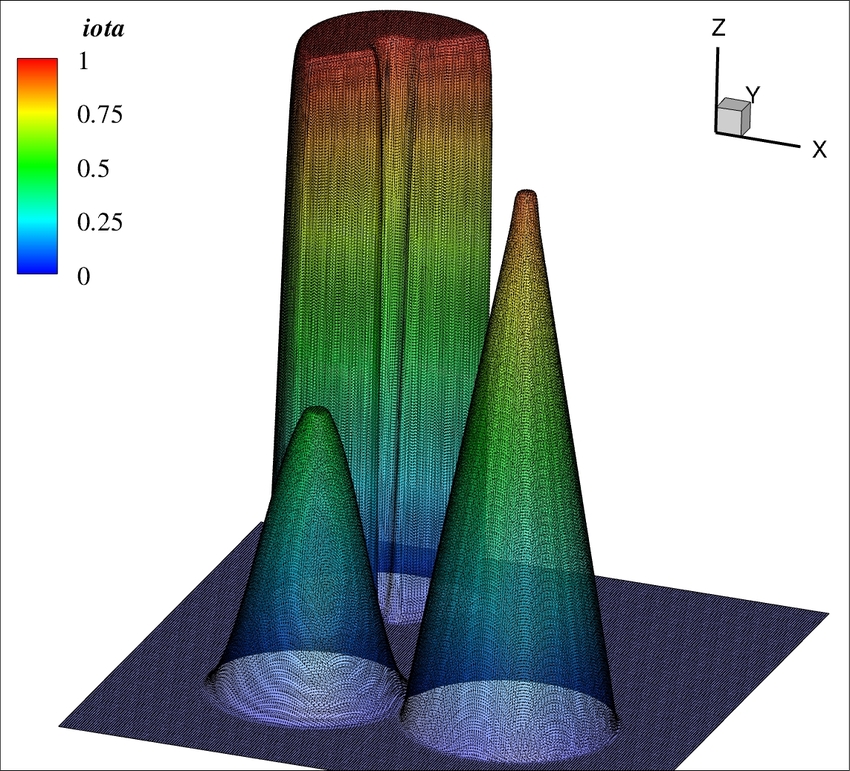}\includegraphics[width=7.2cm]{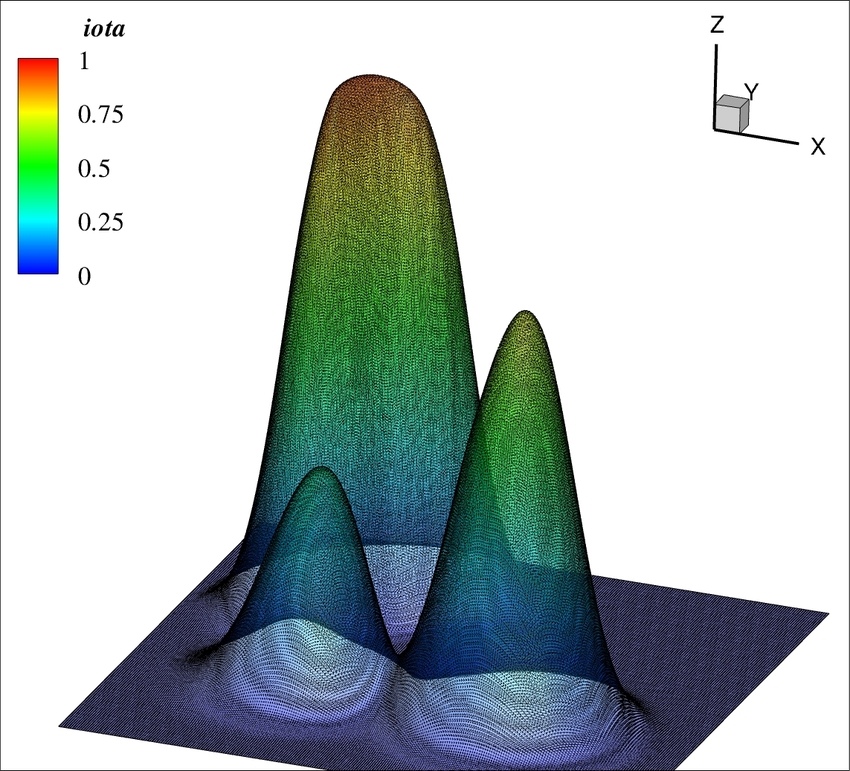}\caption{Here we show the $p=2$ to $p=4$ results from Table \ref{table:errortab1} after one full revolution.  The left column shows the linear restriction of the BDS limiter\textsuperscript{\cite{Bell},\textsection{3.4}} in descending order, while the right column shows the next best limiter, in descending order, \emph{i.e.} at $p=2$, $p=3$ and $p=4$ the hierarchical reconstruction\textsubscript{ENO}\textsuperscript{ \cite{Abgrall2,LSTZ},\textsection{3.3}}.}\label{fig:rotconst2}\end{figure}

For our numerical experiments, we follow a similar case to that presented in \cite{Kuzmin} setting our mesh width to $h=1/128$ and $\Delta t = 1\times 10^{-3}$ in keeping with the CFL condition on hyperbolic transport (\emph{e.g.} see \cite{ThomasJW})   ).  Let us briefly discuss the results shown in Figure \ref{fig:rotconst1} and Figure \ref{fig:rotconst2} and Table \ref{table:errortab1}.  We note that we have run all of our experiments on a regular structured triangular grid.

In Figure \ref{fig:rotconst1} we see the results for linears.  The Durlofsky--Engquist--Osher\textsuperscript{\cite{DEO}} limiter, the vertex limiter \cite{Kuzmin} and the adapted vertex limiter seem to show qualitatively similar behaviors.  The Barth--Jespersen limiter \cite{BarthJesp} is slightly more diffuse here at linears (where the adapted Barth--Jespersen shows only slight improvement over the native Barth--Jespersen limiter as well), while the BDS limiter \cite{Bell} described in \textsection{3.4} shows by far the best $L^{2}$--error behavior and clearly maintains the best signature behavior of the solution everywhere but at the points of discontinuity, where these values are tightly redistributed.  As previously suggested \cite{Kuzmin} the vertex limiter and the Barth--Jespersen limiter are both quite sensitive to mesh geometries, where the former is better suited in some sense to geometries with ``sharp angles,'' and the latter (the Barth--Jespersen limiter) is well-suited for regular structured meshes (\emph{e.g.} Delaunay triangulations).  However, because of the so--called ``blind diffusion'' of both these regimes caused by local extrema --- as discussed in \textsection{3.2.1} --- this behavior is not entirely predictable or monotone with respect to mesh regularity, as we see below.  The  hierarchic linear recombination from \textsection{3.5}, and the hierarchical reconstruction from \textsection{3.3} are both equivalent by construction to the BDS limiter at $p=1$.

When $p>1$ we see an immediate and substantial degradation in the limiting behavior of all the regimes, with the single exception of the linear restriction of the BDS limiter from \textsection{3.4}.  This is immediately prevalent at $p=2$, where the hierarchic linear recombination method from \textsection{3.3} is the next best limiting regime and yet has an $L^2$-error more than four times that of the linear restriction.  In fact, the hierarchical reconstruction method from \textsection{3.3} may be the most natural extension of the BDS limiter to order $p$, where the choice of linearization is the most direct application of the BDS scheme in the Taylor basis.  But even here, where at $p=2$ we have added only three more degrees of freedom to the polynomial hierarchical basis, we see that performing the limiter on the linear reconstructions --- which amounts to performing the limiting procedure on only two more components (\emph{i.e.} the linear components which are limited with respect to their respective slopes) --- shows a substantial loss locally in the sharpness of the resolution along the discontinuities.  

The reason for this loss of resolution is not entirely mysterious or unexpected, though previous work \cite{Kuzmin} has demonstrated geometries where this degradation is not immediately observable at $p=2$, and this behavior seems related to the mass lumping strategy previously discussed in \textsection{3.2.1} (which deserves closer analysis).  Nevertheless, here we see that as $p$ increases the number of applications of the limiter to the solution increases as a function of the degrees of freedom at the $(p-1)$-st degree (\emph{i.e.} $(p-1)(p-2)/2$).   In fact this is true for each of the limiting regimes, with the exception of the  hierarchical reconstruction methods from \textsection{3.3}, which actually perform yet another iteration of the limiter by employing one of the minmod functions at top order.  However, the hierarchical reconstruction methods also seem to benefit from the fact that they utilize information coming from nonlinearities present in the solution at \emph{every level} by linearizing with respect to these nonlinearities (\emph{e.g.} equation (\ref{linavg})) --- in contrast to the vertex-based schemes which linearize about a single monomial component (\ref{higher}) of the expansion, and then utilize a regularizing constraint such as (\ref{alphasec}).  It turns out that the addition of this nonlinear signature behavior at higher order seems to allow the hierarchical reconstruction methods to capture the profile more completely, even with the additional application of the limiting regime at each timestep.  

However, by far the most effective limiting regime for $p>1$ is the linear restriction of the BDS limiter from \textsection{3.4}, where in $\mathfrak{R}$ the $\varepsilon$ has been set to $10^{-4}$.  Again, this result is not entirely unexpected, since slope limiting, as its name suggests, finds its roots in limiting the slopes of lines with respect to some linear basis \cite{vl1}.  That having been said, it then seems unlikely that one should be able to expect an improvement in the accuracy of a solution near a sharp front simply by applying the slope limiter more frequently to the linearization of its respective monomial components.  Since, for example, if one assumes (fairly realistically) that the top order component has an approximately fixed order error which is introduced upon application of the limiter to the FEM solution, then each subsequent application of the limiter to the lower \emph{level} $\mathfrak{l}$ components should only be able to increase the subsequent error introduced over all.  In the hierarchical reconstruction methods of \cite{Abgrall2,LSTZ}, on the other hand, the componentwise minmod function attenuates this effect somewhat, as does the fact that all of the limited higher order components serve to help limit the lower order components at every \emph{level} $\mathfrak{l}$. 

Before discussing this further, let us first confirm that this result is not simply a special case of (\ref{rota}) which demonstrates a pathological behavior with respect to (\ref{rotain}).  Below we take a solution with admits a number of additional types of singular submanifolds that help to further explicate each limiter's behavior.

\subsection{\texorpdfstring{$\S 4.3$ Steady state convective torque}{$\S 4.3$ Steady state convective torque}}

Now we show a steady state solution to equation (\ref{rota}), which effectively isolates the error present in the form of torque away from the steady (discontinuous) state in a rotating constant frame solution.  Our goal here is to present a more difficult set of singular submanifolds $\mathcal{B}_{i}\subset \mathcal{B}$ present with respect to a steady state solution (where the solution here is thought of as the base manifold $\mathcal{B}$) in order to more completely isolate the error explicitly introduced by the limiting regimes over varying order $p$.  

Here we work over the Cartesian domain $\Omega = [0,2]\times[-1,1]$, given the same boundary conditions from \textsection{4.1}, and where the exact steady solution is characterized by a velocity field satisfying $\boldsymbol{u} = (y,1-x)$ and a steady state scalar field $\iota$ given by: \begin{equation}\label{convt} \iota = \left\{\begin{matrix} 2 -\frac{2}{3}r, & \mathrm{if} \ r\leq a_{1} \\  2 a_{1}\left( 1+ \cos{[(r - a_{2})\pi]}\right), & \mathrm{if} \  a_{1} < r \leq 3.5 a_{3}  \\ 3a_{1}, & \mathrm{if} \ 4 a_{3}\leq r\leq 2 a_{1} \\    3 a_{3}\left( 1+ \cos{[(r - a_{2})\pi]}\right), & 6 a_{3} \leq r\leq 7 a_{3} \\  a_{1}, & \mathrm{if} \ 8 a_{3} \leq r\leq 9 a_{3} \\  0, & \mathrm{otherwise}\end{matrix}\right.\end{equation} where  \[ r = \sqrt{(x - 1)^{2} + (y)^{2}}, \quad a_{1} = \frac{1}{4}, \quad a_{2}=\frac{13}{3}\quad \mathrm{and}\quad a_{3}=\frac{1}{10}.\] 

\begin{table}[h]
\centering
\begin{tabular}{|c | c | c | c | c |}
%\toprule[0.1em]
\hline
$p$ &  \bf{Limiter type} &  $\frac{L^{2}\mathrm{error}}{L^{\infty}\mathrm{error}}$  & \bf{Limiter type} &  $\frac{L^{2}\mathrm{error}}{L^{\infty}\mathrm{error}}$ \rule{0pt}{3ex} \rule[0ex]{0pt}{0pt} \\ 
\hline\hline
1 & BJ limiter\textsuperscript{\cite{BarthJesp},\textsection{3.2}}  & $\left(\frac{1.2\times 10^{-2}}{0.49}\right)$ & Vertex\textsuperscript{\cite{Kuzmin,LBL},\textsection{3.2}} & $\left(\frac{1.2\times 10^{-2}}{0.51} \right)$  \rule{0pt}{3ex} \rule[0ex]{0pt}{0pt}  \\ 
\hline
1 & DEO limiter\textsuperscript{\cite{DEO}} & $\left(\frac{1.0\times 10^{-2}}{0.47} \right)$
 & BDS limiter\textsuperscript{\cite{Bell},\textsection{3.4}} & $\left(\frac{6.8\times 10^{-3}}{0.40}\right)$ \rule{0pt}{3ex} \rule[0ex]{0pt}{0pt}  \\
\hline
1 &  Recombination\textsuperscript{\textsection{3.5}} &  $\left(\frac{6.8\times 10^{-3}}{0.40}\right)$  & Reconstruction\textsubscript{ENO}\textsuperscript{ \cite{Abgrall2,LSTZ},\textsection{3.3}} &  $\left(\frac{6.8\times 10^{-3}}{0.40}\right)$ \rule{0pt}{3ex} \rule[0ex]{0pt}{0pt} \\
\hline\hline
2 & BJ limiter\textsuperscript{\cite{BarthJesp},\textsection{3.2}}  & $\left(\frac{1.9\times 10^{-2}}{0.50}\right)$  & Restriction \textsuperscript{\cite{Bell},\textsection{3.4}} & $\left(\frac{6.6\times 10^{-3}}{0.38}\right)$   \rule{0pt}{3ex} \rule[0ex]{0pt}{0pt} \\ 
\hline
2 & Vertex \textsuperscript{\cite{Kuzmin,LBL},\textsection{3.2}} &  $\left(\frac{1.9\times 10^{-2}}{0.50} \right)$  & Adapted vertex\textsuperscript{\textsection{3.2.1}} &  $\left(\frac{1.9\times 10^{-2}}{0.50}\right)$    \rule{0pt}{3ex} \rule[0ex]{0pt}{0pt} \\ 
\hline
2 &  Recombination\textsuperscript{\textsection{3.5}} &  $\left(\frac{1.9\times 10^{-2}}{0.50}\right)$  & Reconstruction\textsubscript{ENO}\textsuperscript{ \cite{Abgrall2,LSTZ},\textsection{3.3}} &  $\left(\frac{1.8\times 10^{-2}}{0.52}\right)$ \rule{0pt}{3ex} \rule[0ex]{0pt}{0pt} \\

\hline\hline

3 & BJ limiter\textsuperscript{\cite{BarthJesp},\textsection{3.2}}  & $\left(\frac{2.2\times 10^{-2}}{0.50}\right)$  & Restriction \textsuperscript{\cite{Bell},\textsection{3.4}} & $\left(\frac{7.7\times 10^{-3}}{0.39}\right)$   \rule{0pt}{3ex} \rule[0ex]{0pt}{0pt} \\ 
\hline
3 & Vertex \textsuperscript{\cite{Kuzmin,LBL},\textsection{3.2}} & $\left(\frac{2.3\times 10^{-2}}{0.50}\right)$  & Adapted vertex\textsuperscript{\textsection{3.2.1}} & $\left(\frac{2.2\times 10^{-2}}{0.50}\right)$   \rule{0pt}{3ex} \rule[0ex]{0pt}{0pt} \\ 
\hline
3 &  Recombination\textsuperscript{\textsection{3.5}} &  $\left(\frac{2.3\times 10^{-2}}{0.50}\right)$  & Reconstruction\textsubscript{ENO}\textsuperscript{ \cite{Abgrall2,LSTZ},\textsection{3.3}} &  $\left(\frac{2.2\times 10^{-2}}{0.51}\right)$ \rule{0pt}{3ex} \rule[0ex]{0pt}{0pt} \\
\hline\hline
4 & BJ limiter\textsuperscript{\cite{BarthJesp},\textsection{3.2}}  & $\left(\frac{2.3\times 10^{-2}}{0.50}\right)$  & Restriction \textsuperscript{\cite{Bell},\textsection{3.4}} & $\left(\frac{7.7\times 10^{-3}}{0.38}\right)$   \rule{0pt}{3ex} \rule[0ex]{0pt}{0pt} \\ 
\hline
4 & Vertex \textsuperscript{\cite{Kuzmin,LBL},\textsection{3.2}} & $\left(\frac{2.3\times 10^{-2}}{0.50}\right)$  & Adapted vertex\textsuperscript{\textsection{3.2.1}} & $\left(\frac{2.3\times 10^{-2}}{0.50}\right)$  \rule{0pt}{3ex} \rule[0ex]{0pt}{0pt} \\ 
\hline
4 &  Recombination\textsuperscript{\textsection{3.5}} &  $\left(\frac{2.3\times 10^{-2}}{0.50}\right)$  & Reconstruction\textsubscript{ENO}\textsuperscript{ \cite{Abgrall2,LSTZ},\textsection{3.3}} &  $\left(\frac{2.2\times 10^{-2}}{0.51}\right)$ \rule{0pt}{3ex} \rule[0ex]{0pt}{0pt} \\
\hline 
\end{tabular}
\caption{We give the $L^{2}$ and $L^{\infty}$-errors of the approximate solutions after $T$ corresponding to a $1/4$ rotation with respect to (\ref{convt}), setting $h=1/128$, $\Delta t = 5\times 10^{-4}$ and using Runge--Kutta SSP$(5,3)$.  The error ratio for the solution with no limiter at $p=1$ is $L^{2}/L^{\infty}= 3.9\times 10^{-3}/0.40$, at $p=2$ is  $L^{2}/L^{\infty}= 2.9\times 10^{-3}/0.34$, at $p=3$ is  $L^{2}/L^{\infty}= 2.4\times 10^{-3}/0.23$, and for $p>3$ is unstable.  Again, as in Table 2, the unlimited solutions are dominated by local overshoots and undershoots along the discontinuities.}
\label{table:errortab2}
\end{table}

\begin{figure}[t!]
\centering\includegraphics[width=7.3cm]{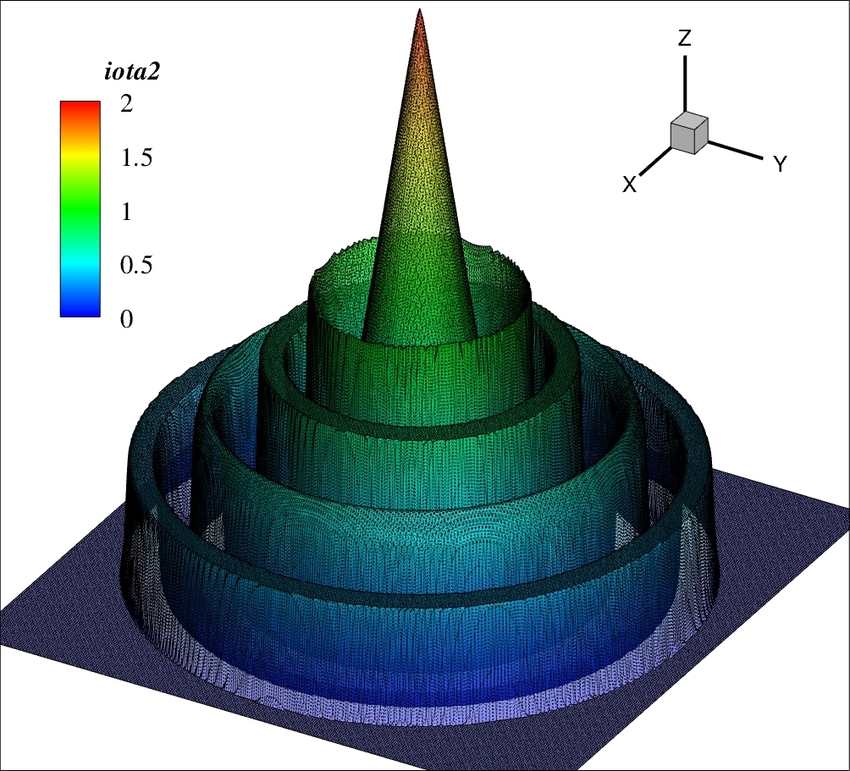}\includegraphics[width=7.3cm]{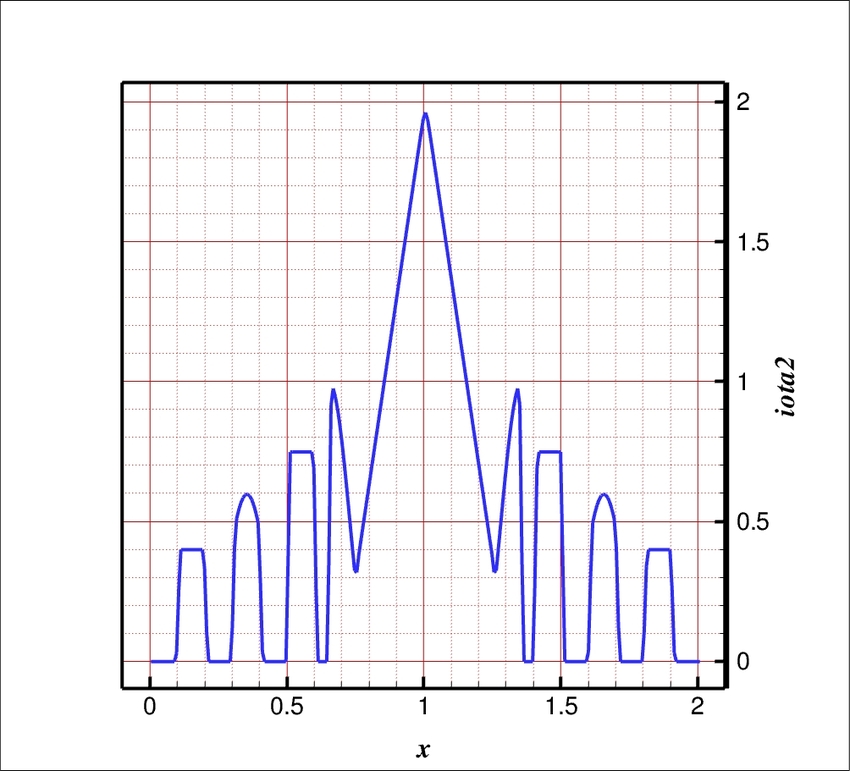} \\ \centering\includegraphics[width=7.3cm]{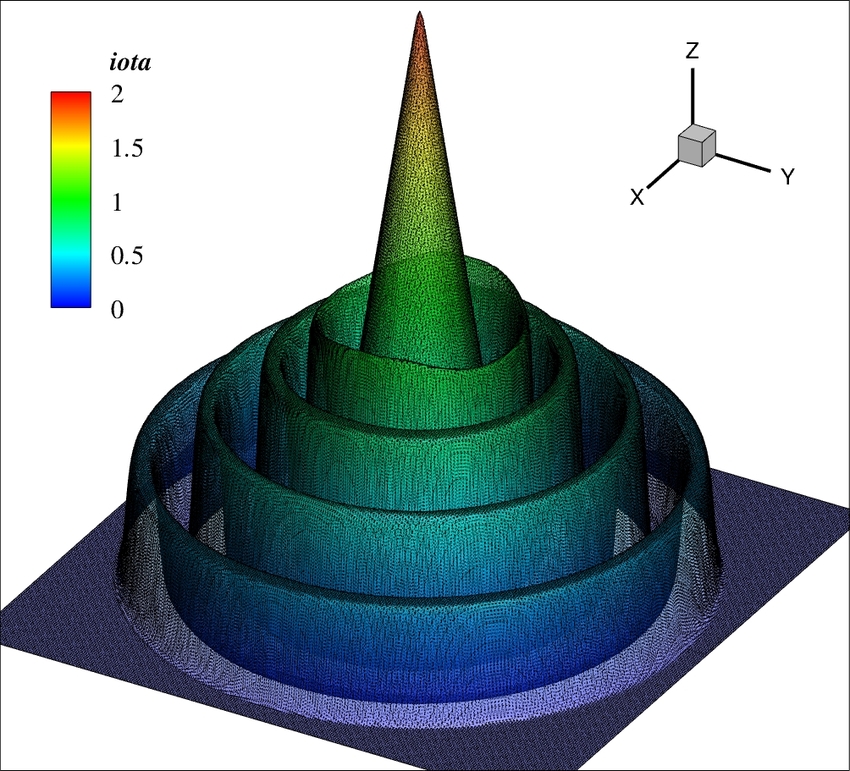}\includegraphics[width=7.3cm]{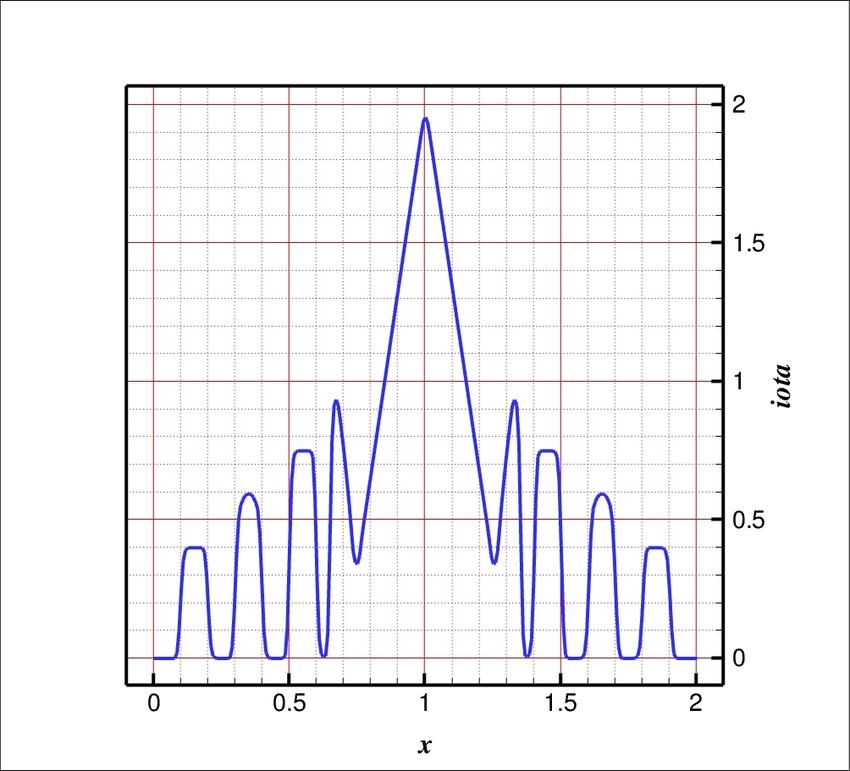} \\ \centering\includegraphics[width=7.3cm]{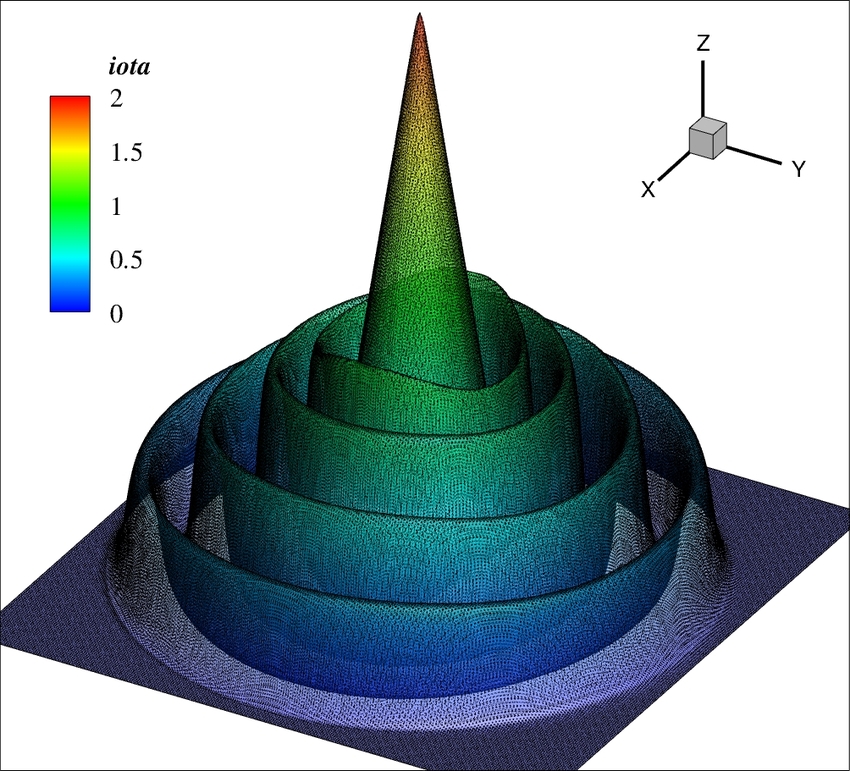}\includegraphics[width=7.3cm]{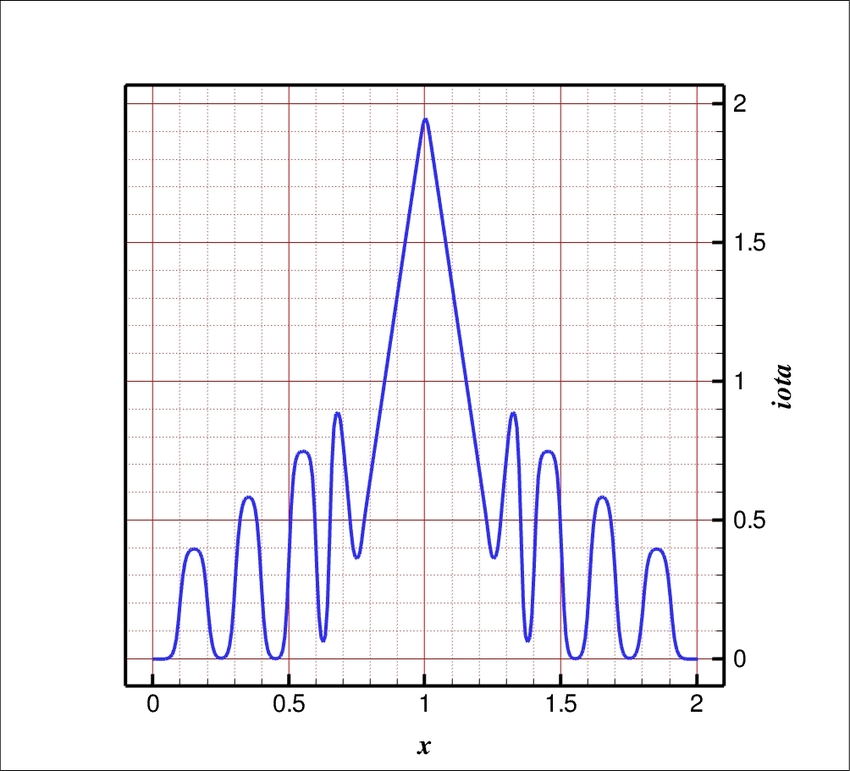} \caption{Here at top we show the $L^2$--projection of the exact solution at $p=7$, with the $xz$--plane slice on the right after $1/4$ turn.  The middle shows the $p=1$ case of the linear restriction \textsuperscript{\cite{Bell},\textsection{3.4}}, and the bottom shows the $p=1$ DEO limiter\textsuperscript{\cite{DEO}}.}\label{fig:ODE21}  \end{figure}

The solution $\mathcal{B}$ as shown in Figure \ref{fig:ODE21} is augmented from the relatively well-behaved circular convection case analyzed in \cite{Kuzmin}.  Here we have similar outer rings (though substantially ``thinned''), but have supplemented a pair of inner ring submanifolds that have a thickness of no more than a single point that similarly intersects an inner cone along a line of singular points, and with a very thin island outer ring.  These initial conditions are not particularly well-behaved, as can be seen in Figure \ref{fig:ODE21}, where even in the $L^2$--projected exact solution at $p=7$ there are variations (jagged lines) at the mesh resolution along the lines of singular points.  To compound this, we use a larger domain than that of  \cite{Kuzmin}, which effectively doubles the velocity of the pseudo-timestepping in the $y$-direction, providing for even more instability in the solution space.

\begin{figure}[t!]
\centering\includegraphics[width=7.3cm]{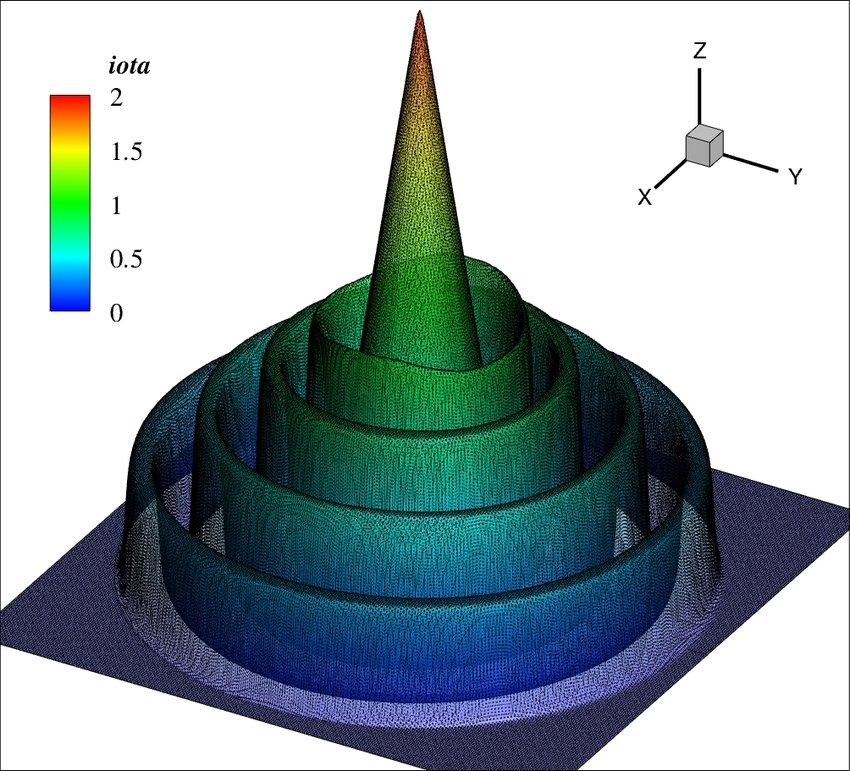}\includegraphics[width=7.3cm]{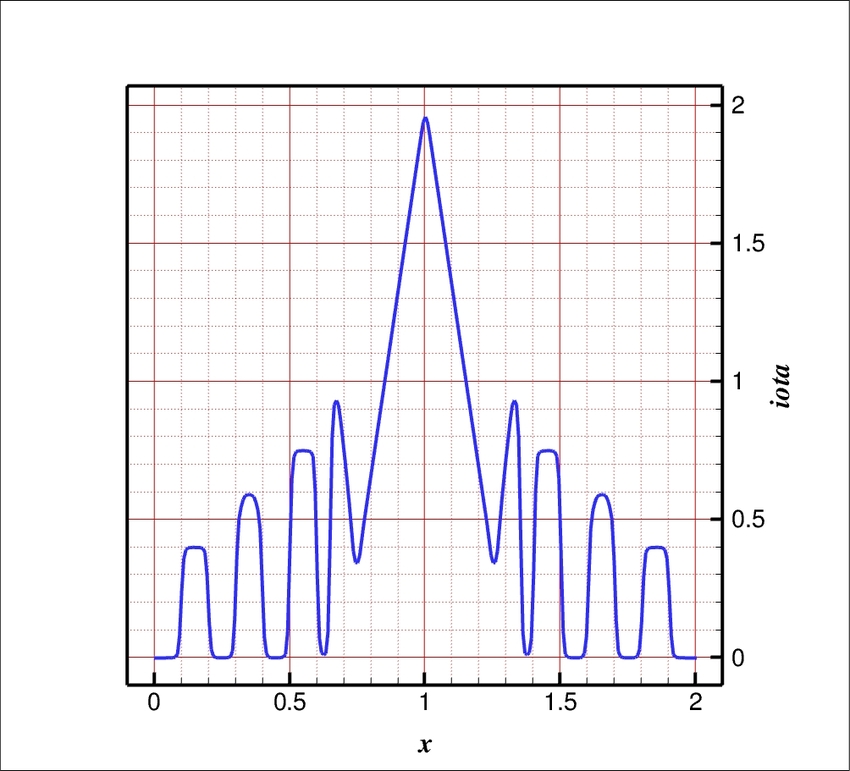} \\ \centering\includegraphics[width=7.3cm]{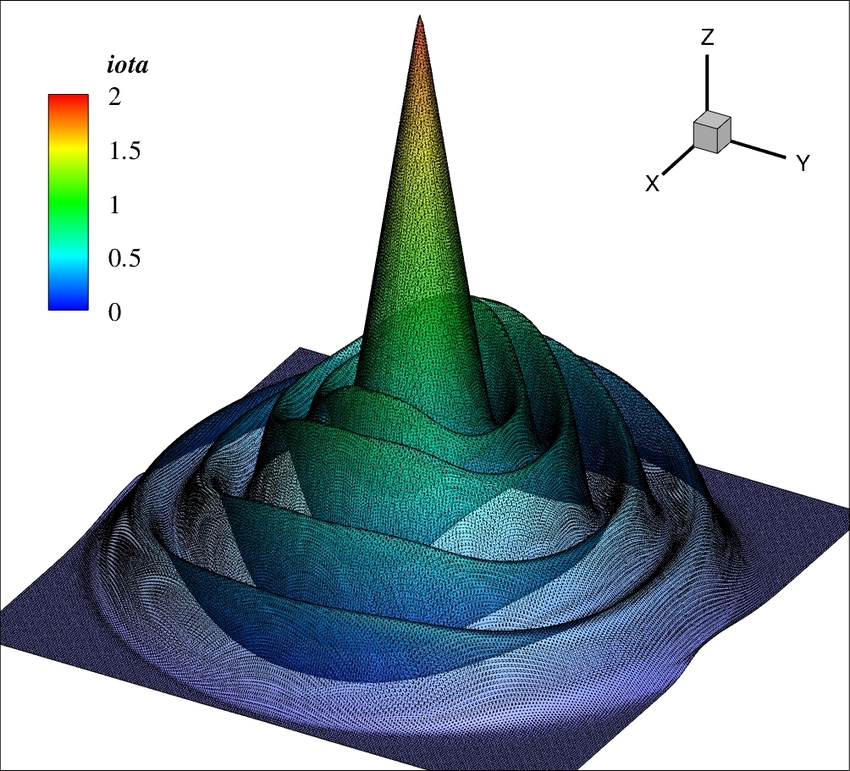}\includegraphics[width=7.3cm]{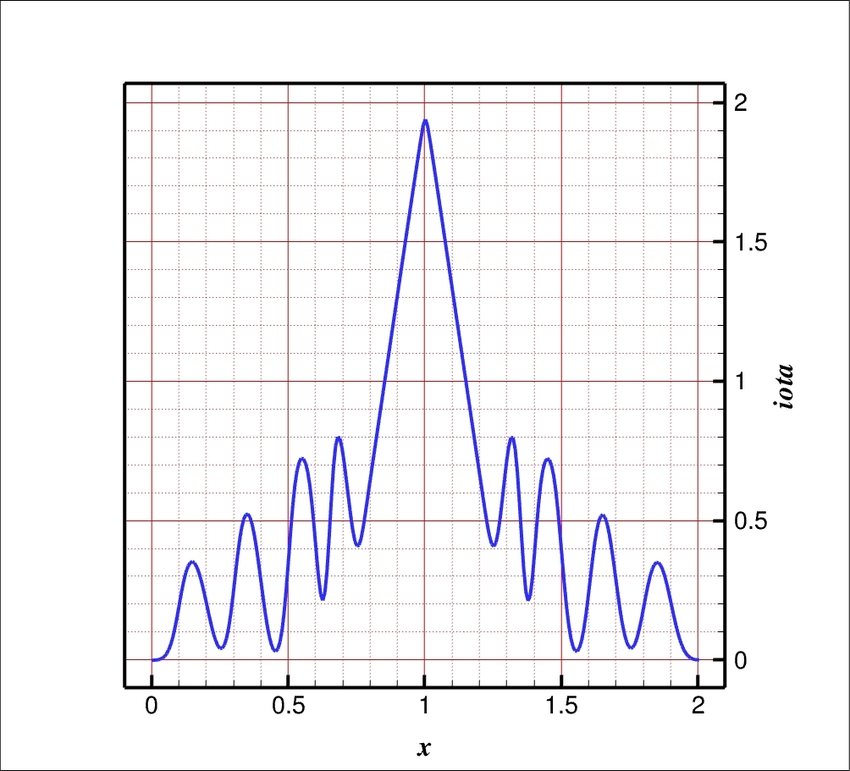} \\ \centering\includegraphics[width=7.3cm]{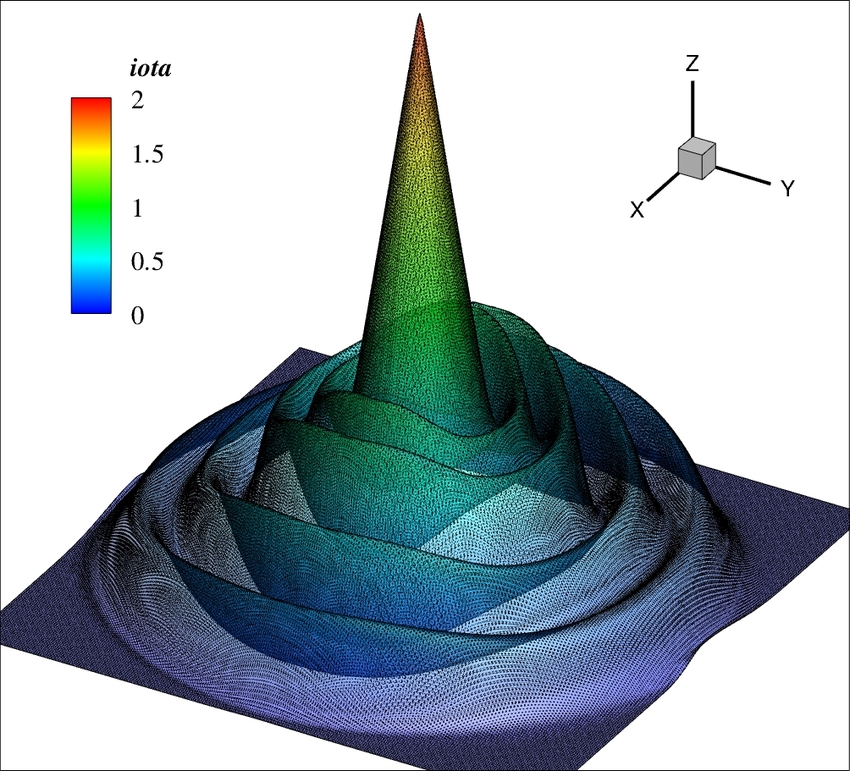}\includegraphics[width=7.3cm]{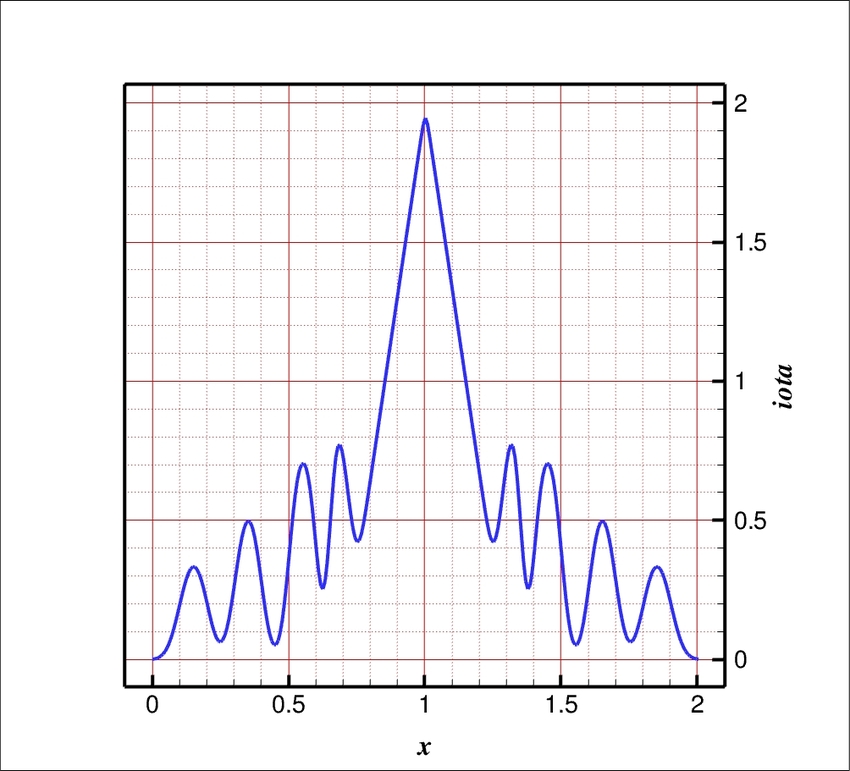} \caption{At top we show the $p=4$  linear restriction after $1/4$ turn.  The middle shows the $p=3$ linear reconstruction\textsuperscript{ \cite{Abgrall2,LSTZ},\textsection{3.3}}, and the bottom the $p=2$ linear recombination\textsuperscript{\textsection{3.5}}.}\label{fig:ODE22}  \end{figure}

Note that in Figure \ref{fig:ODE21} and Figure \ref{fig:ODE22}, the asymmetry in the solution is merely due to that fact that we have only gone a quarter turn, thus the diffusive signature of each limiter has only been advected a quarter turn, and accumulates or dissipates according to the local behavior of the advective flux.

Now, notice that the adapted limiters from \textsection{3.2.1} are not well--suited to handle (\ref{convt}) at all.  In fact (\ref{alphabigadapted}) is, in particular, adapted to represent a case which almost always leads to problems, since it does not deal differentially with the special case of $U^{\max}_{i,b}=U^{\min}_{i,b}$, which in (\ref{alphabigadapted}) up to the resolution $h$ is the case for nearly every element in the domain, leading to an almost globally uniform ``blind diffusion.''  In fact the adapted cases are almost identical to the native cases at low $p$ --- when not explicitly dealing with $U^{\max}_{i,b}=U^{\min}_{i,b}$ --- even though the native vertex and Barth--Jespersen limiters do not recognize local extrema at all, while the adapted cases do recognize local extrema up to, but not including, the degenerate case of  $U^{\max}_{i,b}=U^{\min}_{i,b}$.  As $p$ increases the repeated iterations of the limiter swamps this behavior in both the native and adapted limiters, and thus the solutions converge to the same value.  It is possible that a mass lumping strategy might mitigate some of these affects (see \cite{Kuzmin} for more information on this technique).

Moreover,  in this example (\ref{convt}) the Barth--Jespersen limiter is clearly initially more diffuse than the native vertex limiter, which is primarily due here to the fact that the singular submanifolds are chosen such that they --- again  up to the mesh resolution $h$ --- spatially oscillate on a local neighborhood which is larger than the characteristic length of the \emph{edge neighborhood}, and so the \emph{focal neighborhood} is a more appropriate area to ``sense'' in order to capture this semi-localized signature behavior.  Moreover, the problem of local extrema as discussed in \textsection{3.2.1} is of lesser importance in this case, since up to the set of codimension one submanifolds of $\Omega_{h}$ in  (\ref{convt}), the entire domain is characterized and dominated by extremely sharp profiles, making the diffusion --- which is potentially ``blind'' near smooth regions --- more appropriate here.  However, again as $p$ increases this behavior gets swamped  by the repeated iterations.

The linear restriction of the BDS limiter\textsuperscript{\cite{Bell},\textsection{3.4}} once again demonstrates the best limiting behavior as a function of increasing $p$, which again seems to emphasize the fact that limiting a solution for $p>1$ must somehow account for the implicit nonlinearity present internal to the cell in a relatively explicit way; or, at least, a way which is fully functionally coupled to the entire solution as it exists everywhere on the local cell.

Nevertheless, the linear restriction still substantially outperforms all of the competing limiting regimes.  There seems to be some indication here that, at least presently, one may expect that near areas dominated by shocks the best accuracy that one can hope for is linear accuracy, while still hoping to preserve physically important characteristics of the solution (\emph{e.g.} positivity perserving, local conservation of mass, \emph{etc.}).  Of interest, is that this observation falls very neatly in line with the state of the art in $hp$--adaptive numerical schemes, where a general heuristic follows that for potentially discontinuous solutions, in areas of high cell--wise variability, the local order of $p$ is only increased if inter--element jumps are small or bounded and controlled, and the internal cell-wise variation is strictly bounded above by the cell(s) (usually a subset of cells) containing the global maximum \cite{Dem,MES}.   

We explore this issue some in the subsequent section as it applies to $p$--enrichment, though we also note that at present we are not aware of any formal results which come anywhere near to formulating a theorem that subsumes this observational fact (which may in general prove to be only one part of the story).  Nevertheless, such a result would be of substantial importance to the field, as would a counter example, which here could simply be the development of a fully $p$ convergent slope limiting regime that limits at all \emph{levels} $\mathfrak{l}$ while still preserving the important physical features of the solution (and of course does so without relying on prior knowledge, such as the existence of an exact solution).

\section{\texorpdfstring{\protect\centering $\S 5$  Adjoining the dynamic $p$-enrichment}{\S 5 Adjoining the dynamic $p$-enrichment}}

Here we present a number of generalizable $p$-enrichment/de--enrichment schemes based on local data and apply them to the problems from \textsection{4}. These $p$-enrichment schemes may be viewed as alternatives to, for example, the specific energy methods presented in \cite{MES} which rely upon the variational global entropy of the system of equations, and those discussed in \cite{demk2} and \cite{Bangerth,Kanschat}, which, as in \cite{MES}, try to maximally enrich the domain based on global solution behavior taken with respect to the available computational resources and either \emph{a priori} or \emph{a posteriori} estimates. 

\subsection{A general approach based on local data}
\label{A general approach based on local data}

We implement a dynamic $p$-enrichment scheme that utilizes a number of different methodologies in order to capture higher order structure in areas of ``permissible variability.''  This scheme is built with respect to our collection of $p$-adaptive slope limiters from \textsection{3}, such that we inherently arrive with a dynamically limited $p$-enriched solution.  

The nuance of implementing such a scheme in the generalized formulation is that the solution must demonstrate a minimal smoothness condition in areas of $p$-enrichment, while in areas approaching discontinuity, $p$-enrichment must be suppressed in order to maintain stability (especially in the absence of a limiter).  This issue is not a concern of course when one is able to make smoothness assumptions \emph{a priori} about the entire solution space over $\Omega\times (0,T)$ (\emph{viz}. the formalism of \cite{BurbeauS} and \cite{KubBDW}), and has been shown to demonstrate very nice behavior especially in solution spaces which are not only smooth, but where in particular one would like to resolve stable areas of maximal variation (\emph{e.g.} as are applicable in some storm surge model applications \cite{KubBDW}).  

Nevertheless, in the context of a slightly more generalized system of equations with, for example, a coupled hyperbolic equation (or possessing a hyperbolic character in a system of equations) such as (\ref{rota}), such assumptions cannot generally be made over the entire discrete solution space over $\Omega_{h}\times(0,T)$, since areas demonstrating strong local gradients $\nabla_{x}\boldsymbol{U}_{h}$ may indicate the presence or formation of numeric shock fronts (even given smooth initial data), in which case local $p$-enrichment has a destabilizing effect on the solution (that is, the weak approximation to a discontinuity becomes more ill-behaved with respect to increasing $p$).  

Here we are concerned with dynamically $p$-adapted solutions to the generalized formulation of (\ref{aprox}) and (\ref{SSPRK}) in conjunction with the slope limiters presented in \textsection{3}.  We implement a very simple set of $p$--enrichment strategies, which as we will see, generally tend to undersample the variational space (\emph{e.g.} in contrast to, for example, the \emph{poor man's} or \emph{poor man's greedy} algorithm of \cite{demk2} which always adapts based on some percentage of a global relative bound).   The reason for this simplification here is to reduce the number of varying parameters in the scheme, in order to isolate the stability of the solution with respect to the limiting schemes of \textsection{3}.  Hence, we simply set hard tolerances which do not depend on, for example, the available computational resources or global bounds on the solution.

Now, in order to additionally deal with both smooth and discontinuous initial--boundary data (as well as smooth and discontinuous solutions in $(0,T)$) we implement the following two distinct dynamic $p$-enrichment schemes --- namely we designate them: Type I and Type II $p$-enrichment schemes. We also note that in this section all functions are defined with respect to the master element $\mathcal{M}$ representation.

The first type of enrichment scheme  (\emph{i.e.} Type I ) applies to solutions in which smoothness may be assumed \emph{a priori} over the entire domain $\Omega\times(0,T)$.   That is, taking the approximate solution vector $\boldsymbol{U}_{h}$ we compute the auxiliary sensor over each $i$-th component of of the state variable $\boldsymbol{U}$ (having $m$ components, as in \textsection{2}):  \begin{equation}\label{smooth}\Pi^{i}_{j}=\bigg|\frac{\boldsymbol{U}_{h}^{i}|_{\omega_{j}} - \boldsymbol{U}_{h}^{i}|_{c}}{\chi_{j}}\bigg|,\end{equation} where $c$ is the centroid of element $\Omega_{e}$ and $\omega_{j}$ is the midpoint of the $j$--th edge of $\Omega_{e}$, and the solution $\boldsymbol{U}_{h}$ is evaluated at these two points, respectively. For smooth solutions, the function $\chi_{j}$ may be set to either the distance $\chi_{j} = |\omega_{j} - c|$ as in \cite{KubBDW}, or the product $\chi_{j} = \omega_{j}c$ as in \cite{BurbeauS}. In either case, over each timestep $n$ the following $p$-enrichment functional $\mathfrak{E}_{e_{l}}= \mathfrak{E}_{e_{l}}(\mathscr{P}^{k}(\Omega_{e_{l}}^{n}))$ is evaluated over each cell $\Omega_{e_{l}}$:  \begin{center}\setlength{\fboxsep}{15pt}\setlength{\shadowsize}{2pt}\shadowbox{\begin{minipage}{5.5in}\begin{center}\underline{\bf  Type I $p$-enrichment}\end{center} \begin{equation}\label{disc}\mathfrak{E}_{e_{l}} = \left\{\begin{matrix}  \mathscr{P}^{k+1}(\Omega_{e_{l}}^{n}) & \mathrm{if} \ \left((\sup_{i}\sup_{j}\Pi^{i}_{j} \geq \epsilon) \land (k+1\leq p_{\max})\right) \lor \left( \tau_{0}\geq t^{w} \right), \\  \mathscr{P}^{k-1}(\Omega_{e_{l}}^{n}) & \mathrm{if} \ (\inf_{i}\sup_{j}\Pi^{i}_{j} < \epsilon) \land (k-1 \geq p_{\min})\land (\tau_{0}\geq t^{w}),  \\  \mathscr{P}^{k}(\Omega_{e_{l}}^{n}) & \ \mathrm{otherwise,} \end{matrix}\right.\end{equation}\end{minipage}}\end{center} where $\tau_{0}$ is a counter that restricts the $p$ enrichment/de-enrichment such that it may only occur every $t^{w}$ timesteps, and where $k\in\{1,\ldots,p\}$.

For solutions demonstrating approximately nonzero local approximate gradients $\nabla_{x}\boldsymbol{U}_{h}\neq 0$, wherein we might expect local discontinuities we must find an estimate of the local relative ``smoothness'' of $\boldsymbol{U}_{h}$.  One way of doing this is by setting the auxiliary sensor equal to the following Van Leer minmod function across elements (as used in \cite{BurbeauS}): \begin{equation}\label{minmod3}\Pi^{i}_{j}=\mathrm{minmod}(\boldsymbol{U}^{i}_{h}|_{v_{j}^{+}}-\boldsymbol{U}^{i}_{h}|_{c},\boldsymbol{U}^{i}_{h}|_{v_{j}^{-}}-\boldsymbol{U}^{i}_{h}|_{c}), \end{equation} where $v_{j}$ is the $j$--th vertex of $\Omega_{e_{l}}$.  As $\Pi_{i}^{j}\to 0$ the solution becomes smoother, and one may subsequently employ (\ref{disc}).

A slightly simpler method of dealing with discontinuous solutions simply using local information is to define a local smoothness estimator (as discussed in  \cite{WanMa} and \cite{PMH}) such that we again may calculate an elementwise version of (\ref{smooth}) depending only on the the interior of $\Omega_{e_{l}}$, such that: \begin{equation}\label{smoothdisc}\Pi_{i}^{e_{l}} = \left(\frac{\|\boldsymbol{U}^{i}_{h}-\breve{\boldsymbol{U}}^{i}_{h}\|_{L^{q}(\Omega_{e_{l}})}}{\|\boldsymbol{U}^{i}_{h}\|_{L^{q}(\Omega_{e_{l}})}}\right),\end{equation} for the $L^{q}$ norms (except when $q=2$ in which case we take the standard inner product, as used in our examples below), where $\breve{\boldsymbol{U}}_{h}$ is the elementwise projected solution  $\mathscr{P}^{k-1}(\Omega_{e_{l}}^{n})$, such that in our mixed version (\ref{disc}) becomes:  \begin{center}\setlength{\fboxsep}{15pt}\setlength{\shadowsize}{2pt}\shadowbox{\begin{minipage}{5.5in}\begin{center}\underline{\bf  Type II $p$-enrichment}\end{center} \begin{equation}\label{disc2}\mathfrak{E}_{e_{l}} = \left\{\begin{matrix}  \mathscr{P}^{k+1}(\Omega_{e_{l}}^{n}) & \mathrm{if} \ (\sup_{i}\log_{10}\Pi^{e_{l}}_{i}\leq A) \land (k+1\leq p_{\max}), \\  \mathscr{P}^{k-1}(\Omega_{e_{l}}^{n}) & \mathrm{if} \ (\inf_{i}\log_{10}\Pi^{e_{l}}_{i} \geq  A) \land (k-1 \geq p_{\min})\land (\tau_{0}\geq t^{w}),  \\  \mathscr{P}^{k}(\Omega_{e_{l}}^{n}) & \ \mathrm{otherwise,} \end{matrix}\right.\end{equation}\end{minipage}}\end{center}  where the bound satisfies \begin{equation} A = \left\{ \begin{matrix} \log_{10}\tilde{c}k^{-q^{2}}+c, & \mathrm{for} \ p> p_{\min} \\ \sup_{i}\log_{10}\Pi^{e_{l}}_{i}, & \mathrm{otherwise} \end{matrix}\right.\end{equation} such that $\tilde{c},c\in\mathbb{R}^{+}$ are user defined constants, where $\tilde{c}\in(0,10)$ is recommended (see for example \cite{WanMa}) for resolving discontinuities in the context of $hp$-adaptivity, and where we have found $c\in (-2,2)$ optimal.  The basic intuition that underpins the use of (\ref{smoothdisc}) is the observation that discontinuous basis functions are assumed to decay, for smooth solutions, at a rate comparable to that of the Fourier coefficients in a standard expansion of the solution --- which clearly decay at a rate of  $1/k^{4}$ for $q=2$ (see \cite{WanMa,PMH,PPe}), to which we obtain an indicator of the relative local regularity of the solution, \emph{i.e.} the faster the coefficients decay, the smoother the local solution.  Thus we obtain equation (\ref{smoothdisc}), which approaches zero as the solution becomes smoother, where setting $c>0$ is a sharper restriction than the more permissive (\emph{i.e.} less stable) condition $c<0$.  

\begin{figure}[t!]
\centering\includegraphics[width=8cm]{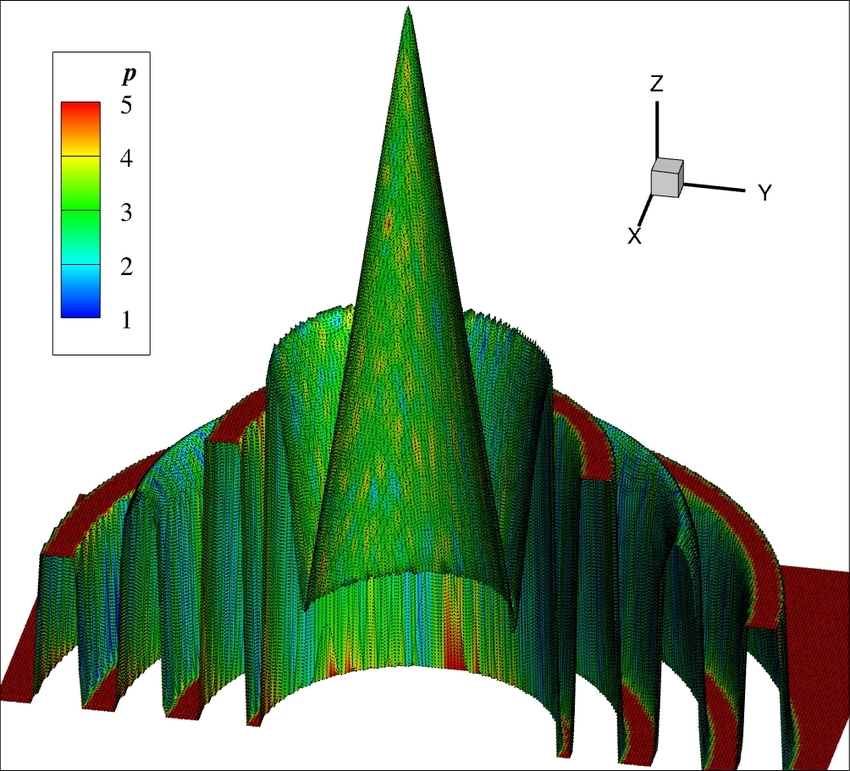}\includegraphics[width=8cm]{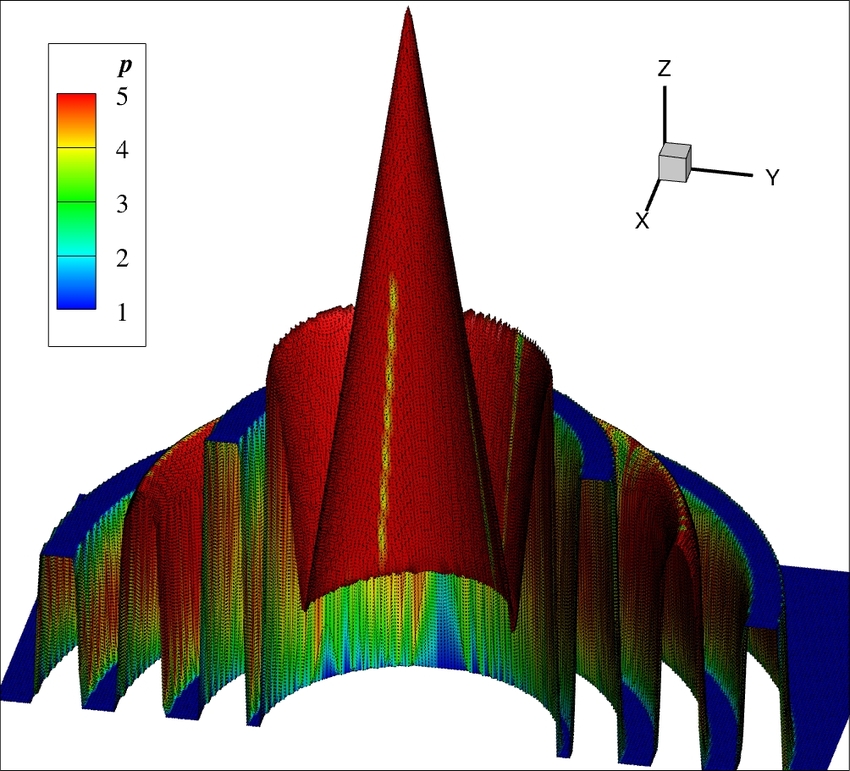} \\ \centering\includegraphics[width=8cm]{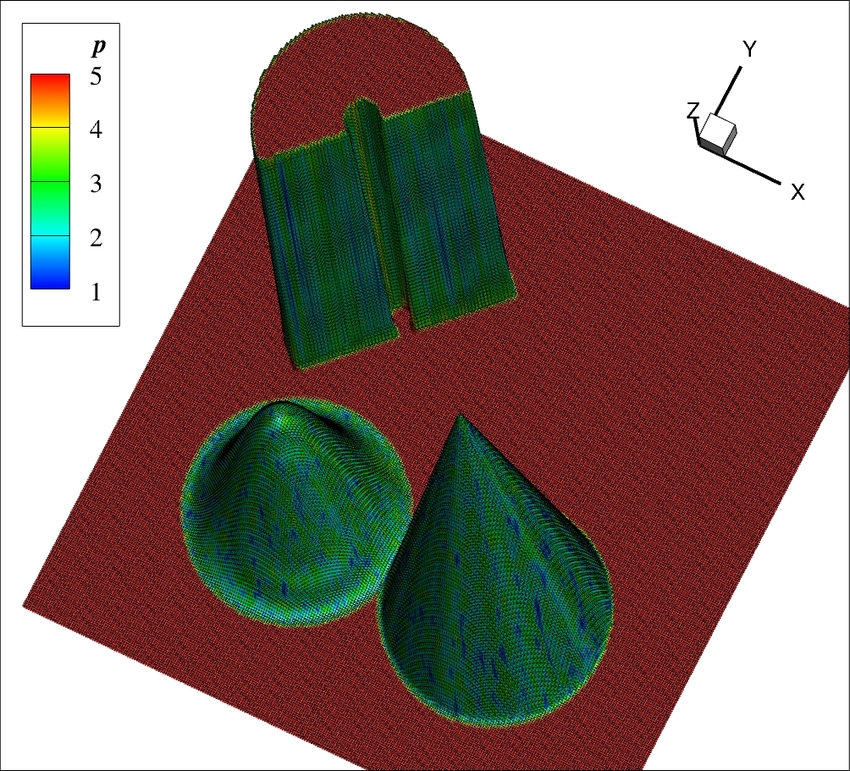}\includegraphics[width=8cm]{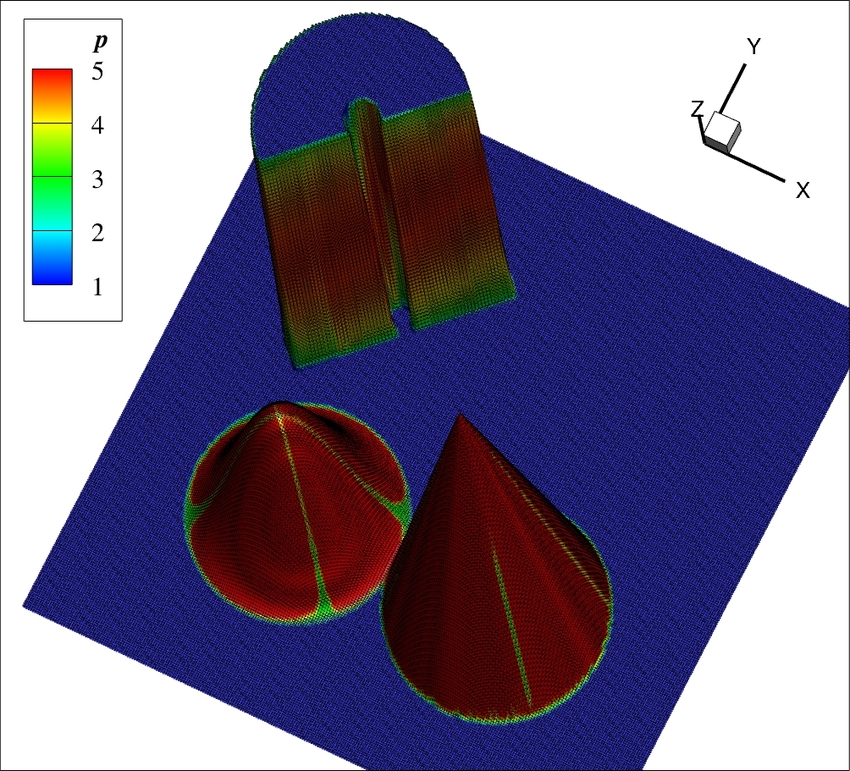}  \caption{Here we show the $p$ values mapped over the $p=1$, $L^2$-projected profiles at $T=0.003$ on the top solutions, and at $T=0.06$ on the bottom solutions using the settings given in Table \ref{table:errortab4}.  The solutions on the left use the Type II $p$-enrichment, and those on the right use the Type I $p$-enrichment.}\label{fig:pad1}  \end{figure}

\begin{table}[t!]
\centering 
\begin{tabular}{|c | c | c | c | c | c | c | c | c |}
%\toprule[0.1em]
\hline
$p$ &  \bf{Limiter type} &  Type I, $L^{2}$ &  Type II, $L^{2}$ & $\epsilon$ & $c$, & $\tilde{c}$ & $t^{w}$ & $q$ \rule{0pt}{3ex} \rule[0ex]{0pt}{0pt} \\ 
\hline\hline
1--5 & BJ limiter\textsuperscript{\cite{BarthJesp},\textsection{3.2}}  & $3.66\times 10^{-3}$ &   $2.18\times 10^{-3}$ & 0.1 & -1 & 0.1 & 0 & 2  \rule{0pt}{3ex} \rule[0ex]{0pt}{0pt}  \\
\hline
1--5 &  Vertex\textsuperscript{\cite{Kuzmin,LBL},\textsection{3.2}} & $3.10\times 10^{-3}$  &  $2.10\times 10^{-3}$ & 0.1 & -1 & 0.1 & 0 & 2 \rule{0pt}{3ex} \rule[0ex]{0pt}{0pt}\\
\hline 
1--5 & Restriction \textsuperscript{\cite{Bell},\textsection{3.4}} & X &  $8.03\times 10^{-4}$ & 0.1 & -1 & 0.1 & 0 & 2 \rule{0pt}{3ex} \rule[0ex]{0pt}{0pt}  \\
\hline
1--5 &  Recombination\textsuperscript{\textsection{3.5}} &  $2.17\times 10^{-3}$  & $2.09\times 10^{-3}$ & 0.1 & -1 & 0.1 & 0 & 2 \rule{0pt}{3ex} \rule[0ex]{0pt}{0pt} \\
\hline
1--5 &  Reconstruction\textsubscript{ENO}\textsuperscript{ \cite{Abgrall2,LSTZ},\textsection{3.3}} & $1.98\times 10^{-3}$ & $1.91\times 10^{-3}$ & 0.1 & -1 & 0.1 & 0 & 2 \rule{0pt}{3ex} \rule[0ex]{0pt}{0pt} \\
\hline 
\end{tabular}

\caption{We give the $L^{2}$-errors of the approximate solutions after $T$ corresponding to a $1/4$ rotation with $p$-enrichment on (\ref{convt}), setting $h=1/128$, $\Delta t = 5\times 10^{-4}$ and using Runge--Kutta SSP$(5,3)$.}
\label{table:errortab4}
\end{table}

The results are shown in Table \ref{table:errortab4} and Figure \ref{fig:pad1}.  As expected from before, the linear restriction from \textsection{3.4} is again by far the most accurate of the choice of limiters when it is stable, where it is important to note that in the $p$-enrichment case the restriction function $\mathfrak{R}$ from \textsection{3.4} is calculated using $\varepsilon =10^{-4}$, which has the effect of passing cells containing steep gradients to the dynamic $p$-enrichment functions $\mathfrak{E}$.  This is enough, it turns out, to make the Type I $p$-enrichment regime unstable with respect to the dynamically adaptive linear restriction limiting regime from \textsection{3.3} due in part to the function of $\mathfrak{R}$, which creates an unstable $p$-flickering along sharp profile edges.  However, even turning off the $p$-de-enriching functionality of $\mathfrak{R}$ does not help in this case, since the linear restriction still zeros out the higher order components, which in the Type I case effectively still allows $p$ to flicker locally, leading to the formation of instabilities along sharp edges.  We also note that in Table \ref{table:errortab4} we have suppressed the $L^{\infty}$-error, as numerical experimentation suggests that very small changes in the $p$-enrichment settings $\epsilon,c,\tilde{c},$ and $t^{w}$ can cause big shifts in $L_{loc}^{\infty}$, which make the $L^{\infty}$-error a deceptive measure in the discontinuous $p$-enrichment case.  

Finally, we emphasize that the $p$-enriched slope limited solutions show substantially better accuracy than the constant-in-$p$ solutions from \textsection{4.2}.  This can be attributed in large part to the observation that the majority of error in the solutions is accumulated along the discontinuities, which is precisely where the $p$-enrichement schemes $p$-transition between \emph{levels} (see Figure \ref{fig:pad1}).  Hence, in both Type I and Type II cases (\emph{i.e.} spatially, from either side of the discontinuity) the $p$-enrichement strongly attenuates (by explicit truncation) the oscillatory instabilities present in these regions, as long as the solution does not flicker unstably between them.

\section{\texorpdfstring{\protect\centering $\S 6$ Conclusion}{\S 6 Conclusion}}

We have presented a discontinuous Galerkin finite element method for solving dynamically $p$-enriched solutions with consistent slope limiting to arbitrary order in two spatial dimensions over generalized coupled systems of PDEs.  We have provided a formalism for transforming between the polynomial basis of different regimes in order to move between representation spaces.  We then introduced, up to but not including a substantial choice of minmod functions, seven dynamic-in-$p$ slope limiting regimes, and performed numerical experiments on these regimes in order to develop a sense of their strengths and weaknesses.  We found that our numerical results suggest that, given discontinuous initial data, slope limiting over fixed order solutions when $p>1$ is most effectively accomplished by restricting back to the linear case and using a sharp limiter in that regime, rather than keeping the higher order data and trying to limit it in a consistent way --- which we found introduces more numerical diffusion (\emph{i.e.} error) on average over time.

We then presented two types of $p$-enrichment schemes, fully coupled to the above slope limiting regimes.  These schemes are designed to exploit certain properties of the solution, and simple algorithms were implemented.  We then tested these coupled systems on the same model problem in order to develop a sense of how dynamic-in-$p$ systems perform relative to fixed-in-$p$ systems.  Here again, we found that restricting to the linear case seems to be the most effective (and also, incidentally, efficient) way of limiting a dynamically $p$-adapting solution.  Moreover, we found that in general using the Type I and Type II methods of $p$-enrichment the accuracy of the solution was substantially improved (\emph{i.e.} by an order of magnitude) with respect to the native solution using only the dynamic-in-$p$ slope limiters of \textsection{3}.

Future directions include taking the slope limited solution from \textsection{3} coupled to the $p$-enrichment scheme from \textsection{5} and adding dynamic $h$-adaptivity to it, in order to fully exploit the power of $hp$-adaptive convergence.

\section{\texorpdfstring{\protect\centering $\S 7$ Acknowledgements}{\S 7 Acknowledgements}}

The first author would like to thank P.G.~Schmitz, Wenhao Wang, Troy Butler, Corey Trahan, Nishant Panda and Jennifer Proft for helpful conversations.  The authors would also like to aknowledge the support of the National Science Foundation grants OCI-0749075 and OCI-0746232.

{\setlength\parskip{0pt} 

\bibliographystyle{plainnat}
\def\cprime{$'$} \def\cprime{$'$}
  \def\polhk#1{\setbox0=\hbox{#1}{\ooalign{\hidewidth
  \lower1.5ex\hbox{`}\hidewidth\crcr\unhbox0}}}
  \def\polhk#1{\setbox0=\hbox{#1}{\ooalign{\hidewidth
  \lower1.5ex\hbox{`}\hidewidth\crcr\unhbox0}}}
  \def\polhk#1{\setbox0=\hbox{#1}{\ooalign{\hidewidth
  \lower1.5ex\hbox{`}\hidewidth\crcr\unhbox0}}} \def\cprime{$'$}
  \def\cprime{$'$} \def\polhk#1{\setbox0=\hbox{#1}{\ooalign{\hidewidth
  \lower1.5ex\hbox{`}\hidewidth\crcr\unhbox0}}}
  \def\polhk#1{\setbox0=\hbox{#1}{\ooalign{\hidewidth
  \lower1.5ex\hbox{`}\hidewidth\crcr\unhbox0}}}
  \def\polhk#1{\setbox0=\hbox{#1}{\ooalign{\hidewidth
  \lower1.5ex\hbox{`}\hidewidth\crcr\unhbox0}}} \def\cprime{$'$}
  \def\cprime{$'$} \def\polhk#1{\setbox0=\hbox{#1}{\ooalign{\hidewidth
  \lower1.5ex\hbox{`}\hidewidth\crcr\unhbox0}}}
  \def\polhk#1{\setbox0=\hbox{#1}{\ooalign{\hidewidth
  \lower1.5ex\hbox{`}\hidewidth\crcr\unhbox0}}}
  \def\polhk#1{\setbox0=\hbox{#1}{\ooalign{\hidewidth
  \lower1.5ex\hbox{`}\hidewidth\crcr\unhbox0}}} \def\cprime{$'$}
  \def\cprime{$'$} \def\polhk#1{\setbox0=\hbox{#1}{\ooalign{\hidewidth
  \lower1.5ex\hbox{`}\hidewidth\crcr\unhbox0}}}
  \def\polhk#1{\setbox0=\hbox{#1}{\ooalign{\hidewidth
  \lower1.5ex\hbox{`}\hidewidth\crcr\unhbox0}}}
  \def\polhk#1{\setbox0=\hbox{#1}{\ooalign{\hidewidth
  \lower1.5ex\hbox{`}\hidewidth\crcr\unhbox0}}}

}

\end{document}